\documentclass[prf,superscriptaddress,showkeys,longbibliography]{revtex4-1}

\usepackage{amsmath,amssymb}
\usepackage[titletoc]{appendix}
\usepackage{todonotes}
\usepackage{graphicx}
\usepackage{dcolumn}
\usepackage{bm}
\usepackage{stmaryrd}

\renewcommand{\AA}{\mathcal{A}}
\newcommand{\BB}{\mathcal{B}}

\newcommand{\ff}{\mathbf{f}}
\newcommand{\nn}{\mathbf{n}}
\renewcommand{\tt}{\mathbf{t}}

\newcommand{\RR}{\mathbb{R}}

\renewcommand{\SS}{\mathcal{S}}
\newcommand{\TT}{\mathcal{T}}
\newcommand{\uu}{\mathbf{u}}

\newcommand{\xx}{\mathbf{x}}
\newcommand{\xxi}{\boldsymbol{\xi}}
\newcommand{\yy}{\mathbf{y}}

\newcommand{\subfigimg}[3][,]{%
  \setbox1=\hbox{\includegraphics[#1]{#3}}
  \leavevmode\rlap{\usebox1}
  \rlap{\hspace*{0pt}\raisebox{\dimexpr\ht1-0\baselineskip}{\bf
  \normalsize #2}}
  \phantom{\usebox1}
}

\begin{document}

\preprint{APS/123-QED}

\title{Hydrodynamics and rheology of a vesicle doublet suspension}

\author{Bryan Quaife}
\affiliation{Department of Scientific Computing, Florida State University, Tallahassee, Florida, USA}
\author{Shravan Veerapaneni}%
\affiliation{Department of Mathematics, University of Michigan, Ann Arbor, Michigan, USA}%
\author{Y.-N.~Young}%
 \email[Corresponding author: ]{yyoung@njit.edu}
\affiliation{Department of Mathematical Sciences, Newark, NJ, USA}%

\date{\today}

\begin{abstract}
The dynamics of an adhesive two-dimensional vesicle doublet under
various flow conditions is investigated numerically using a high-order,
adaptive-in-time boundary integral method. In a quiescent flow, two
nearby vesicles move slowly towards each other under the adhesive
potential, pushing out fluid between them to form a vesicle doublet at
equilibrium. A lubrication analysis on such draining of a thin film
gives the dependencies of draining time on adhesion strength and
separation distance that are in good agreement with numerical results.
In a planar extensional flow we find a stable vesicle doublet forms only
when two vesicles collide head-on around the stagnation point.  In a
microfluid trap where the stagnation of an extensional flow is
dynamically placed in the middle of a vesicle doublet through an active
control loop, novel dynamics of a vesicle doublet are observed.
Numerical simulations show that there exists a critical extensional flow
rate above which adhesive interaction is overcome by the diverging
stream, thus providing a simple method to measure the adhesion strength
between two vesicle membranes. 
In a planar shear flow, numerical simulations reveal that a vesicle
doublet may form provided that the adhesion strength is sufficiently
large at a given vesicle reduced area.  Once a doublet is formed, its
oscillatory dynamics is found to depend on the adhesion strength and
their reduced area.  Furthermore the effective shear viscosity of a
dilute suspension of vesicle doublets is found to be a function of the
reduced area.  Results from these numerical studies and analysis shed
light on the hydrodynamic and rheological consequences of adhesive
interactions between vesicles in a viscous fluid.
\end{abstract}

\pacs{87.16.D-,83.80.Lz,83.60.-a,47.11.Hj}
\keywords{Vesicle hydrodynamics, membrane adhesion, drop and bubble phenomena/drop interactions, interfacial flows/thin fluid films, rheology/flow confinement, biological fluid dynamics/collective behavior/clustering, emergence of patterns}
\maketitle



\section{Introduction}
Vesicles (closed fluid-filled phospholipid bilayer membranes) have been
widely utilized as biological cell mimics in biophysics and material
engineering~\cite{sackmann1996, FenzSengupta2012_IntegrBiol,
Barthes-Biesel2016_ARFM}.  The ever expanding applications of vesicles
have encouraged detailed experimental, theoretical, and numerical
investigations of vesicle dynamics in applied flow, electric fields, and
magnetic fields.  Experimental investigations of vesicles have uncovered
vesicle properties and various novel
applications~\cite{Dobereiner2000_CurrentOpinionCIS,
EvansRawiczSmith2013_FaradayDiscussions, SugiyamaToyota2018_Life}.
Theoretical investigations are often limited to small-deformation
analysis on a nearly spherical vesicle or spheroidal analysis on a
spheroidal vesicle~\cite{Barthes-BieselRallison1981_JFM, Misbah2006_PRL,
Vlahovska2007_PRE, Finken2008_EPL, ZhangZahnTanLin2013_PoF,
Nganguia2013_PRE}.  On the other hand, various numerical methods have
been developed for simulating the transient hydrodynamics of vesicle
suspensions~\cite{BagchiJohoson2005_JBE, Biben2005_EJP,
Veerapaneni2009_JCP, SeolHuKimLai2016_JCP}.  

Hydrodynamics of a single vesicle in Stokes flow has been extensively
investigated. In a planar shear flow, the vesicle hydrodynamics is
characterized by the reduced volume (reduced area in two dimensions),
viscosity contrast between interior and exterior fluids, and shear rate
of the imposed far-field fluid flow. In addition, a vesicle with a rigid
particle inside is also investigated as a biological mimic of a
eukaryotic cell with a nucleus that occupies nearly half of the
intracellular volume~\cite{Veerapaneni2011_PRL}. Small-deformation
analysis shows that a vesicle tank-treads in a planar shear flow for low
viscosity contrast and shear rate. At high viscosity contrast the
tank-treading dynamics transitions to tumbling
dynamics~\cite{Misbah2006_PRL, Vlahovska2007_PRE}, and this leads to a
transition in the effective shear viscosity of the vesicle
suspension~\cite{Misbah2006_PRL,Vitkova2008_BJ} that is also validated
by direct numerical simulations~\cite{GhigliottiBibenMisbah2010_JFM} and
experiments~\cite{Vitkova2008_BJ, DeschampsKantsler2009_PNAS,
KantslerSegreSteinberg2008_EPL, ZabuskySegreDeschamps2011_PoF}.  Between
tank-treading and tumbling vesicle dynamics, a breathing (tremble) mode
is also observed~\cite{Misbah2006_PRL, KantslerSegreSteinberg2008_PRL,
ZhaoShaqfeh2011_JFM, SpannZhaoShaqfeh2014_PoF} to alter the effective
shear viscosity.  In an extensional flow (planar or uniaxial), vesicle
shape dynamics depends sensitively on the vesicle reduced volume and the
elastic capillary number~\cite{KantslerSegreSteinberg2008_PRL,
ZhaoShaqfeh2013_JFM, Narsimhan2014_JFM,
DahlNarsimhanGouveia2016_SoftMatt}: Asymmetric shape and oscillatory
undulation of the vesicle membrane are two examples of the complex
vesicle hydrodynamics in an extensional flow.

Collective hydrodynamics of vesicles is dictated by the vesicle-vesicle
interactions.  In a quiescent flow, vesicle-vesicle adhesion leads to
the formation of vesicle doublets~\cite{Ziherl2007_PRL,
ZiherlSvetina2007_PNAS} or
clusters~\cite{SvetinaZiherl2008_Bioelectrochemistry,
GuWangGunzburger2016_MathBiol, FlormannAouane2017_SciReports}.  As a
model for red blood cell (RBC) aggregates, a simplified model for
adhesive vesicle-vesicle interactions can reproduce vesicle shapes
similar to those observed in experiments of fibrinogen-induced RBC
aggregates~\cite{SvetinaZiherl2008_Bioelectrochemistry,
GuWangGunzburger2016_MathBiol, FlormannAouane2017_SciReports,
HooreYayaPodgorski2018_SoftMatt,
bru-aou-thi-flo-ver-Kae-las-sel-ben-pod-cou-mis-wag2014,
cla-aou-thi-abk-cou-mis-joh-wag2017}.  Using the Lennard-Jones
(L.-J.)~potential for point-point interaction between two vesicle
membranes without the molecular details of adhesive interactions between
RBCs, Flormann {\em et al.}~\cite{FlormannAouane2017_SciReports} found
that, under strong adhesion, the membrane may buckle to a sigmoidal
shape in the contact region.  Such a sigmoidal vesicle shape is also
observed in RBC doublets and explained in a slightly different
model~\cite{ZiherlSvetina2007_PNAS}.  It remains unknown how adhesive
interactions between vesicles might lead to different vesicle
hydrodynamics that results in different rheological properties of a
vesicle suspension. Such studies will provide useful insight into the
rheology of blood flow~\cite{chi-usa-del-gre-nan-gue1967} and how to use
nano solutes to control the hypertension by adjusting the adhesive
interactions between RBCs.

In numerical simulations of a vesicle suspension, the adhesive
interaction between two vesicles is often
ignored~\cite{Veerapaneni2009_JCP,  RahimianVeerapaneniBiros2010_JCP}
for numerical convenience, mostly because a small time step is often
needed to resolve the dynamics of the lubrication thin film between two
membranes with adhesion.  However, adhesive interactions between RBCs
lead to RBC clusters that are expected to alter the rheological
properties of the RBC suspension~\cite{NeuMeiselman2002_BJ,
SvetinaZiherl2008_Bioelectrochemistry}. Hydrodynamic simulations of
adhesive vesicles in a quiescent flow have revealed physical insights to
the equilibrium shapes of RBC aggregates in both
experiments~\cite{FlormannAouane2017_SciReports} and
theory~\cite{ZiherlSvetina2007_PNAS}.  The main goal of this work is to
investigate the hydrodynamics of adhesive vesicles in flow conditions
common in microfluidics such as a planar shear flow and an extensional
flow.

The electrostatic nature of lipid molecules (a hydrophobic tail and a
hydrophilic head with an electric dipole) complicates the interactions
between a lipid bilayer membrane and another bilayer
membrane~\cite{EvansMetcalfe1984_BJ, Book_PhysicalBasisCellAdhesion,
Book_IntermolecularSurfaceForces, PerutkovaFrank-Bertoncelij2013_CSB} or
a solid (such as a glass substrate or a nano particle).  Adhesion
between a vesicle and a solid has been extensively studied with more
focus on the static equilibrium shapes~\cite{Seifert1990_PRA,
BernardGuedeau-Boudeville2000_Langmuir, ShiFengGao2006_ActaMechSin,
LinFreund2007_IntJSolidsStructures, GruhnFrankeDimova2007_Langmuir,
das2008adhesion,KehWalterLeal2014_PoF,zhang2009phase,
Agudo-Canalejo2015_ACSNano, Agudo-CanalejoLipowsky2015_NanoLett,
SteinkuhlerAgudo-Canalejo2016_BJ, KehLeal2016_PRF,
Agudo-CanalejoLipowsky2017_SoftMatt} than the transient adhesion
process~\cite{cantat1999lift, suk-sei2001, BlountMiksisDavis2013_PRSa}.
Adhesive interactions between lipid membranes are essential in many
biomedical, biological, and biophysical processes.  For example, vesicle
adhesion is crucial to initiate membrane fusion and fission in
endocytosis, exocytosis, and the transport of small vesicles through
membrane surfaces.  In the absence of an external electric field and
ions in the solvents, it is reported that the adhesive interactions
between two membranes can be well approximated by the L.-J.-type
potential~\cite{FlormannAouane2017_SciReports}, that consists of a
long-range attraction component (truncated at some finite distance
beyond which membrane interaction becomes negligible) and a short-range
repulsion component~\cite{Book_IntermolecularSurfaceForces}, and the
combination of the two gives rise to an equilibrium distance where the
interaction potential is at its
minimum~\cite{Book_IntermolecularSurfaceForces}. 

The strength of membrane-membrane adhesion can be estimated through the
membrane contact angle at the edge of contact
zone~\cite{RamachandranAndersonLealIsraelachvili2010_Langmuir,
MaresDanielIglic2012_SciWorldJ}.  Alternatively, researchers have used
the micropipette to measure the adhesive force between two vesicles
bound by strong
adhesion~\cite{EvansMetcalfe1984_BJ,FrostadSethBernasekLeal2014_SoftMatt}:
Evans and Metcalfe~\cite{EvansMetcalfe1984_BJ} were able to measure the
reduction in the free energy per unit area of membrane-membrane contact
formation due to the van der Waals' attraction. For the adhesive
interaction between two lipid bilayer membranes in buffer solutions, a
typical range for the energy density is between $1$ to $10$ $\mu J/m^2$,
similar to the adhesion energy density of $3$ $\mu J/m^2$ between two
RBCs~\cite{FlormannAouane2017_SciReports}.  Frostad {\it et
al.}~\cite{FrostadSethBernasekLeal2014_SoftMatt} investigated the
depletion-attraction induced adhesion between two vesicles, and
identified the dynamic role of membrane tension during membrane
adhesion/peeling.

For adhesion interaction between a vesicle and a
substrate~\cite{GruhnFrankeDimova2007_Langmuir}, the boundary between
weak and strong adhesion is around $1$ $\mu J/m^2$: the
vesicle-substrate interaction is ``strong" when the adhesion energy
density is larger than $1$ $\mu J/m^2$ and ``weak" when the energy
density is less than $1$ $\mu J/m^2$. The ratio of total adhesion energy
to the bending energy (bending modulus of lipid bilayer membranes $\sim
10^{-19}$ $J$) gives a measure of the vesicle deformation in the
presence of
adhesion~\cite{RamachandranAndersonLealIsraelachvili2010_Langmuir}. 

In this work we focus on regimes where such ratio is of order one,
between the weak adhesion (adhesion energy/bending energy $\ll 1$) and
strong adhesion (adhesion energy/bending energy $\gg 1$) regimes.  In
this regime the membrane deformation may increase the ``contact area''
between two vesicles and enhance the adhesion effects.  The
equilibration of vesicle membranes in the strong adhesion regime has
been well-studied and documented (see~\cite{Agrawal2011_MathMechSolids,
RamachandranAndersonLealIsraelachvili2010_Langmuir,
SteinkuhlerAgudo-Canalejo2016_BJ, FlormannAouane2017_SciReports} and
references therein). However it is unclear how the adhesion couples to
the vesicle hydrodynamics in this regime.  This paper aims to address
this question with quantitative characterization in terms of physical
parameters.

In a quiescent environment, sub-micron size vesicles are found to form a
doublet due to their van der Waals' attractive
interactions~\cite{RamachandranAndersonLealIsraelachvili2010_Langmuir}.
Due to the strong van der Waals' adhesive force, the vesicles are far
from spherical shape and the membrane is almost flat in the ``contact"
region. Gires {\em et al.}~used small-deformation analysis to
investigate the hydrodynamic interactions between two vesicles in a
planar shear flow with a long separation
distance~\cite{GiresDankerMisbah2012_PRE}.  They found that the vesicle
interaction could be either attractive or repulsive depending on the
organization of the two vesicles relative to the shear
flow~\cite{gir-sri-mis-pod-cou2014}. To the best of authors' knowledge,
the effects of close-range vesicle adhesion on their hydrodynamics under
an external flow have not been studied and quantified. The goal of this
work is to use state-of-the-art boundary integral simulations to
numerically investigate the dynamics of two vesicles in both planar
shear flow and extensional flow for a wide range of vesicle shapes and
adhesive strength and distance.

Boundary integral equation (BIE) approaches are well-suited for solving
the low Reynolds flow problems considered here as they lead to reduction
in dimensionality and achieve high-order accuracy even for moving
geometry problems. When the vesicles adhere, one major issue for BIE
solvers is to resolve the vesicle-vesicle hydrodynamic forces which
become {\em nearly singular}.  We use an interpolation-based quadrature
rule~\cite{qua-bir2014} to maintain high-order accuracy for all
vesicle-vesicle separation distances.  To overcome the numerical
stiffness induced by the membrane bending forces and to control the
global error, we employ a second-order spectral deferred
implicit-explicit adaptive time stepping
scheme~\cite{quaife2016adaptive}. 

This paper is organized as follows. In Section~\ref{sec:ge} we formulate
our model for two-dimensional vesicle hydrodynamics with adhesive
interactions between membranes of distinct vesicles.  We simulate the
adhesion process of two vesicles in a quiescent flow in
Section~\ref{sec:qflow}, where we also present a simple lubrication
model to estimate how long it takes for two nearby vesicles to reach the
separation distance set by the adhesion potential.  In
Section~\ref{sec:eflow} we first study the hydrodynamics of two vesicles
in a planar extensional flow. We then numerically investigate the
hydrodynamics of a vesicle doublet in a fluid trap where the stagnation
point is actively controlled to be at the center of the vesicle doublet.
From these results we propose a novel application of the microfluidic
fluid trap to probe the adhesion strength between lipid bilayer
membranes in solutions. In Section~\ref{sec:sflow} we investigate how
two vesicles may form a doublet as they move toward each other under a
planar shear flow.  We map out the parameter regions for bound/unbound
vesicles under a planar shear flow, and we also investigate the effects
of adhesive interactions on the rheological properties of a dilute
suspension of vesicle doublets.  Finally in
Section~\ref{sec:conclusions} we discuss the implications of our results
and point out a few potential future directions.

\section{Governing Equations \label{sec:ge}}
We consider a suspension of locally inextensible vesicles in an
unbounded two-dimensional viscous fluid.  For simplicity, we assume the
fluid viscosity both inside and outside the vesicles is the same;
however, incorporating viscosity contrast is rather straightforward in
our numerical algorithm.  While we focus on a vesicle doublet suspension
for the rest of the paper, here we provide the most general formulation
for multiple vesicles interacting with each other through both
hydrodynamic and adhesive forces.  Individual vesicles are denoted as
$\gamma_j$, $j=1,\ldots,M$, and they are parameterized in arclength as
$\xx_j(s,t)$.  The union of all vesicles is denoted by $\gamma$.  Given
a background velocity $\uu_\infty$, the governing (dimensionless)
equations are
\begin{equation}
\begin{aligned}
  \mu \Delta \uu = \nabla p, \quad &\xx \in \RR^2,
    &&\mbox{\em conservation of momentum} \\
  \nabla \cdot \uu = 0, \quad &\xx \in \RR^2, 
    &&\mbox{\em conservation of mass} \\
  \uu \rightarrow \uu_\infty, \quad &\|\xx\| \rightarrow \infty,
    &&\mbox{\em far-field condition} \\
  \uu(\xx,t) = \dot{\xx}, \quad &\xx \in \gamma,
    &&\mbox{\em velocity continuity} \\
  \xx_s \cdot \uu_s =0, \quad &\xx \in \gamma,
    &&\mbox{\em local inextensibility} \\
  \llbracket T \rrbracket \nn = \xxi, \quad &\xx \in \gamma,
    &&\mbox{\em stress balance on membranes}
\end{aligned} \label{eq:gov}
\end{equation}
where $\uu$ is the fluid velocity field, $p$ is the pressure, and the
scaled viscosity $\mu=1$ inside the vesicles. Outside the vesicle
$\mu=\mu_e/\mu_i$ with $\mu_i$ ($\mu_e$) the viscosity of the interior
(exterior) fluid.  We set the viscosity ratio $\mu_e/\mu_i=1$ for the
rest of the paper.  $T = - p I + \mu\left(\nabla \uu + \nabla
\uu^T\right)$ is the stress tensor, $\llbracket T \rrbracket$ is the
jump across the membrane, and $\xxi$ is the traction that is the sum of
the bending, stretching, and adhesion forces defined in
equation~(\ref{eqn:traction}) of Appendix~\ref{sec:AppendixB}.  The
bending force, arising from the Helfrich energy model, is $\BB \xx =
-\kappa_b \xx_{ssss}$, where the subscript $s$ denotes derivative with
respect to arclength $s$. $\kappa_b$ is the bending modulus which we set
to be 1 for all examples. The stretching force is $\TT \sigma = (\sigma
\xx_s)_s$, where the tension, $\sigma$, acts as a Lagrange multiplier to
satisfy the local extensibility constraint.  The resistance to bending
and stretching are standard assumptions on vesicles.  In this work, we
include an adhesive force, $\AA \xx$, that we now describe.

\begin{figure}[htp]
  \begin{tabular}{@{}p{0.3\linewidth}@{\qquad}p{0.3\linewidth}@{\qquad}p{0.3\linewidth}@{}}
  \subfigimg[height=\linewidth]{(a)}{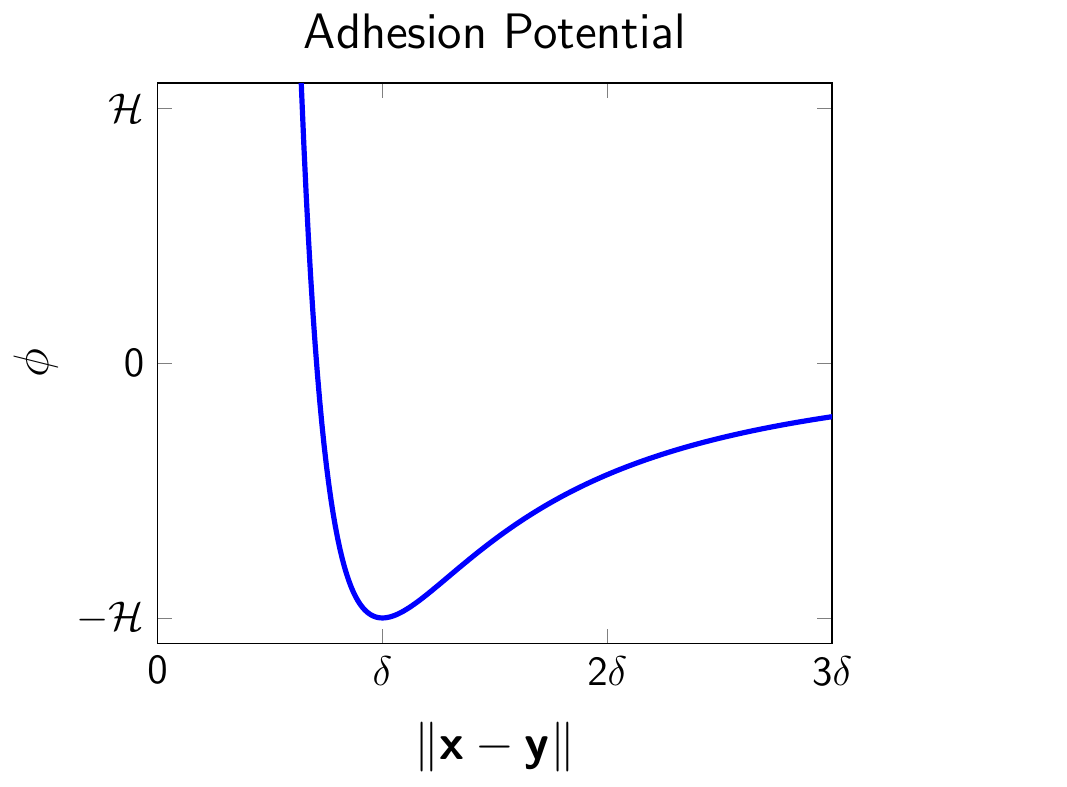} &
  \subfigimg[height=\linewidth]{(b)}{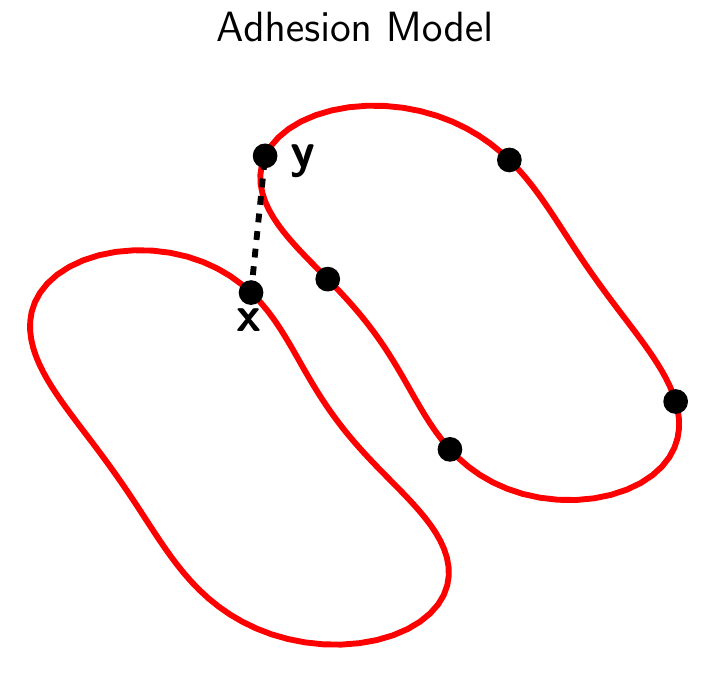} &
  \subfigimg[height=\linewidth]{(c)}{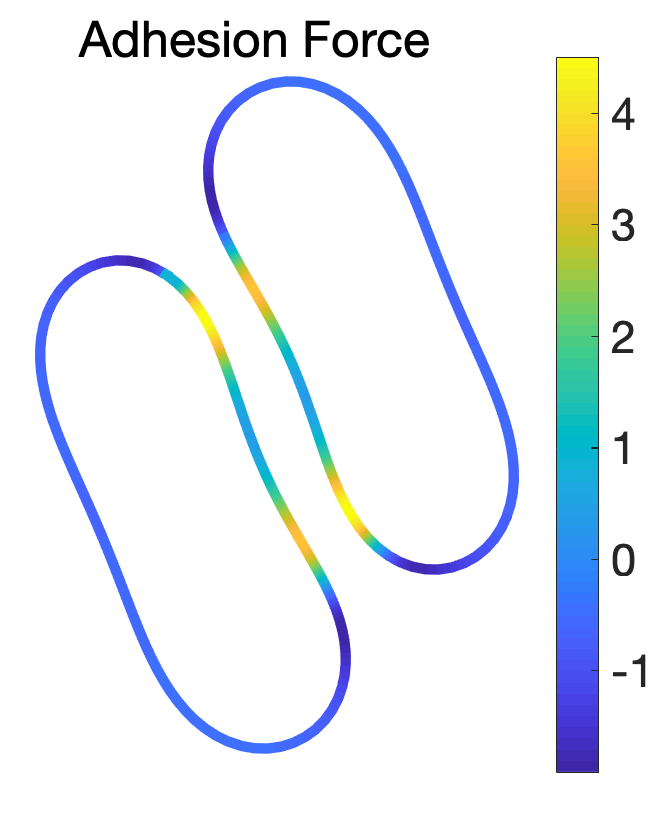}
  \end{tabular}
  \caption{\label{fig:adhesionModel} (a) The adhesion potential
  $\phi(z)$ in equation~\eqref{eq:adhesion_potential}.  Distances less
  than $\delta$ are repulsive and distances greater than $\delta$ are
  attractive.  (b) Each point on a vesicle is repelled or attracted by
  all points on the other vesicles.  (c) The resulting adhesive force is
  found by integrating over all points on all other vesicles
  (equation~\eqref{eqn:adhesionForce}).  The color is the dot product of
  the adhesion force with the vector joining the vesicles.  Therefore,
  the green and yellow regions are repelled by the other vesicle while
  the blue regions are attracted to the other vesicle.}
\end{figure}

The adhesion potential for a vesicle membrane has been modeled by a
L.-J.~type potential
\begin{align}
\label{eq:adhesion_potential}
  \phi(z) = \mathcal{H} \left[ 
    \left(\frac{\delta}{z}\right)^m - \frac{m}{n} \left(\frac{\delta}{z}\right)^n \right],
\end{align}
where $\mathcal{H}$ is the Hamaker constant, $\delta$ is the adhesion
length scale, and $z$ is the distance between a patch on the vesicle
membrane and a point on the other object, which could be another vesicle
membrane or a flat solid wall.  The characteristics of the
L.-J.~potential are summarized in Figure~\ref{fig:adhesionModel}(a): The
adhesive force is zero at the equilibrium distance $z=\delta$.  For
large distances ($z > \delta$) the adhesive potential is attractive
while for small distances ($z < \delta$) the potential is repulsive to
prevent physical contact. The long-range attraction component should
decay sufficiently fast to mimic the finite-range attraction between
RBCs. In this work we are mainly interested in effects of adhesive 
interactions between two vesicles drawn together by the surrounding linear flow.
Therefore we  do not introduce any truncation in the long-range attraction.

The exponents $(m,n)$ in equation~(\ref{eq:adhesion_potential}) depend
on the geometry and the molecular details of the two objects under
adhesion~\cite{Book_IntermolecularSurfaceForces}: $(m,n)=(4,2)$
corresponds to two flat, planar surfaces interacting with each other,
and has been a common choice for studying membrane-solid adhesion
\cite{Seifert1990_PRA,suk-sei2001, ShiFengGao2006_ActaMechSin,
LinFreund2007_IntJSolidsStructures, BlountMiksisDavis2013_PRSa,
YoungStone2017_PRF}.  On the other hand, $(m,n) = (12,6)$ corresponds to
the L.-J.~potential between two molecules, and has been used to model
membrane-membrane adhesion~\cite{FlormannAouane2017_SciReports}.  In our
case, the adhesion potential is between two small patches of lipid
bilayer membrane (because the lipid molecules are coarse-grained in the
continuum modeling).  Thus a reasonable choice for $(m,n)$ between two
coarse-grained membrane patches would be between those for two planar
surfaces and two point molecules.
 
A large value of $m$ corresponds to a sharp increase in the repulsion
force as two objects are within the separation distance $\delta$.  This
poses a numerical challenge since the problem becomes stiff (i.e.,
requires a very small time step) for large $m$.  The adaptive
time-stepping BIE scheme makes it possible to simulate vesicle adhesive
dynamics with specified numerical precision for reasonable computation
time.  We explore several combinations of $(m,n)$ in the simulations of
two vesicles forming a doublet in a quiescent flow and in a shear flow.
We found that, as long as the close-range interaction is well resolved
both in space (using the near-singular evaluations) and time (using the
high-order adaptive time integration), there is very little difference
in both the dynamic evolution and equilibrium configuration between
$(m,n)=(8,6)$ and $(m,n)=(4,2)$.  Therefore, in this work, we use $(m,n)
= (4,2)$ to regulate the numerical stiffness introduced by the adhesive
force. 

Focusing on intermediate adhesive strengths, we use Hamaker constants
ranging from $\mathcal{O}(10^{-1})$ to $\mathcal{O}(10^0)$ times the
bending modulus.  We assume that the adhesion force from
equation~\eqref{eq:adhesion_potential} applies between all pairs of
points on different vesicles, and in Appendix~\ref{sec:appendixA} the
net adhesive force at a point $\xx$ on vesicle $j$ is shown to be
\begin{align}
  \AA\xx:=-\mathcal{H} m \delta^{n}\sum_{\substack{k=1 \\ k \neq j}}^M 
  \int_{\gamma_k} \frac{\xx - \yy}{\|\xx - \yy\|^{m+2}} 
  \left(\delta^{m-n} - \|\xx - \yy\|^{m-n} \right) ds_\yy.
  \label{eqn:adhesionForce}
\end{align}
This summation of adhesive forces is illustrated in
Figure~\ref{fig:adhesionModel}(b).  Figure~\ref{fig:adhesionModel}(c) is
an example of the calculated adhesive force projected onto their
center-of-mass vector. We notice that the adhesive force is repulsive in
the contact region, while for the rest of the membranes there is an
attractive force that acts to pull the two vesicles together. The
summary of boundary integral formulation can be found in
Appendix~\ref{sec:AppendixB}. For the rest of the paper, the
dimensionless bending modulus is set to be one ($\kappa_b = 1$) unless
otherwise specified.


\section{Adhesion of two vesicles in a quiescent flow} 
\label{sec:qflow} 
We consider two identical vesicles suspended in a quiescent flow.
Without any external forcing such as an imposed electric field or a
fluid flow in the far-field, the long-range attraction pulls the
vesicles together until their separation distance is close to $\delta$.
The L.-J.~type potential prohibits physical contact between the two
membranes and instead keeps them close to the separation distance
$\delta$.  In Section~\ref{subsec:qflow_draining_times}, we compute the
expected time for the vesicles to reach an equilibrium configuration and
compare the analysis with numerical simulations.  In
Section~\ref{subsec:qflow_adhesion_parameters}, we characterize the
effects of the adhesion parameters and the vesicles' reduced area on the
shape of the contact region and the bending and adhesive energies.

\subsection{Effects of the adhesion parameters on draining times}
\label{subsec:qflow_draining_times}
When two vesicles move towards (or away from) each other under a
constant force $F$ without any imposed external flow, the height, $h$,
of the thin liquid film between two vesicles follows the draining
dynamics~\cite{RamachandranLeal2010_PoF}
\begin{align}
  \label{eq:dhdt}
  \frac{d h}{dt} \sim \frac{K_a^{2/3} F^{1/3}}{\mu R_0^{10/3}} h^3,
\end{align}
where $K_a$ is the area expansion modulus, $\mu$ is the viscosity of the
exterior fluid, and $R_0$ is the radius of the undeformed vesicle.  For
the case of a constant forcing $F$ (independent of separation distance
$h$ and time $t$), equation~\eqref{eq:dhdt} can be easily integrated to
relate the film thickness $h$ and time $t$:
\begin{align*}
  t \sim \frac{\mu R_0^{10/3}}{K_a^{2/3} F^{1/3}}
    \frac{1}{2 h^2} \bigg|^{h(t)}_{h(0)},
\end{align*}
where $h(0)$ is the vesicle separation at $t=0$.  We note that $F< (>)
0$ for an attractive (repulsive) interaction between two vesicles,  and
consequently $h(t)$ decreases (increases) from  the initial separation
distance $h(0)$.  When the force is attractive, $h$ decreases
monotonically, and $t$ in the above equation is the draining time that
diverges as $h\rightarrow 0$.

When the force on each vesicle is a function of the separation distance
$h$ of the form
\begin{align}
  \label{eq:Fh}
  F(h) \sim \mathcal{H} \frac{m\delta^m}{h^{m+1}}
    \left(1-\left(\frac{\delta}{h}\right)^{n-m}\right)
\end{align}
with integers $m>n\ge 2$, attraction is dominant at ``large"
distances ($h > \delta$) while repulsion is dominant at ``small"
distances ($h<\delta$).  Integration of equation~\eqref{eq:dhdt} with
the $F(h)$ as defined in equation~\eqref{eq:Fh} gives the relationship
between $t$ and $h$.  In this case, the relationship involves an
integral of a function of the dimensionless variable $\bar{h} \equiv
h/\delta$:
\begin{align*}
  t\frac{K_a^{2/3}}{\mu R_0^{10/3}}& = 
    \int^{h(t)}_{h(0)}-\frac{dh}{h^3 F^{1/3}} = 
    -\frac{1}{m^{1/3}\mathcal{H}^{1/3}\delta^{5/3}}\int^{\bar{h}(t)}_{\bar{h}(0)}
      \frac{d \bar{h}}{\bar{h}^{(8-m)/3}(\bar{h}^{m-n}-1)^{1/3}}.
\end{align*}
Solving for $t$,
\begin{align}
\label{eq:teq_scaling}
  t\sim \frac{\mu R_0^{10/3}}{K_a^{2/3}}\frac{1}{ m^{1/3} \mathcal{H}^{1/3}\delta^{5/3}},
\end{align}
and equation~\eqref{eq:teq_scaling} tells us that both the adhesion
strength $\mathcal{H}$ and the separation distance $\delta$ affect the
time it takes for a pair of vesicles to reach the separation distance
$\delta$ under adhesion force $F$ in equation~\eqref{eq:Fh}.  In
particular, the duration $t$ is proportional to (i) the separation
distance $\delta^{-5/3}$ (thus $t\rightarrow \infty$ as
$\delta\rightarrow 0$), and (ii) the adhesion strength
$\mathcal{H}^{-1/3}$ (as in the constant forcing
case~\cite{RamachandranLeal2010_PoF}).  From the above analysis we also
expect that the scaling of $t$ with respect to $\delta$ depends on the
adhesion potential, while the scaling with respect to adhesion strength
is independent of the specific form of the potential.

\begin{figure}
  \begin{tabular}{@{}p{0.45\linewidth}@{\quad}p{0.45\linewidth}@{}}
  \subfigimg[width=\linewidth]{(a)}{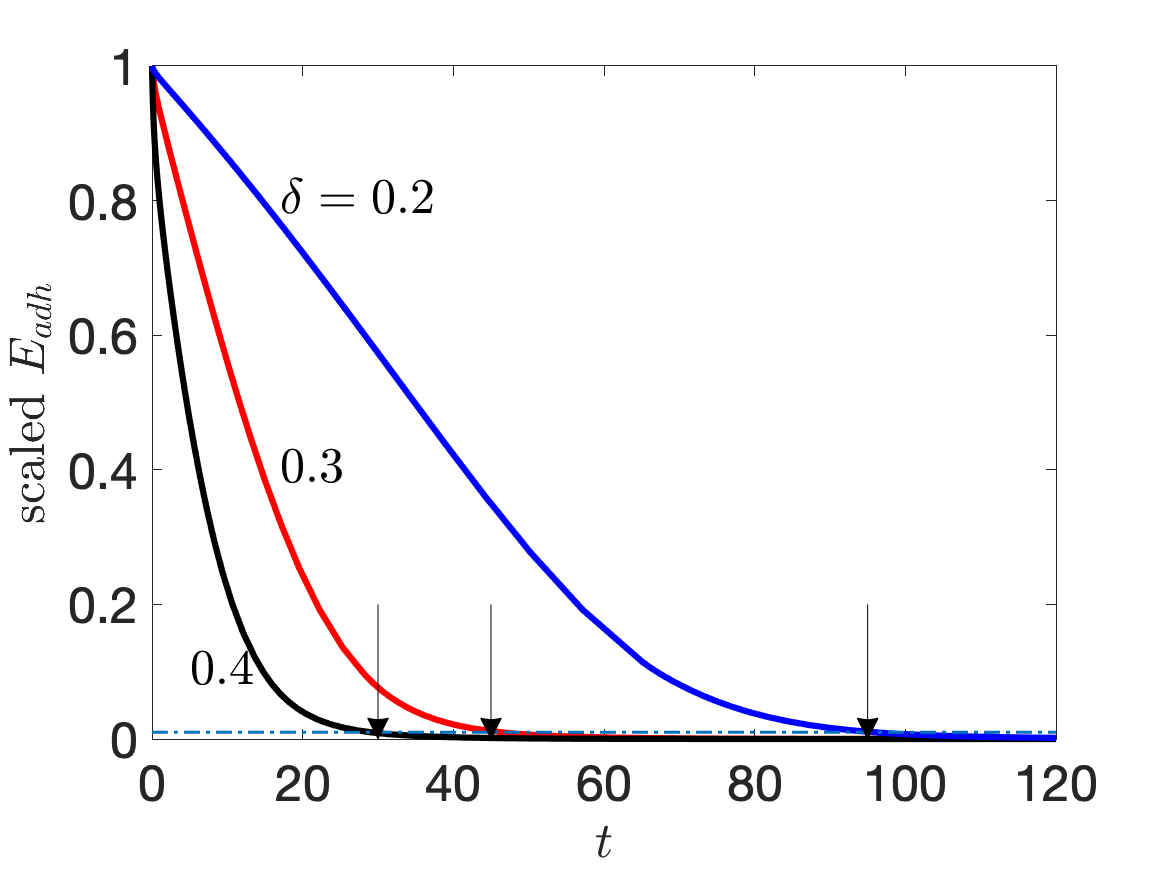} &
  \subfigimg[width=\linewidth]{(b)}{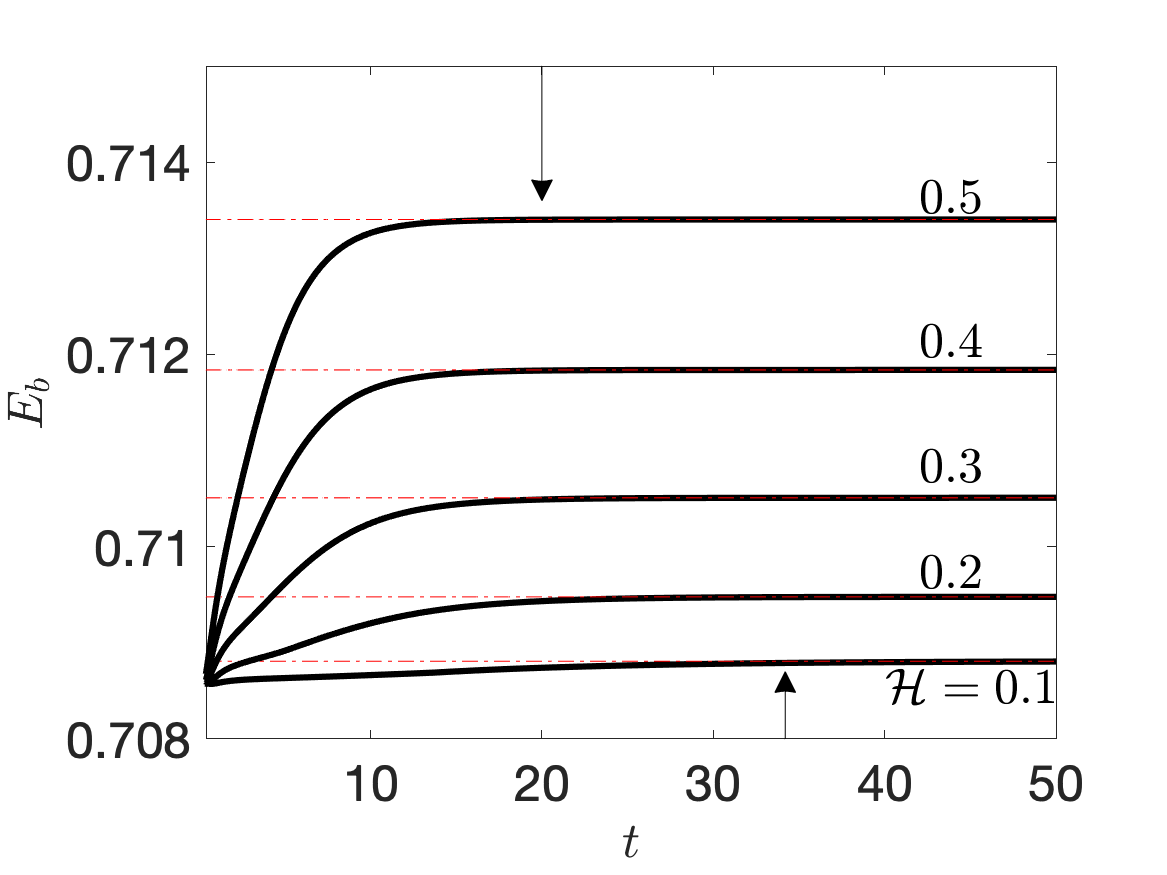}
  \end{tabular}
\caption{\label{fig:qflow00} Scaling of the time it takes for two
vesicles to reach equilibrium. (a) Scaling with respect to $\delta$ with
$\mathcal{H}=0.1$. (b) Scaling with respect to $\mathcal{H}$ with
$\delta = 0.4$. The arrows indicate the time by when the equilibrium is
reached within $1\%$.}
\end{figure}

To test the scaling of the duration $t$ with respect to $\mathcal{H}$
and $\delta$, we simulate the vesicle adhesion dynamics in a quiescent
flow.  Starting with two vesicles at a distance of twice the vesicle
radius, the long-range attraction pulls the vesicles together.
Figure~\ref{fig:qflow00}(a) shows the scaled adhesion energy versus time
with $\mathcal{H}=0.1$ and three values of $\delta$ as labeled.
Adhesion energy reaches minimum at equilibrium, and the scaled $E_{adh}$
evolves towards zero at equilibrium.  The arrows indicate the times when
the scaled adhesion energy reaches within $1\%$ of equilibrium: $t\sim
96$ for $\delta=0.2$, $t\sim 45$ for $\delta = 0.3$, and $t\sim 30$ for
$\delta=0.4$:
\begin{align*}
\frac{t_{\delta=0.2}}{t_{\delta=0.4}} \sim \frac{96}{30}=3.2 
  &\Longleftrightarrow \left(\frac{0.4}{0.2}\right)^{5/3}\sim 3.17,\\
\frac{t_{\delta=0.2}}{t_{\delta=0.3}} \sim \frac{96}{45}=2.1 
  &\Longleftrightarrow \left(\frac{0.3}{0.2}\right)^{5/3}\sim 2.
\end{align*}

The scaling with respect to adhesion strength $\mathcal{H}$ is
illustrated in Figure~\ref{fig:qflow00}(b), where $\delta = 0.4$ and
$\mathcal{H}$ varies from $0.1$ to $0.5$ as labeled.  Again the arrows
indicate the times when the equilibrium is reached within $1\%$:
\begin{align*}
  \frac{t_{\mathcal{H}=0.1}}{t_{\mathcal{H}=0.5}} \sim \frac{34}{20} = 1.71 
  &\Longleftrightarrow \left(\frac{0.5}{0.1}\right)^{1/3}\sim 1.71.
\end{align*}
From the above results, we conclude that the scaling in
equation~\eqref{eq:teq_scaling} captures the adhesion dynamics of two
vesicles that interact with each other via the adhesion force in
equation~\eqref{eq:Fh}.

\subsection{Effects of the adhesion parameters on equilibrium
configuration of a vesicle pair}
\label{subsec:qflow_adhesion_parameters} 
We again simulate two identical adhering vesicles with the adhesion
force in equation~\eqref{eq:Fh}.  The initial vesicle separation is
smaller than in the previous simulations so that the equilibrium
configuration is achieved in a shorter time horizon.  Once a vesicle
doublet is formed, the membrane shape in the contact region depends on
the adhesion strength relative to the membrane bending modulus.  For
weak to moderate adhesion strength, vesicle membranes are flattened in
the contact region while the rest of vesicle maintains a nearly
spherical shape~\cite{EvansMetcalfe1984_BJ,
Book_PhysicalBasisCellAdhesion, Book_IntermolecularSurfaceForces,
RamachandranAndersonLealIsraelachvili2010_Langmuir}.  Under a strong
adhesion, however, the vesicle membranes in the contact region buckle
and form a sigmoidal shape that has also been observed in RBC
doublets~\cite{Ziherl2007_PRL, ZiherlSvetina2007_PNAS,
FlormannAouane2017_SciReports}.  An external electric field is also able
to buckle a vesicle membrane that is adhered to a solid
substrate~\cite{SteinkuhlerAgudo-Canalejo2016_BJ}.

\begin{figure}
   \includegraphics[keepaspectratio=true,scale=0.45]{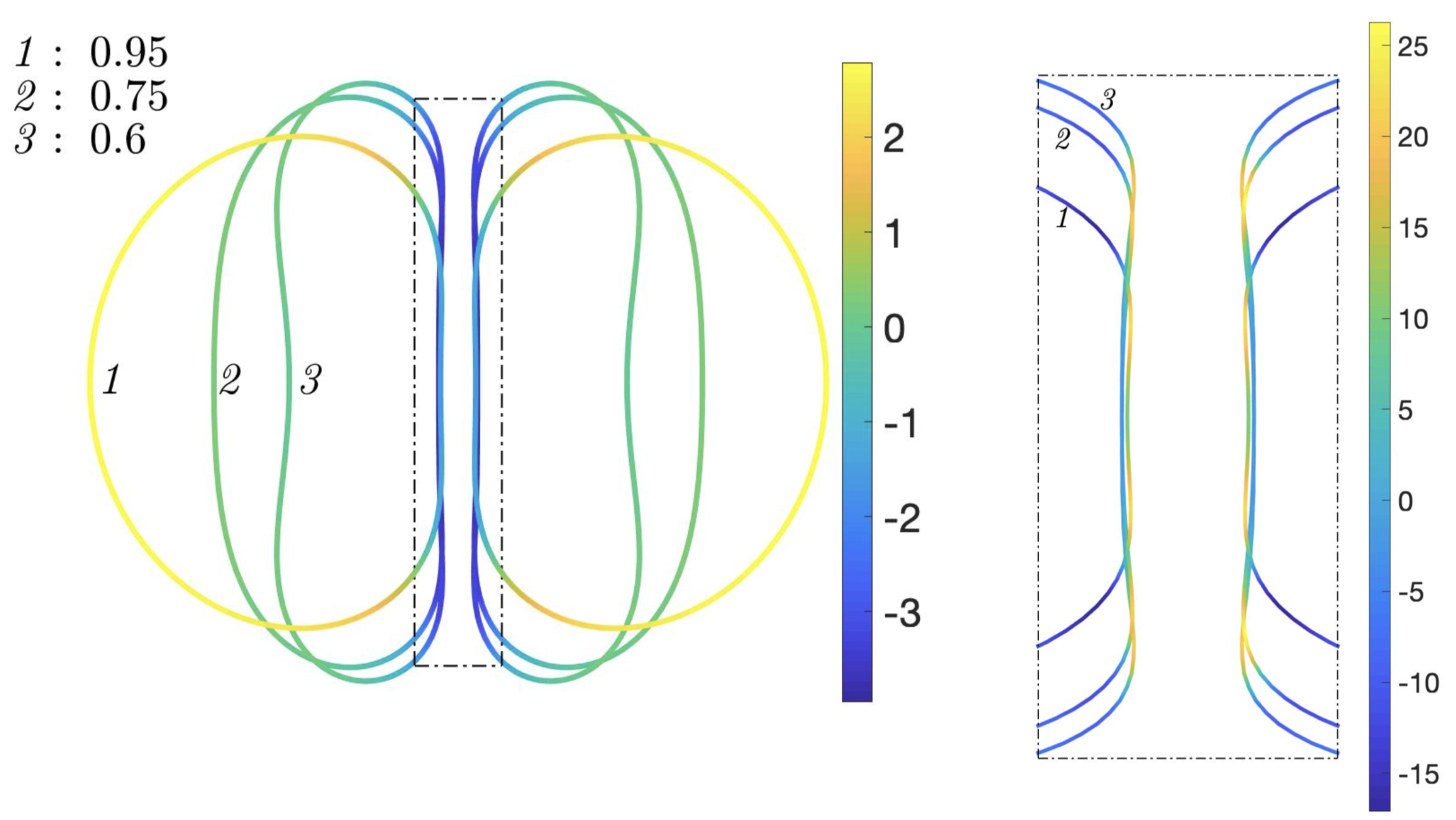}
  \caption{\label{fig:Dec18_vesicle_shape} The equilibrium
  configurations of a doublet of identical vesicles under adhesion in a
  quiescent flow at three values of $\Delta A$.  The vesicle reduced
  areas of $\Delta A=0.95$, $0.75$, and $0.6$ are labeled.  The vesicle
  length is fixed at $2\pi$, the dimensionless bending modulus is
  $\kappa_b=1$, the Hamaker constant is $\mathcal{H}=5$, and the
  separation distance is $\delta=0.2$.  The color coding is the tension
  along the vesicle.  The overall equilibrium shapes of two vesicles
  under adhesive interactions vary with the reduced area.  The right
  figure shows the  membrane shapes in the contact region. The color
  coding is the adhesive force projected along the center-of-mass
  vector.} 
\end{figure}

In this work we focus on weakly adhesive vesicles whose equilibrium
shapes in a quiescent flow are shown in the left plot of
Figure~\ref{fig:Dec18_vesicle_shape} for three reduced areas: $\Delta
A=0.95$, $0.75$, and $0.6$ with $\mathcal{H}=5$ and $\delta = 0.2$. 
The equilibrium vesicle shape for $\Delta A=0.95$ is the circular cap
with a flat contact region.  This is similar to the observed shapes of
two vesicles under strong adhesive interaction
in~\cite{RamachandranAndersonLealIsraelachvili2010_Langmuir}.  When
vesicles are more deflated with a reduced area $\Delta A = 0.75$, the
equilibrium vesicle shape is elongated with a bigger contact region.
This is consistent with the equilibrium shapes of a vesicle doublet
under a $(m,n) = (12,6)$
L.-J.~potential~\cite{FlormannAouane2017_SciReports}.  As the reduced
area decreases further, we observe undulation of the vesicle membrane on
the non-contact side while the contact region remains flat.  The color
coding along each curve is the tension of the vesicle membrane.  We
observe that the membrane tension in the contact region is very
negative, indicating a dominant compression of membrane when the
adhesion force is strong to keep the vesicles bound together.

The right plot of Figure~\ref{fig:Dec18_vesicle_shape} is a zoom (not
to-scale) of the membranes in the contact region, where the membranes
are not perfectly flat at all three values of reduced area. We observe
that the membranes are slightly curved with a dip at the edge.  Such
membrane undulation in the contact region is predicted by lubrication
analyses on an elastic membrane under adhesion with a solid
substrate~\cite{BlountMiksisDavis2013_PRSa, YoungStone2017_PRF}. Results
from the lubrication analysis show that this dip and slightly curved
shape in the contact region are independent of the adhesion strength.
The high membrane curvature at the edge may pose a problem for using the
contact angle there to estimate the adhesion strength. This inspires us
to investigate the possibility of using a dynamic fluid trap to measure
the adhesion strength (see Section~\ref{sec:eflow}).

Also the dip at the edge of the contact region is related to (but
different from) the buckling of membrane under strong adhesion: A closer
inspection on the dip at the edge shows that the membrane distance is
smaller than the neutral separation distance $\delta$ there for more
deflated vesicles.  For $\Delta A =0.6$ and $0.75$  the adhesion force
is mostly attractive in the contact region except at the edge, where the
adhesive force turns repulsive.  For $\Delta A=0.95$, however, the
adhesion force is mostly repulsive, leading to large tension in the rest
of the vesicle membrane.

\begin{figure}
  \begin{tabular}{@{}p{0.45\linewidth}@{\quad}p{0.45\linewidth}@{}}
  \subfigimg[width=\linewidth]{(a)}{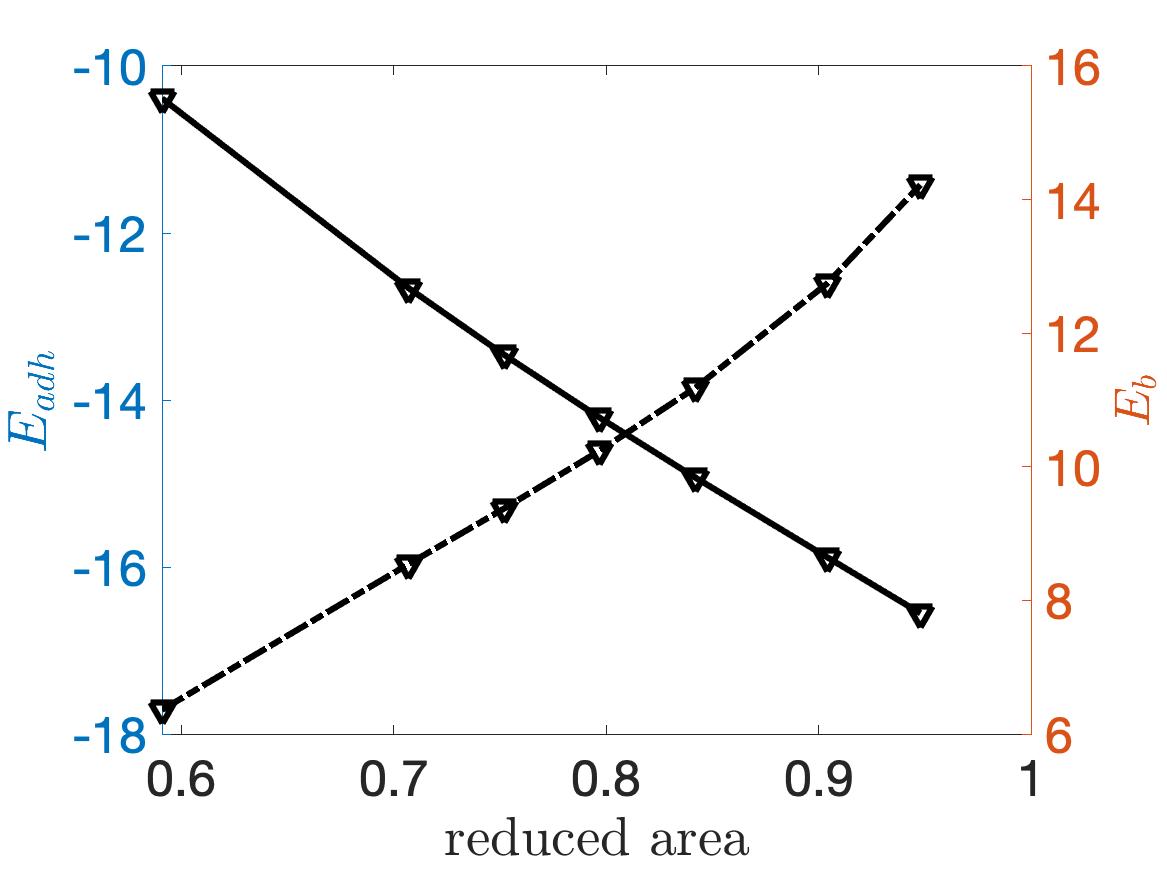} &
  \subfigimg[width=\linewidth]{(b)}{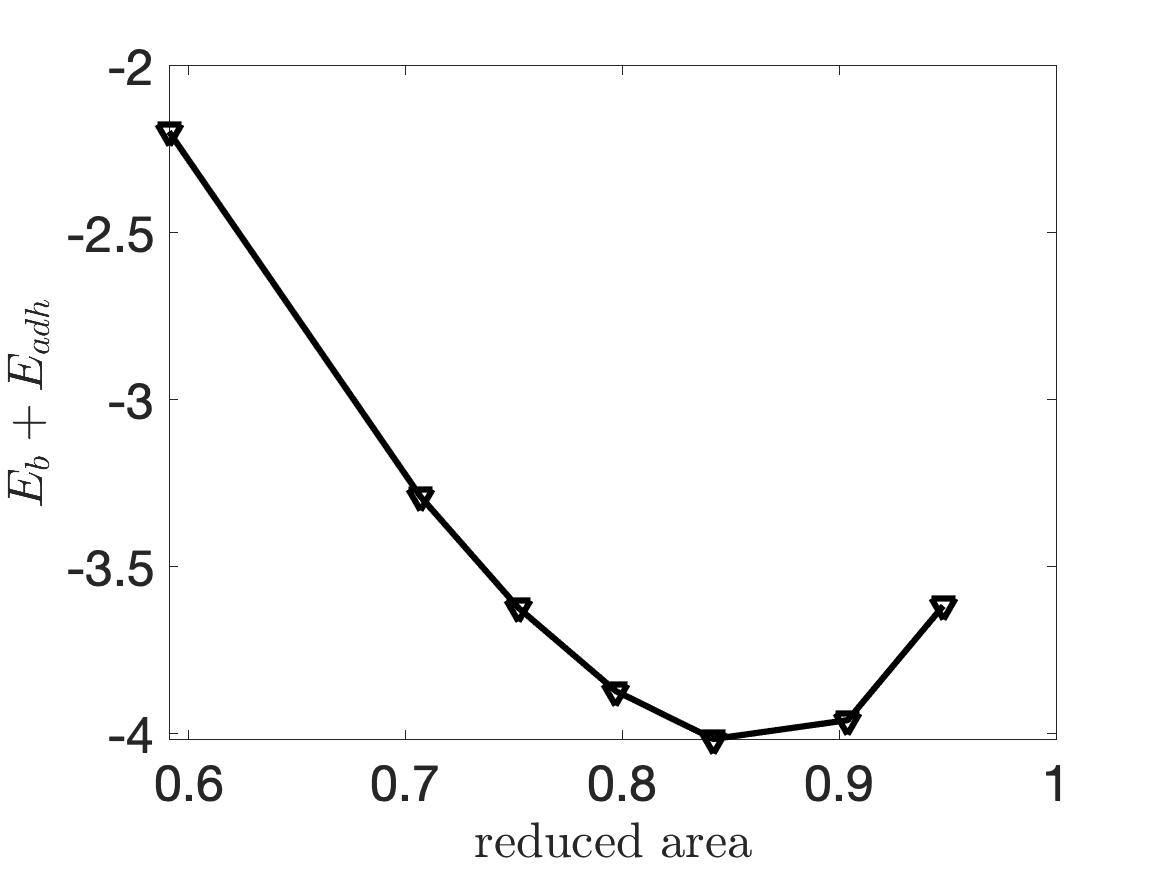}
  \end{tabular}
  \caption{\label{fig:Dec18_vesicle_equilibrium1} (a) The total bending
  energy (solid curve) and adhesion energy (dash-dotted curve) plotted
  against the reduced area $\Delta A$. (b) The sum of the two energies
  plotted against reduced area $\Delta A$. The two vesicles in the
  doublet are of the same length and reduced area.  The Hamaker constant
  is $\mathcal{H} = 5$ and the separation distance is $\delta = 0.2$.}
\end{figure}

Figure~\ref{fig:Dec18_vesicle_equilibrium1}(a) shows the total bending
energy and adhesion energy as a function of the reduced area. The
smaller the reduced area, the more vesicle area is available for
deformation and thus the bending energy is higher.  In contrast, the
total adhesion energy becomes more negative as the reduced area
decreases.  The sum of the two energies is plotted in
Figure~\ref{fig:Dec18_vesicle_equilibrium1}(b), where a local minimum in
the total energy is found around $\Delta A = 0.85$.

\begin{figure}
  \begin{tabular}{@{}p{0.45\linewidth}@{\quad}p{0.45\linewidth}@{}}
  \subfigimg[width=\linewidth]{(a)}{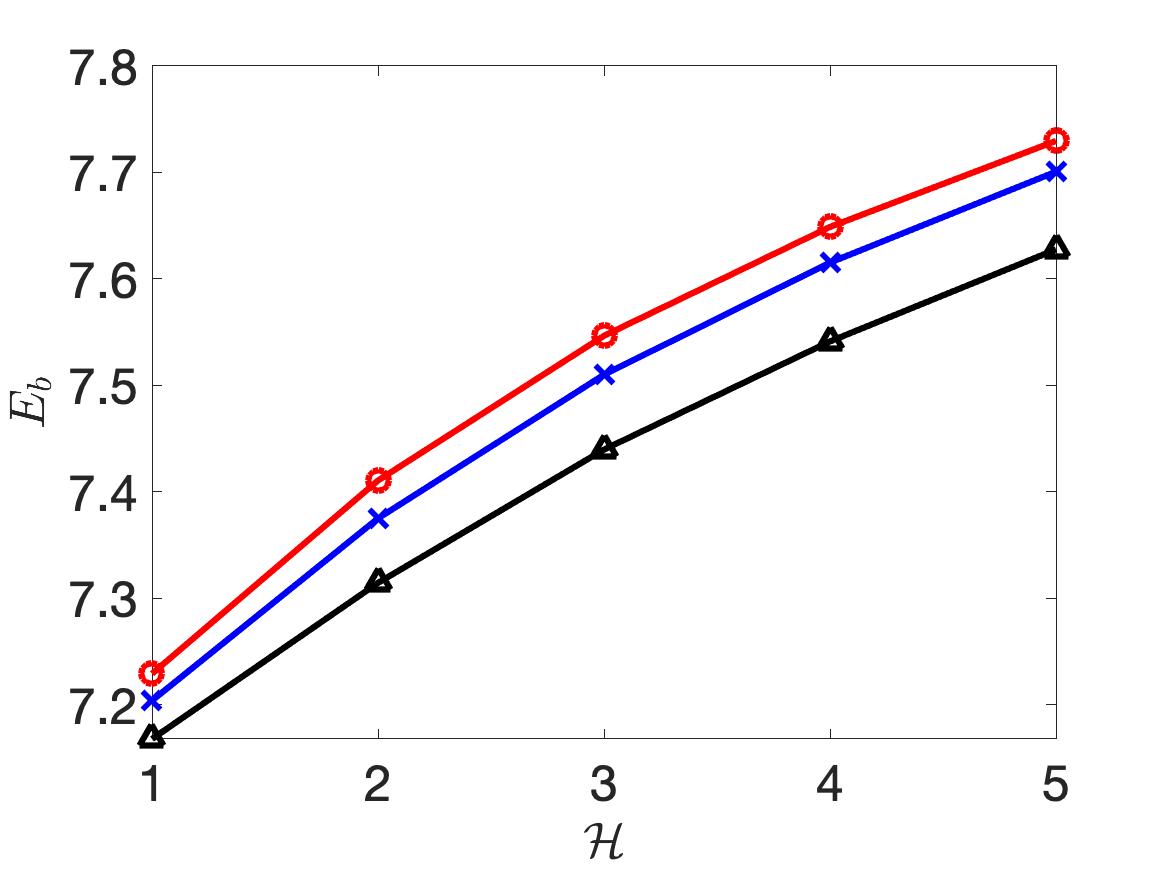} &
  \subfigimg[width=\linewidth]{(b)}{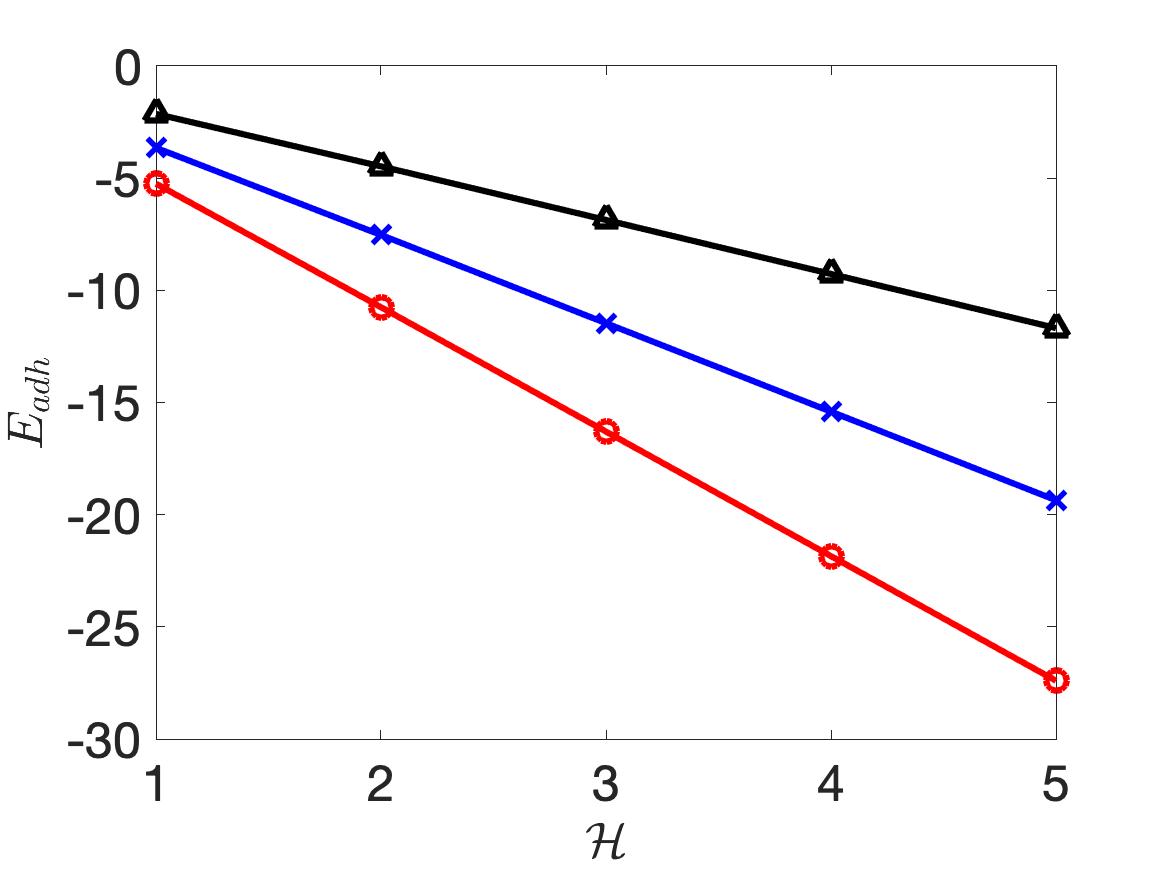}
  \end{tabular}
  \caption{\label{fig:Dec18_equilibrium} The total bending energy (a)
  and adhesion energy (b) of a vesicle doublet at equilibrium versus the
  adhesion strength $\mathcal{H}$.  The two identical vesicles in the
  doublet have a length of $2\pi$ and a reduced area of $\Delta A=0.95$,
  with separation distance $\delta = 0.2$ (triangles), $0.3$ (crosses),
  and $0.4$ (circles).}
\end{figure}

Figure~\ref{fig:Dec18_equilibrium} demonstrates how the adhesion
strength $\mathcal{H}$ and separation distance $\delta$ affect the
equilibrium configuration of two vesicles under adhesive interactions.
Both vesicles have a length of $2\pi$ and a reduced area of $\Delta A =
0.95$.  The total bending (a) and adhesion (b) energies at equilibrium
are plotted against $\mathcal{H}$ for three values of the separation
distance $\delta$.  We observe that for $\Delta A=0.95$ the equilibrium
vesicle shape does not vary much with the adhesion strength
$\mathcal{H}$, while the total adhesion energy varies linearly with
$\mathcal{H}$.

\section{Adhesion of two vesicles in a planar extensional flow} 
\label{sec:eflow} 

\begin{figure}[htp]
  \includegraphics[height = 0.2\textwidth,trim={1cm 0cm 1cm 0cm},clip]
    {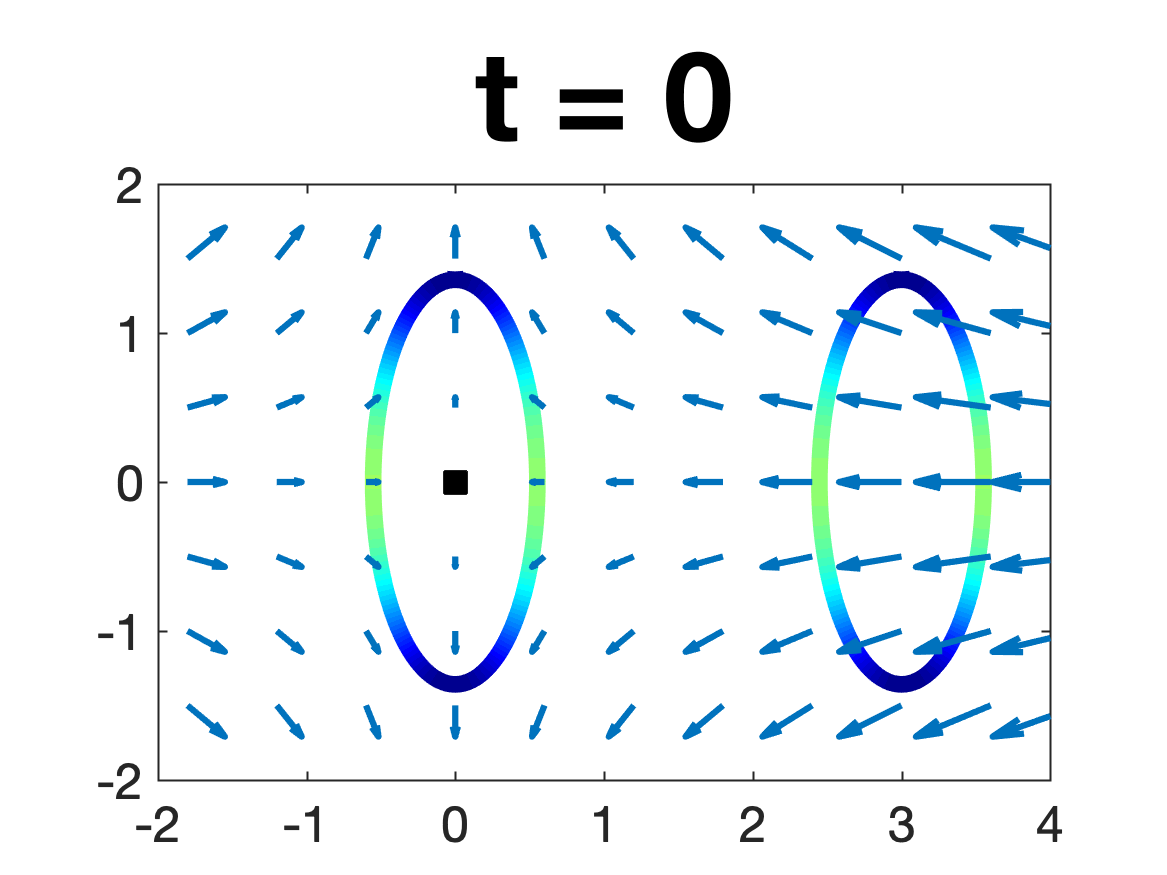}
  \includegraphics[height = 0.2\textwidth,trim={4cm 0cm 4cm 0cm},clip]
    {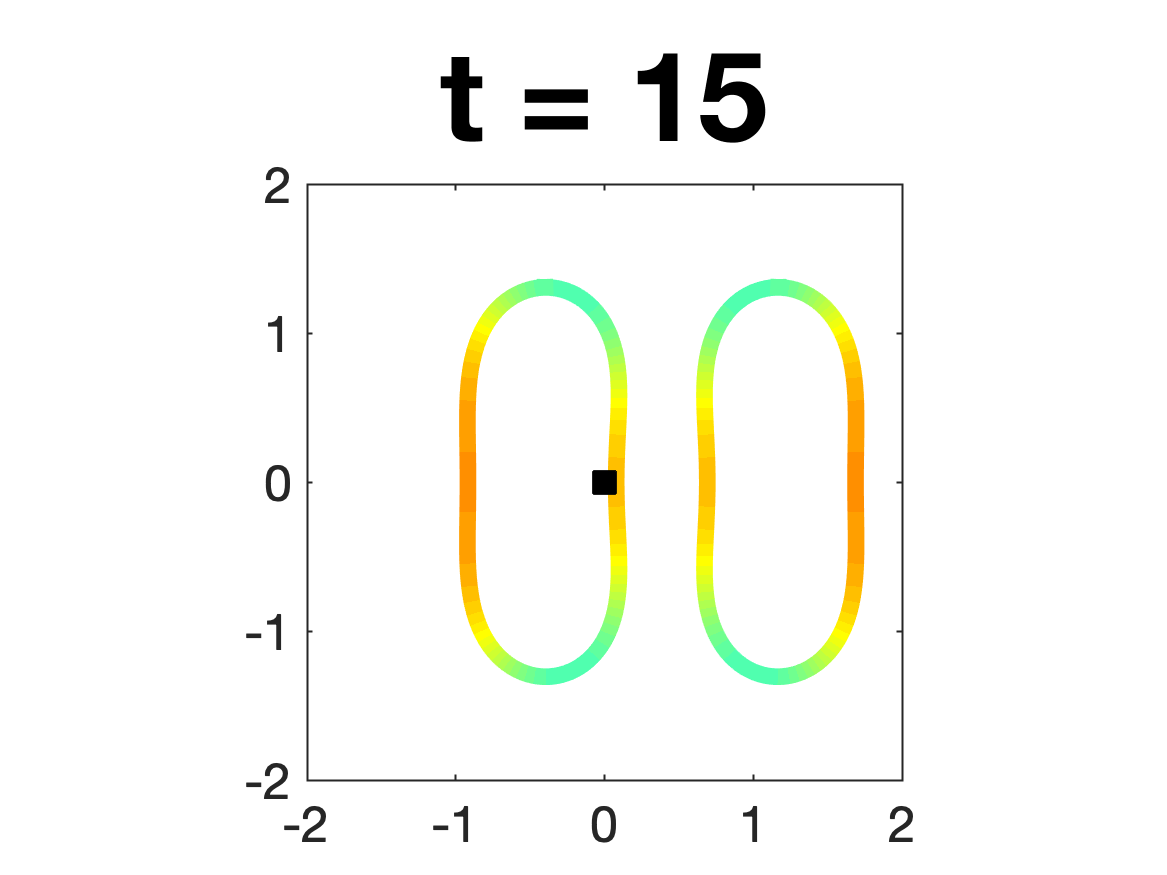}
  \includegraphics[height = 0.2\textwidth,trim={4cm 0cm 4cm 0cm},clip]
    {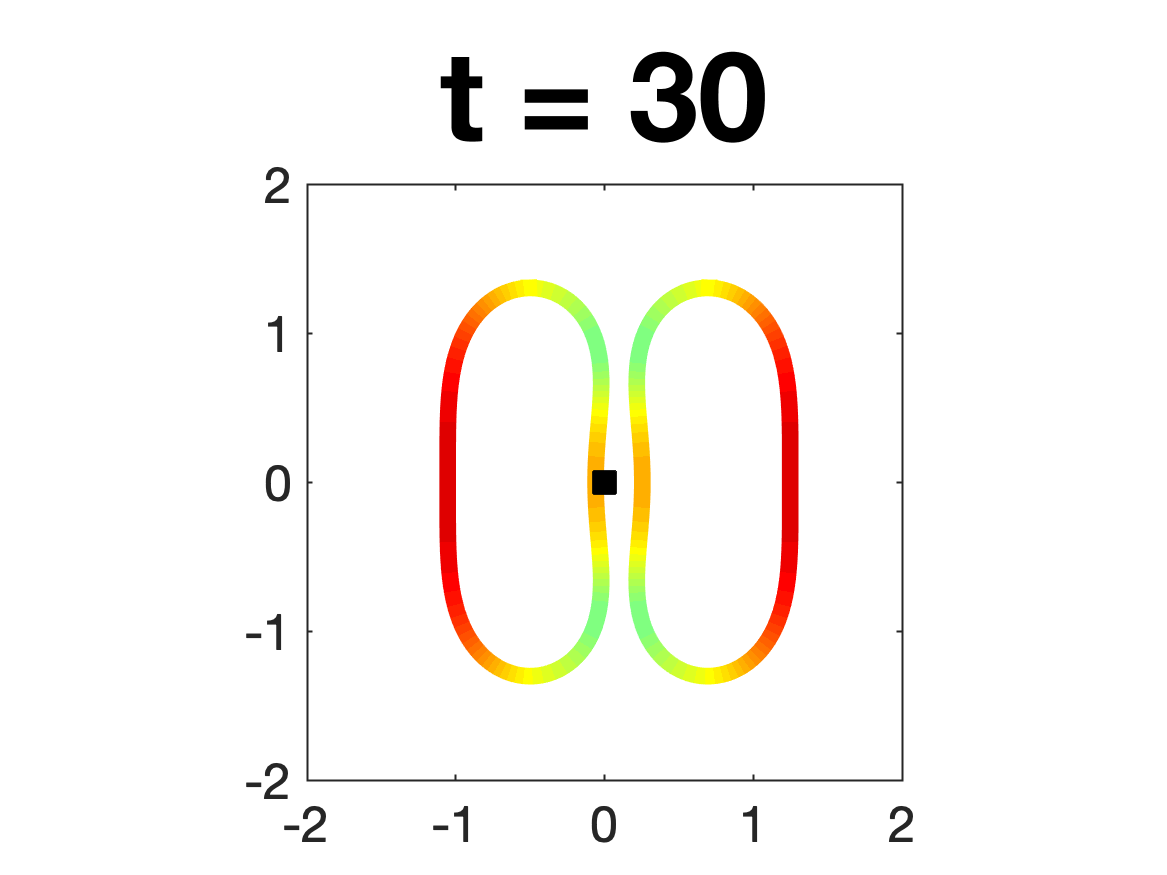}
  \includegraphics[height = 0.2\textwidth,trim={4cm 0cm 4cm 0cm},clip]
    {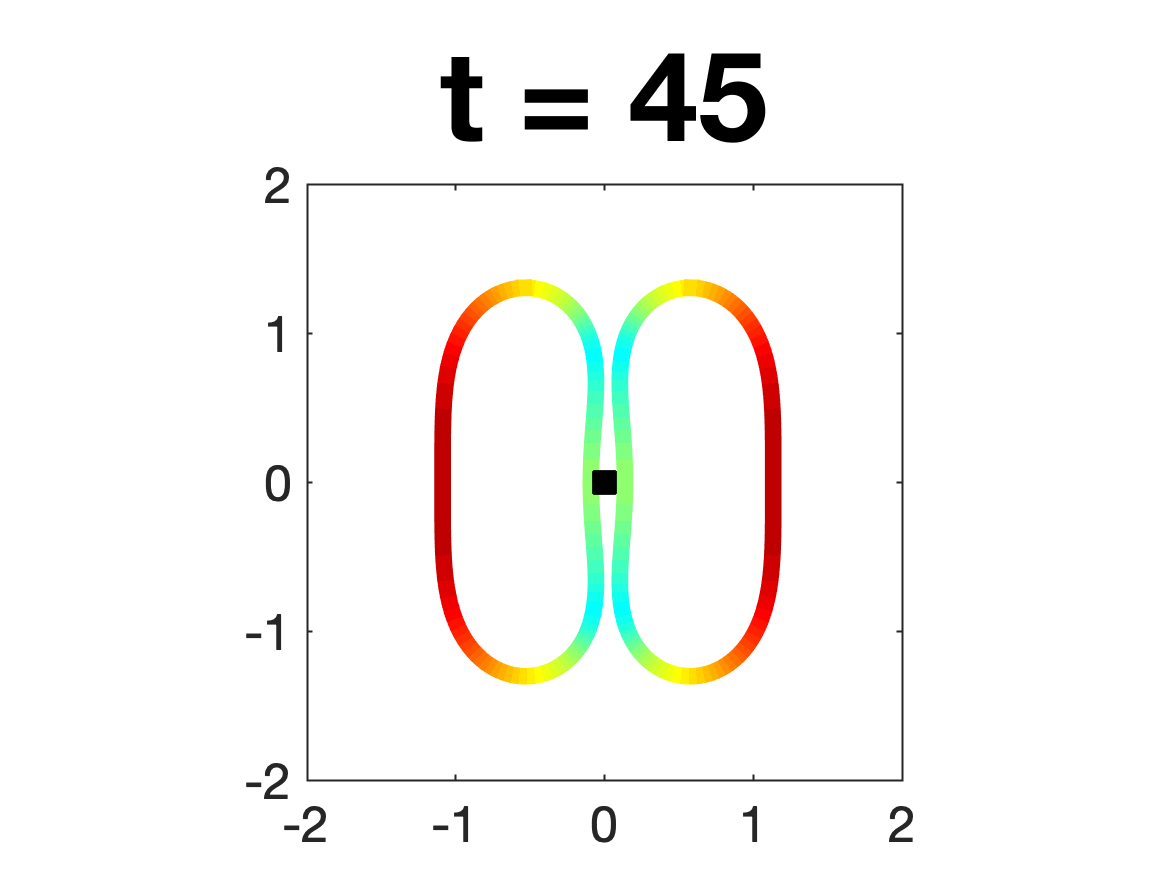}
  \includegraphics[height = 0.2\textwidth,trim={4cm 0cm 0cm 0cm},clip]
    {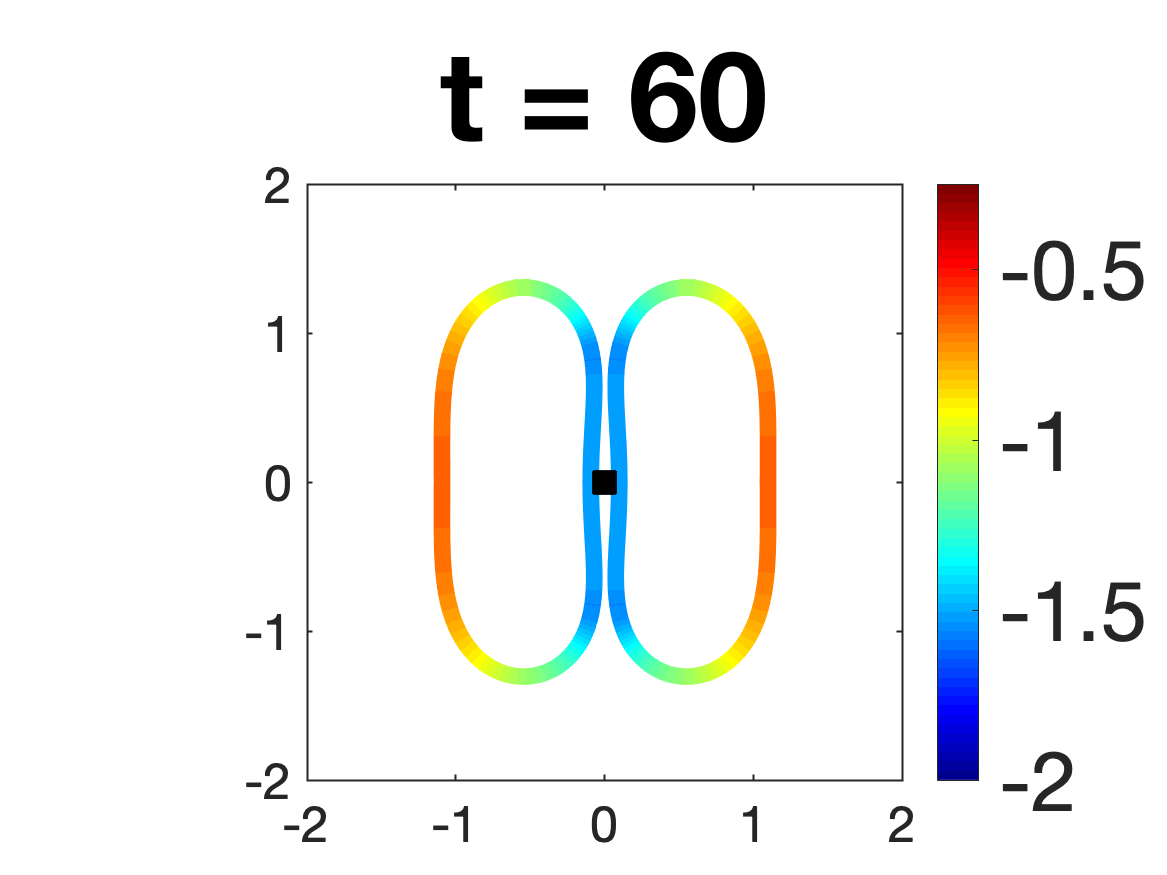} \\

  \includegraphics[height = 0.2\textwidth,trim={1cm 0cm 1cm 0cm},clip]
    {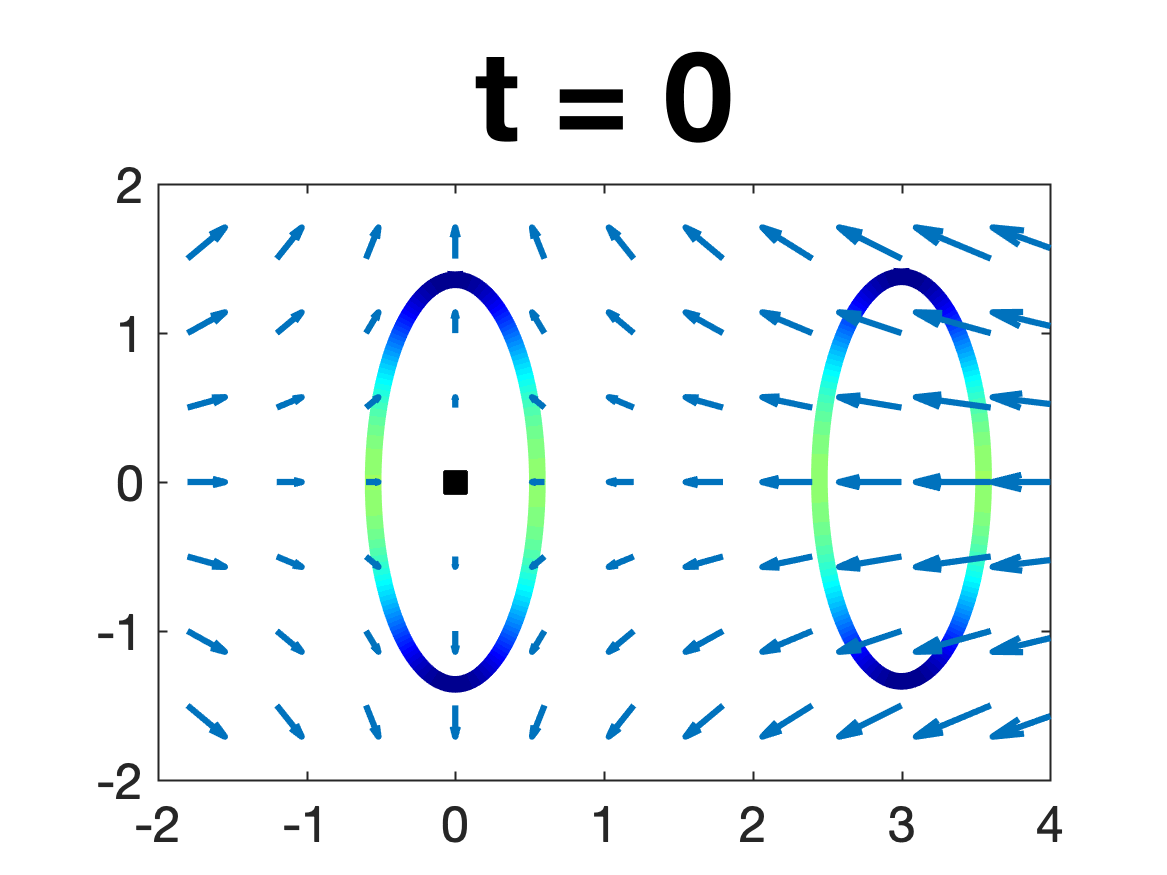}
  \includegraphics[height = 0.2\textwidth,trim={4cm 0cm 4cm 0cm},clip]
    {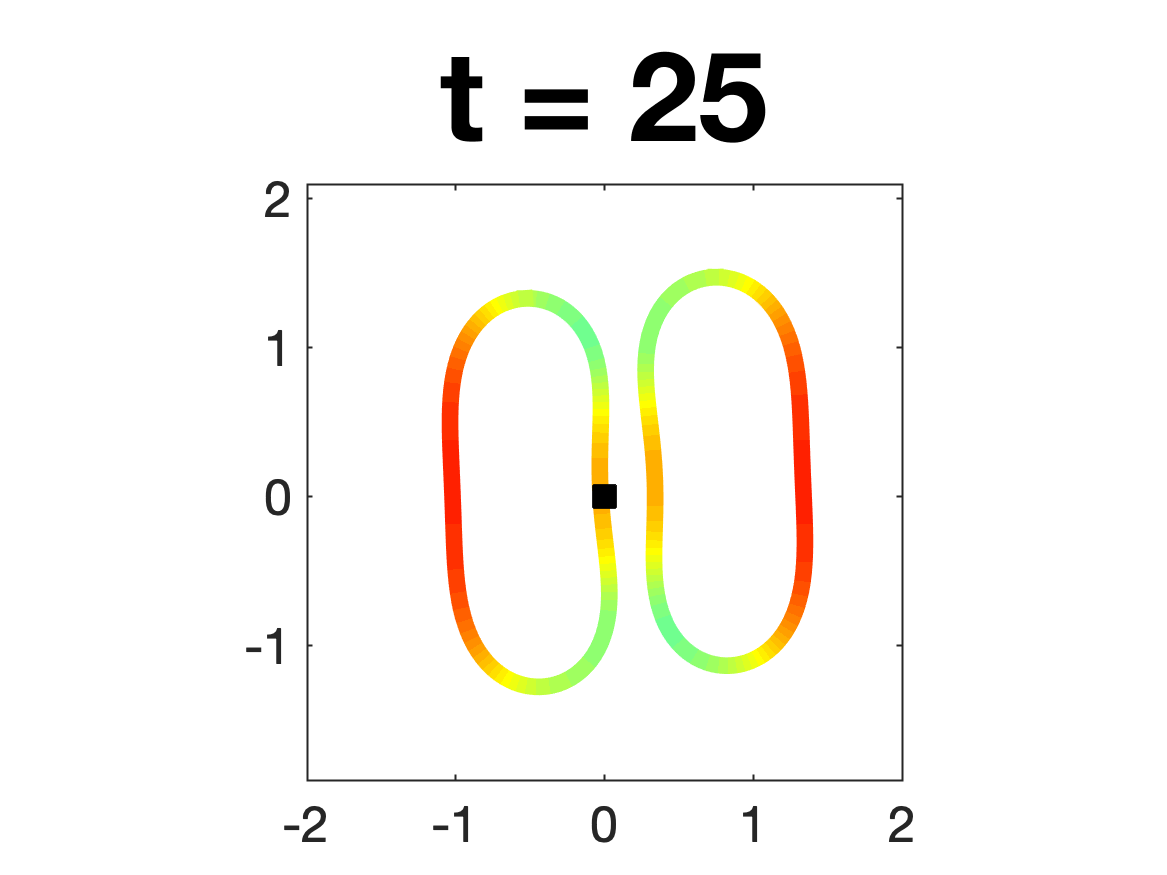}
  \includegraphics[height = 0.2\textwidth,trim={4cm 0cm 4cm 0cm},clip]
    {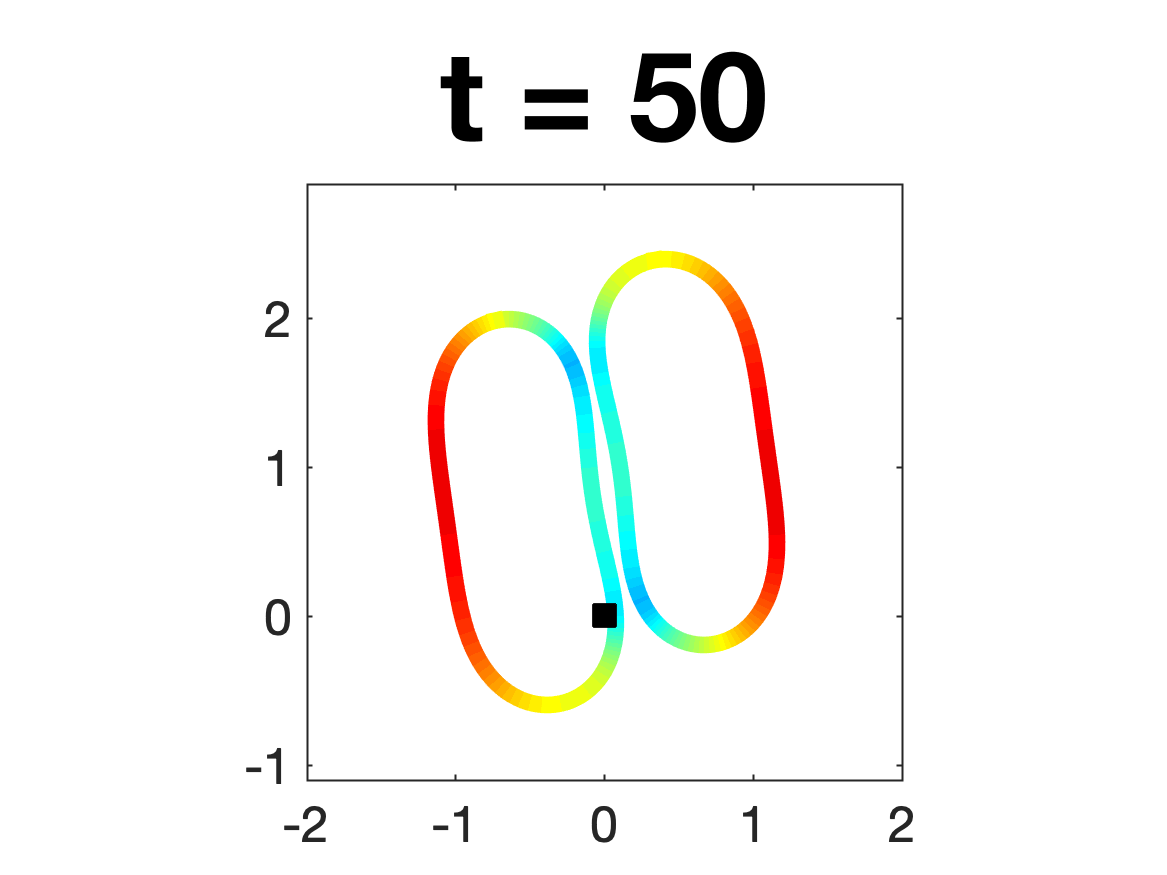}
  \includegraphics[height = 0.2\textwidth,trim={4cm 0cm 4cm 0cm},clip]
    {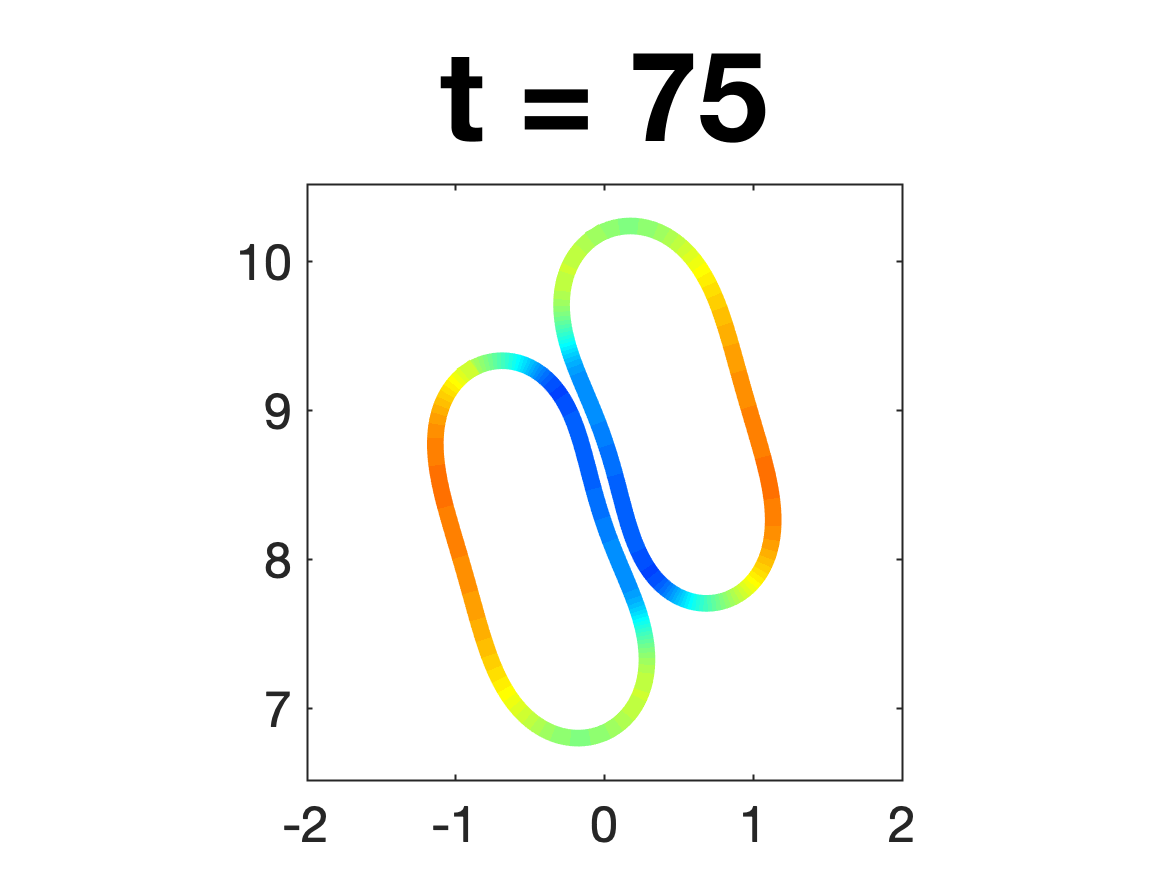}
  \includegraphics[height = 0.2\textwidth,trim={4cm 0cm 0cm 0cm},clip]
    {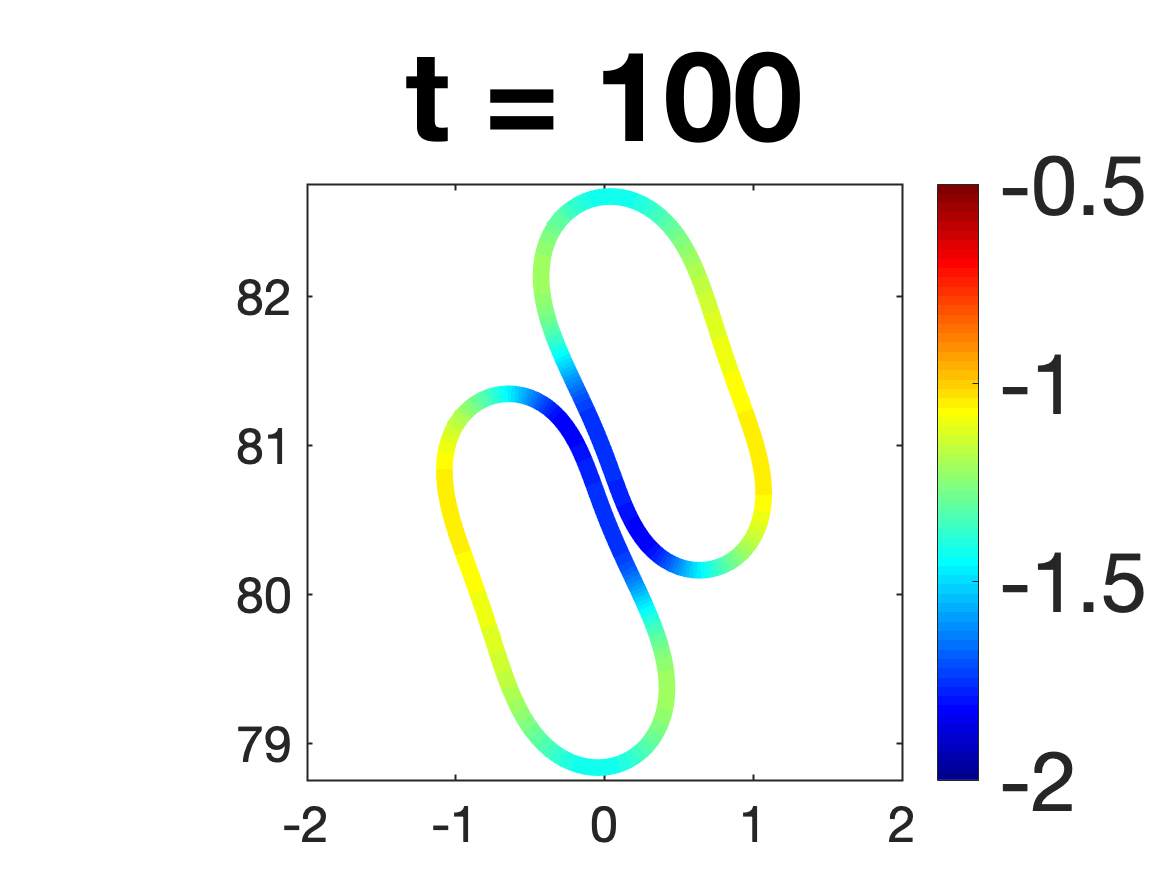} \\

  \includegraphics[height = 0.2\textwidth,trim={1cm 0cm 1cm 0cm},clip]
    {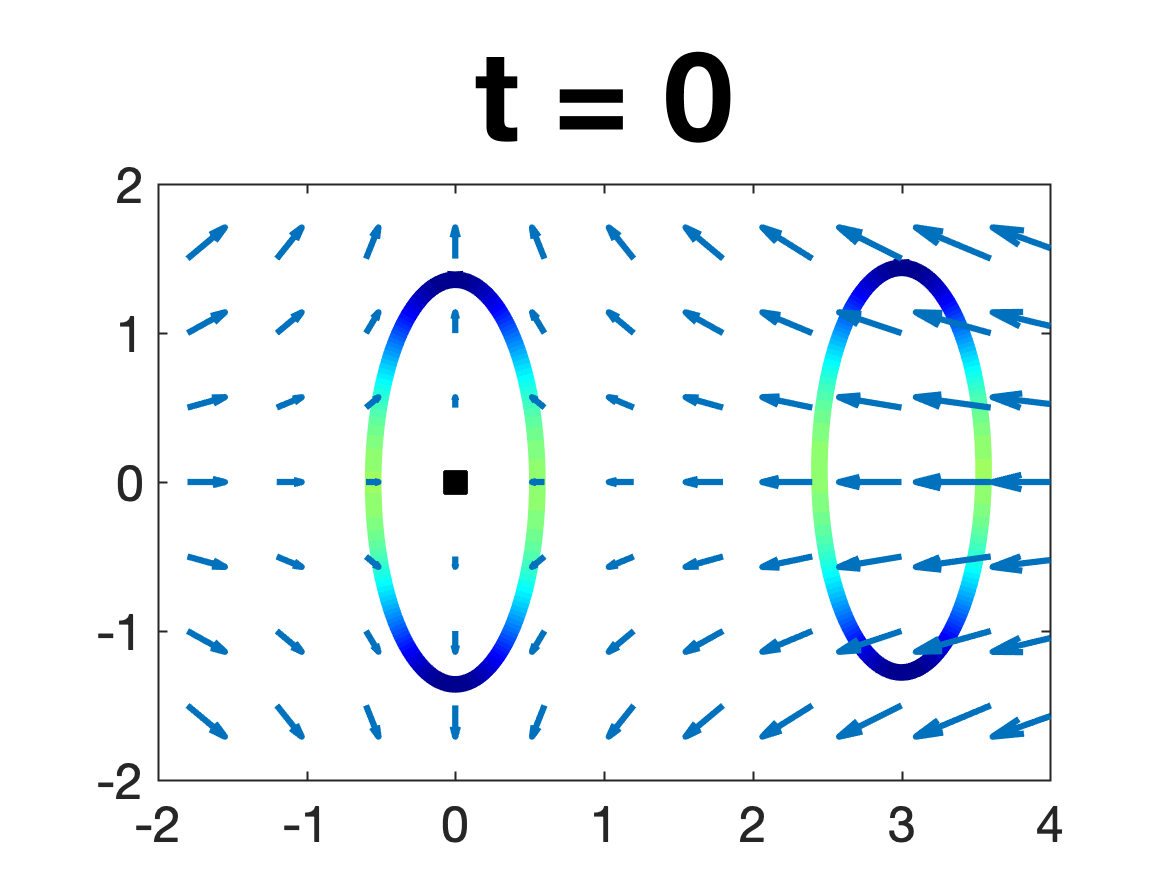}
  \includegraphics[height = 0.2\textwidth,trim={4cm 0cm 4cm 0cm},clip]
    {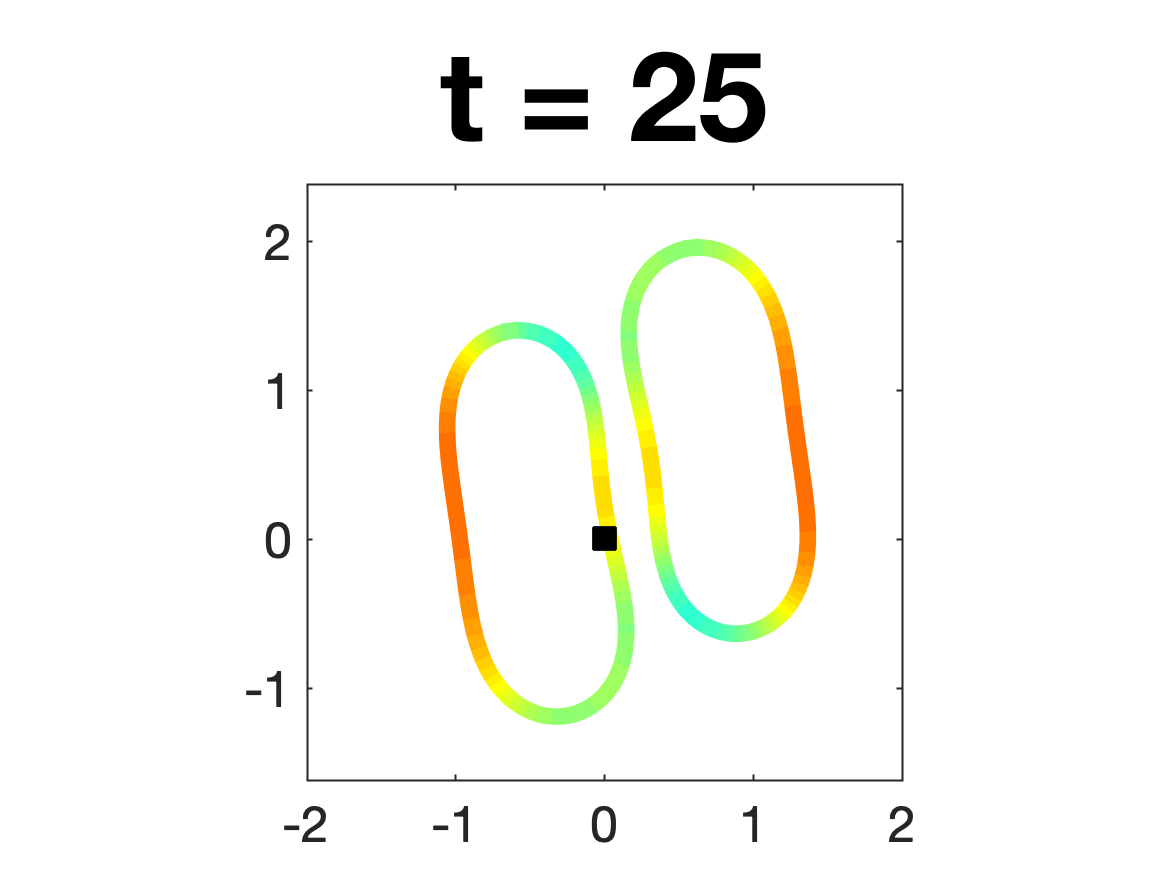}
  \includegraphics[height = 0.2\textwidth,trim={4cm 0cm 4cm 0cm},clip]
    {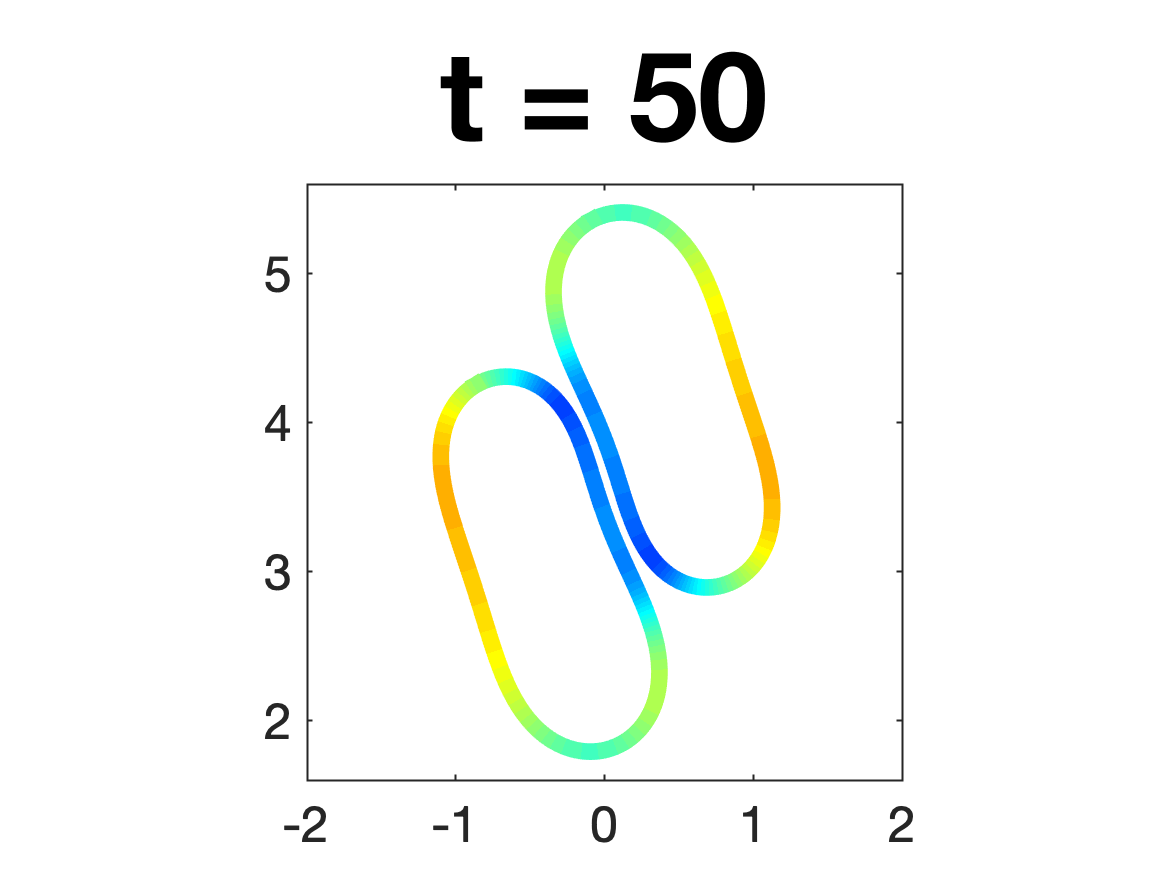}
  \includegraphics[height = 0.2\textwidth,trim={4cm 0cm 4cm 0cm},clip]
    {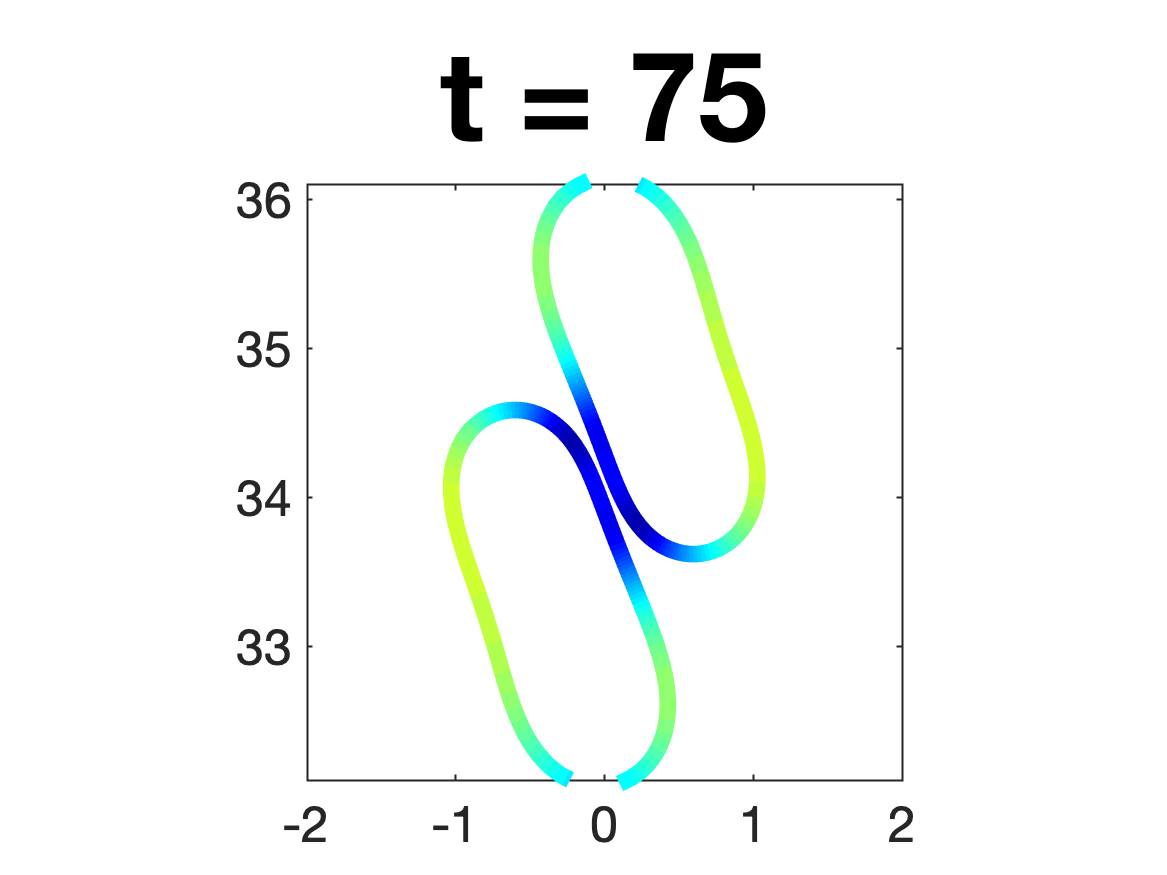}
  \includegraphics[height = 0.2\textwidth,trim={3.5cm 0cm 0cm 0cm},clip]
    {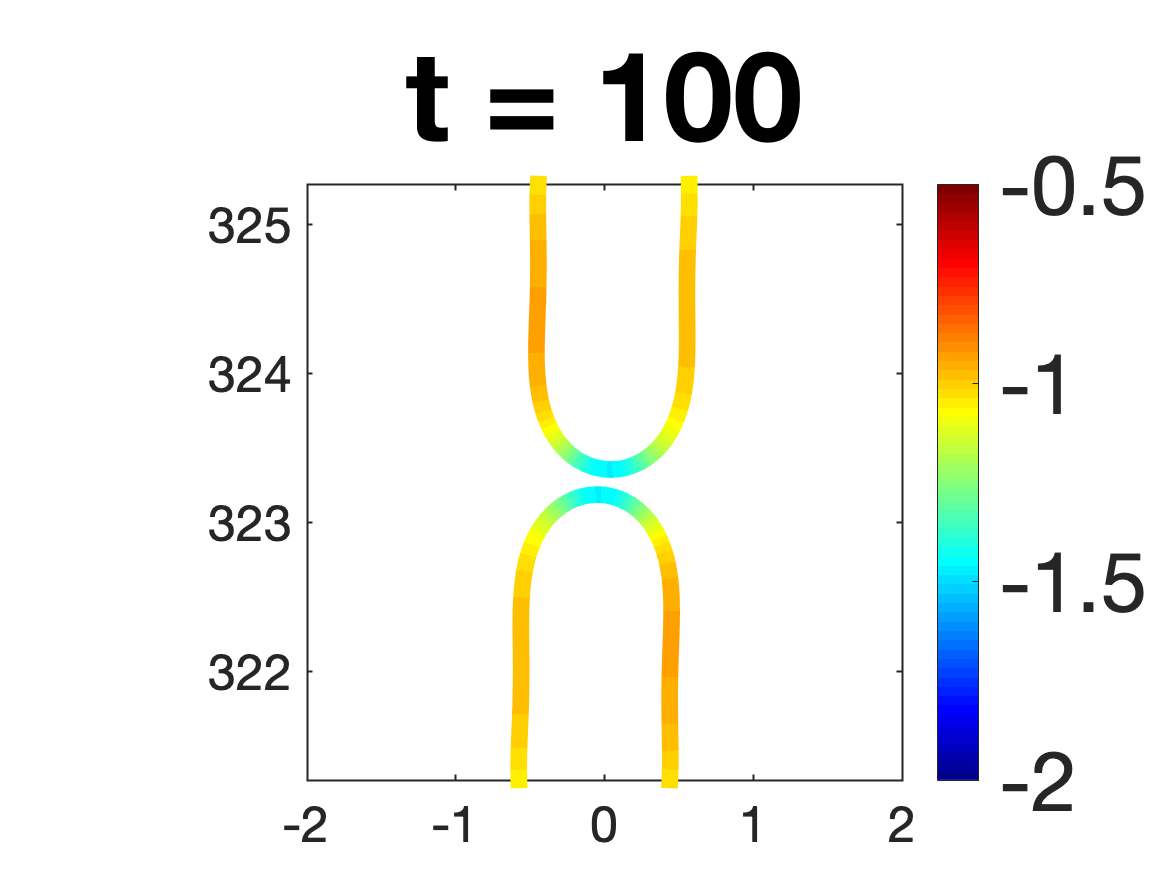} \\

  \includegraphics[height = 0.2\textwidth,trim={1cm 0cm 1cm 0cm},clip]
    {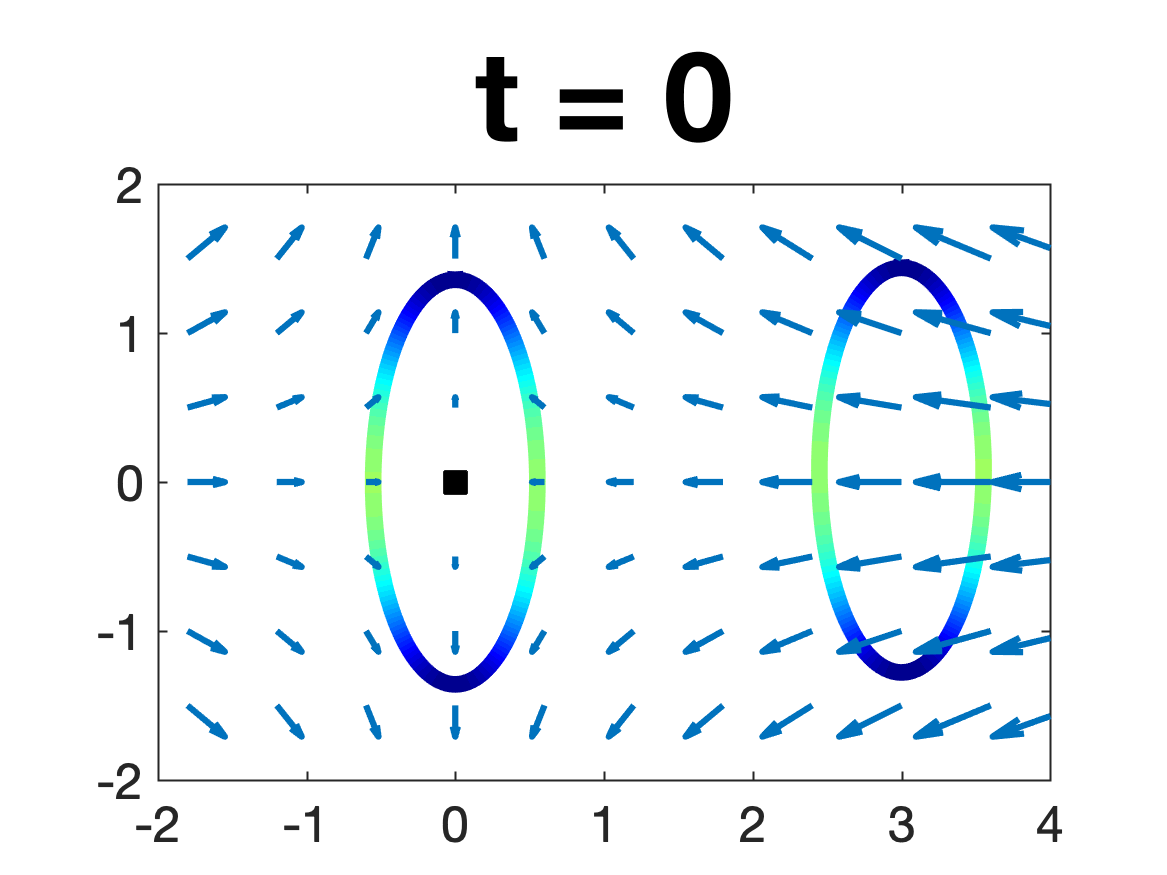}
  \includegraphics[height = 0.2\textwidth,trim={4cm 0cm 4cm 0cm},clip]
    {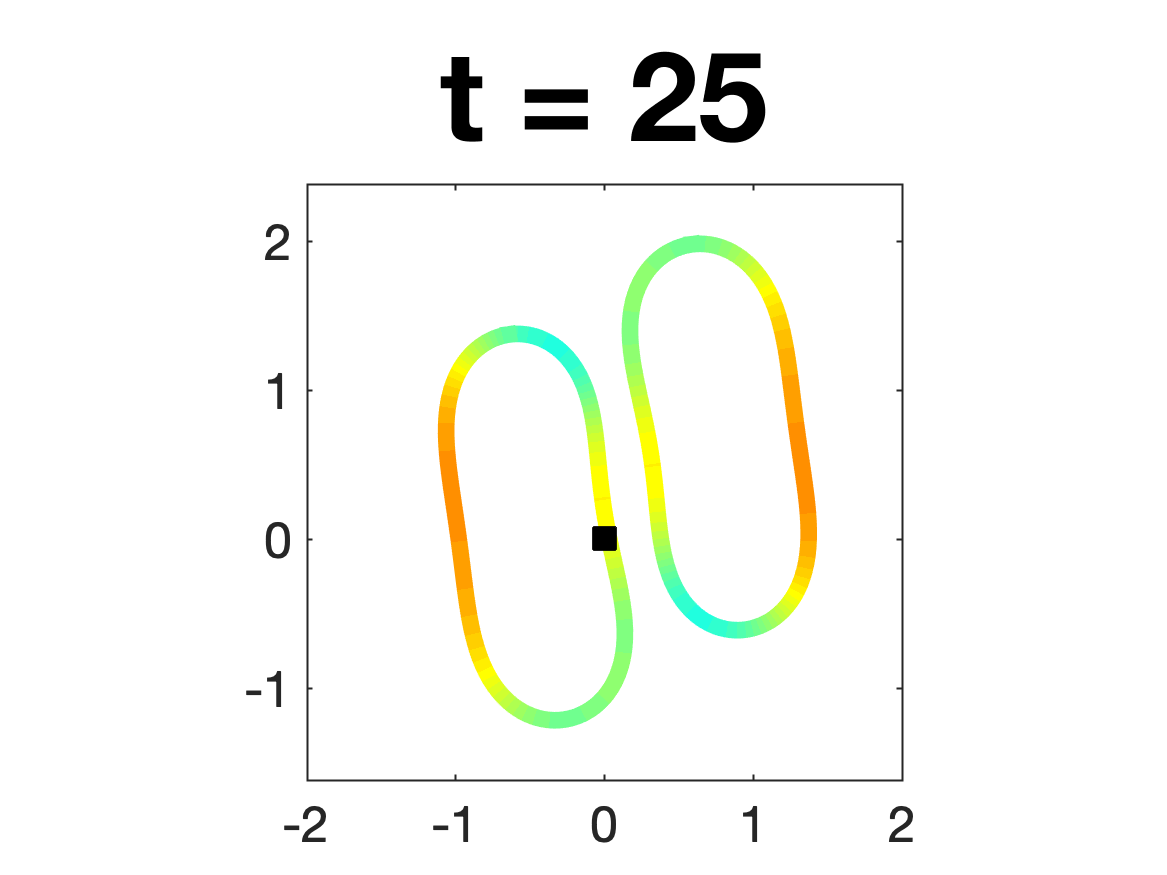}
  \includegraphics[height = 0.2\textwidth,trim={4cm 0cm 4cm 0cm},clip]
    {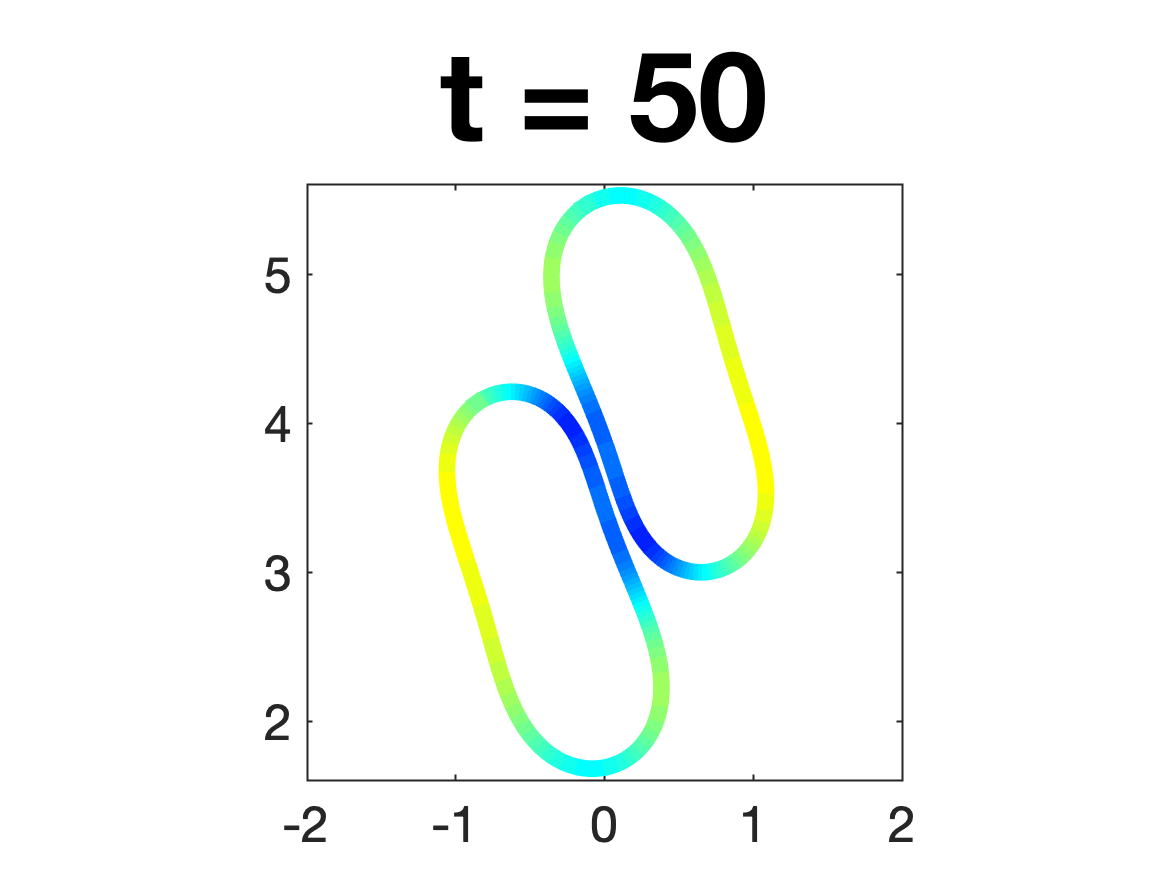}
  \includegraphics[height = 0.2\textwidth,trim={4cm 0cm 4cm 0cm},clip]
    {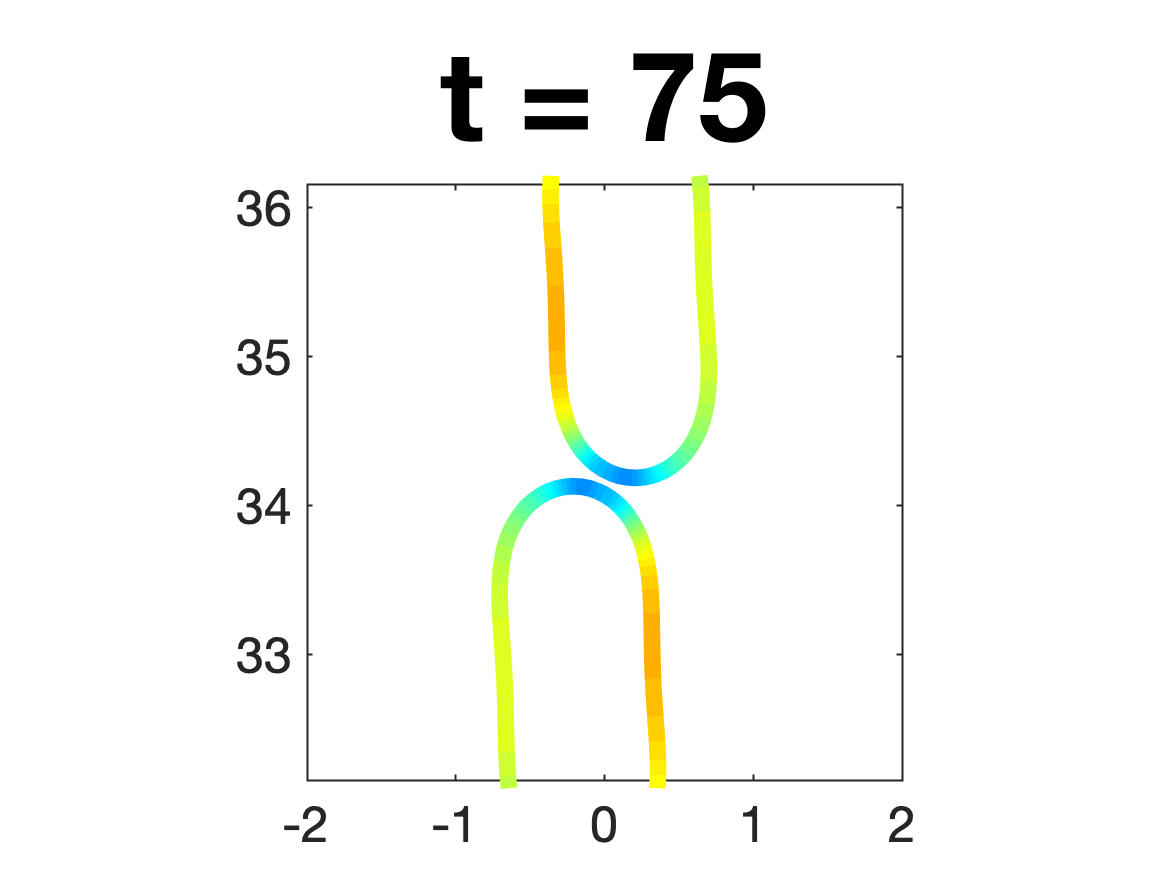}
  \includegraphics[height = 0.2\textwidth,trim={3.5cm 0cm 0cm 0cm},clip]
    {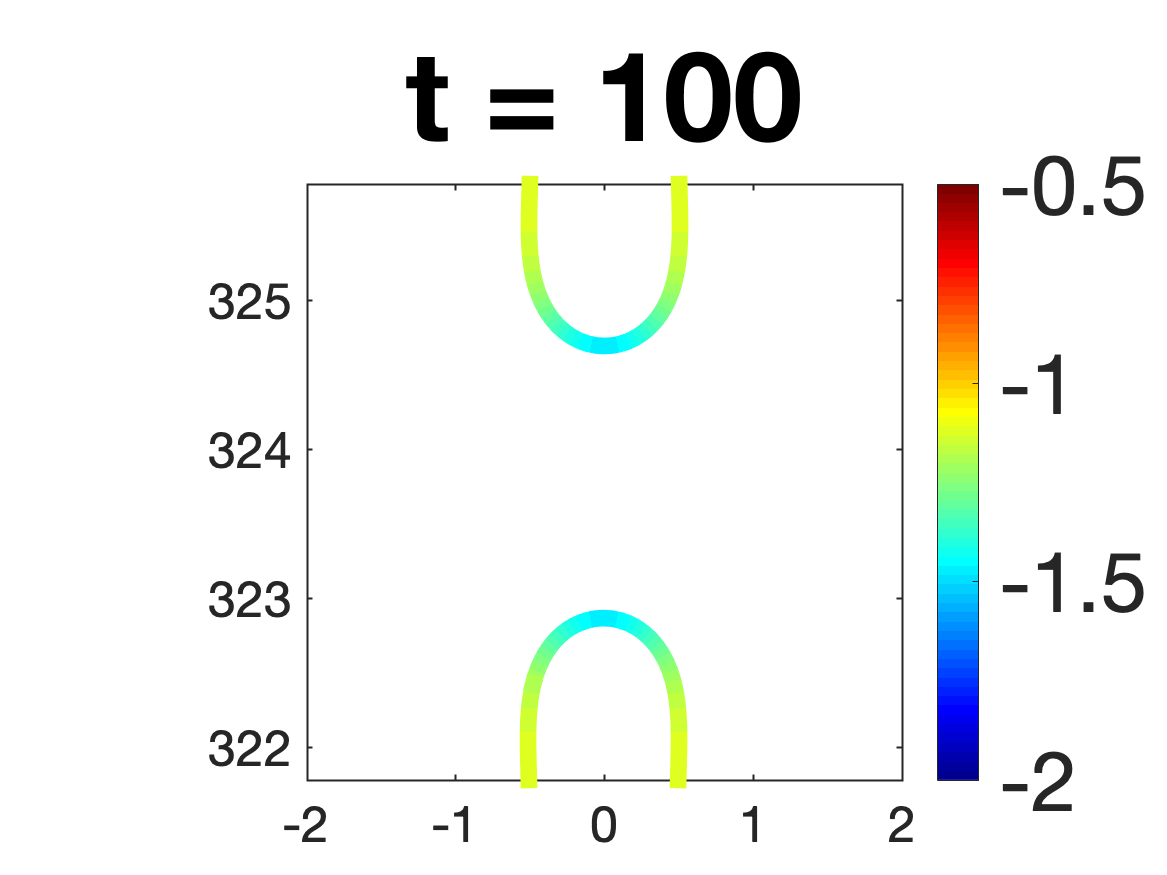}
  \caption{\label{fig:extensionalNonsymmetric} The dynamics of a vesicle
  pair in a planar extensional flow. The Hamaker constant is ${\cal H} =
  0.7$ for the first three rows, and ${\cal H}=0.6$ for the bottom row.
  From top to bottom, the initial vertical displacement is $\Delta y=0$,
  $0.02$, $0.1$, and $0.1$. The reduced area is $\Delta A = 0.75$, the
  extension rate $\dot\epsilon= 0.09$, and the adhesion distance $\delta
  = 0.4$.  The color coding is the tension.}
  \end{figure}

The hydrodynamics of a single vesicle in an extensional flow has
revealed novel nonlinear vesicle dynamics not found for a viscous
drop~\cite{KantslerSegreSteinberg2007_PRL,
KantslerSegreSteinberg2008_PRL, ZhaoShaqfeh2013_JFM, Narsimhan2014_JFM,
DahlNarsimhanGouveia2016_SoftMatt}.  Here we focus on a planar
extensional flow  where the fluid velocity field
$\uu=\dot\epsilon(-x,y)$ is centered at the origin ($(x,y)=(0,0)$), the
stagnation point where the fluid flow converges horizontally and
diverges vertically.  Placed symmetrically at the stagnation point of an
extensional flow, a single vesicle may remain steady and symmetric for
low extension rate $\dot\epsilon$.  For sufficiently large extension
rate $\dot\epsilon$ the vesicle may undergo asymmetric
deformation~\cite{KantslerSegreSteinberg2008_PRL, Narsimhan2014_JFM,
DahlNarsimhanGouveia2016_SoftMatt}, similar to a surfactant-laden
viscous drop under a planar extensional
flow~\cite{JanssenBoonAgterof1997_AIChE, HuPineLeal2000_PoF}.  Frostad
{\it et al.} investigated the draining of thin film between two
identical vesicles as they collide head-on in an extensional flow
\cite{FrostadWalterLeal2013_PoF}, however they did not include adhesive
interaction between vesicles.  In the following we first illustrate how
two vesicles interact with each other under an adhesive potential in a
planar extensional flow. We then illustrate how the adhesion strength
between membranes can be measured non-intrusively by using the fluid
trap.

At the beginning of the first set of simulations (left panels in
Figure~\ref{fig:extensionalNonsymmetric}) we place a vesicle at
$(x,y)=(0,0)$, and the other vesicle at $(x,y) = (3,\Delta y)$ where the
initial vertical displacement $\Delta y$ varies from $\Delta y=0$ for
the top row, $\Delta y = 0.02$ for the second row, to $\Delta y =0.1$
for the bottom two rows in Figure~\ref{fig:extensionalNonsymmetric}.
The Hamaker constant is ${\cal H} = 0.7$ for the top three rows, and is
${\cal H} =0.6$ for the bottom row.  The color coding is the membrane
tension with the color bar on the right of each row. We note that the
negative tension in an extensional flow is consistent with the
destabilizing tension for a vesicle with finite bending
forces~\cite{Narsimhan2014_JFM}.

Once the planar extensional flow is turned on at $t=0^+$, two vesicles
move towards each other as shown in
Figure~\ref{fig:extensionalNonsymmetric}.  For $\Delta y=0$ in the top
row, the two vesicles stay on the $x$-axis as they move closer to form a
doublet.  We observe that the left vesicle (initially centered at
$(0,0)$) moves to the left due to the impinging vesicle from the right.
As time progresses they form a doublet and reach a steady configuration
that is symmetric with respect to the stagnation point (square marks in
the top row of Figure~\ref{fig:extensionalNonsymmetric}).

For $\Delta y>0$ we observe similar vesicle dynamics only at the
beginning when the right vesicle moves towards the stagnation point and
the left vesicle is pushed to the left of the stagnation point.  As both
vesicles come to close vicinity of the stagnation point, the elevated
right vesicle ($\Delta y > 0$) is pulled up by the diverging flow in the
$y$ direction, causing rotation of both vesicles.  We also observe that
the larger the initial vertical displacement, the larger the rotation
and the faster both vesicles move away from the stagnation point. For
sufficiently strong adhesion (${\cal H}=0.7$ for the first three rows in
Figure~\ref{fig:extensionalNonsymmetric}), the two vesicles rotate and
move away from the stagnation point as a doublet until the end of
simulations ($t=100$).  As the two vesicles move farther away from the
stagnation point, the extensional flow becomes more effective in pulling
the two drops apart as shown in the third row of
Figure~\ref{fig:extensionalNonsymmetric}.  For ${\cal H}=0.6$ (bottom
row in Figure~\ref{fig:extensionalNonsymmetric}), the two vesicles are
separated by the diverging flow as early as $t=75$, and are clearly not
a doublet at the end of simulation.

Here we provide two main observations drawn from the results in
Figure~\ref{fig:extensionalNonsymmetric}: (1) The membrane tension is
highly non-uniform on the vesicles, leading to a significant Marangoni
stress along the membrane as in the case of a surfactant-laden drop at
the center of an extensional flow. Thus it is reasonable to expect that
the flow around the vesicles is significantly altered.  (2) In an
extensional flow, vesicles are prone to move away from the stagnation
point.  A small initial vertical displacement $\Delta y$ in one  of the
vesicles causes both vesicles to not only rotate but also move away from
the center. Once vesicles move off the horizontal axis, their
hydrodynamics is dominated by the diverging flow as they move farther
away.  Thus in the microfluidic experiment it is often more desirable to
trap the vesicle around the center of an extensional
flow~\cite{Spjut2010_MSThesis_Chapter3}. With an active control
algorithm to place the stagnation point at a desirable location by
adjusting the flow with a feedback loop~\cite{BentleyLeal1986_JFMa,
Johnson-Chavarria2011_EMJ}, a particle can be trapped at the stagnation
point for long time scales to facilitate image acquisition or other
detailed measures such as particle image velocimetry of flow inside and
around the particle.  

To explore the application of such a fluid trap to measure the adhesion
strength between two bound vesicles, we propose the following
experiment.  Beginning with two identical vesicles in an equilibrium
configuration that form a doublet with a flat contact region
(Figure~\ref{fig:Dec18_vesicle_shape}), we turn on the fluid trap with
the stagnation point placed at the center of the vesicle doublet.
Depending on the Hamaker constant, we expect that either the flow
overcomes the adhesive force and the doublet is broken, or the adhesive
force is sufficiently strong and the doublet reaches a stable stationary
configuration.

\begin{figure}[htp]
  \begin{tabular}{@{}p{0.3\linewidth}@{\quad}p{0.3\linewidth}@{\quad}p{0.3\linewidth}@{}}
  \subfigimg[width=\linewidth]{(a)}{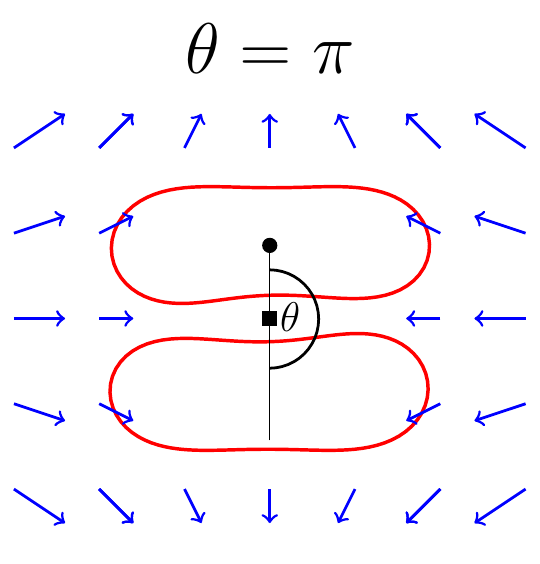} &
  \subfigimg[width=\linewidth]{(b)}{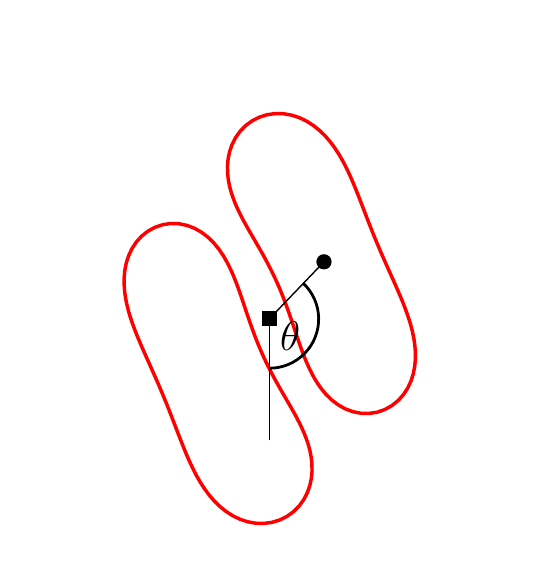} &
  \subfigimg[width=\linewidth]{(c)}{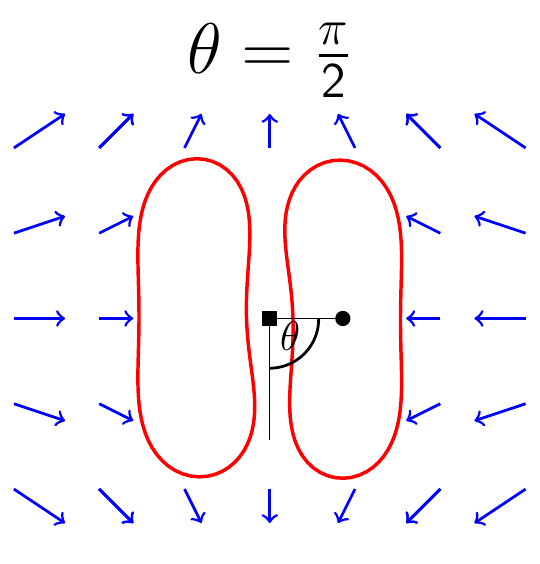}
  \end{tabular}
   \caption{\label{fig:InclinationAngle} The definition of the
   inclination angle of a vesicle doublet at the center of an
   extensional flow.  An inclination angle of $\pi$ indicates a doublet
   whose long axis is orthogonal to the diverging direction, while an
   inclination angle of $\pi/2$ indicates a doublet whose short axis is
   orthogonal to the diverging direction.  The square mark is a
   stagnation point and the round point is located at the center of the
   vesicle.  The initial configurations in this section all have an
   inclination angle of $\theta = \pi$.}
 \end{figure}

Based on how the contact region aligns with the extensional flow, we can
define the inclination angle $\theta$ of the vesicle doublet as
illustrated in Figure~\ref{fig:InclinationAngle}(b). When $\theta=\pi$
(Figure~\ref{fig:InclinationAngle}(a)) the diverging flow pulls the
vesicles apart from each other and the attractive adhesion force is
essential to keep the two vesicles from separating.  In contrast, when
$\theta = \pi/2$ (Figure~\ref{fig:InclinationAngle}(c)), the converging
flow pushes the vesicles towards each other and the repulsive force is
essential to keep the two vesicles at the separation distance.  With the
converging flow pushing the two vesicles towards the stagnation point
dynamically placed at the doublet center, we expect the $\theta=\pi/2$
configuration to be more stable than the $\theta=\pi$ configuration. 

\begin{figure}[htp]
  \includegraphics[height = 0.18\textwidth,trim={4cm 1cm 4cm 1cm},clip]
    {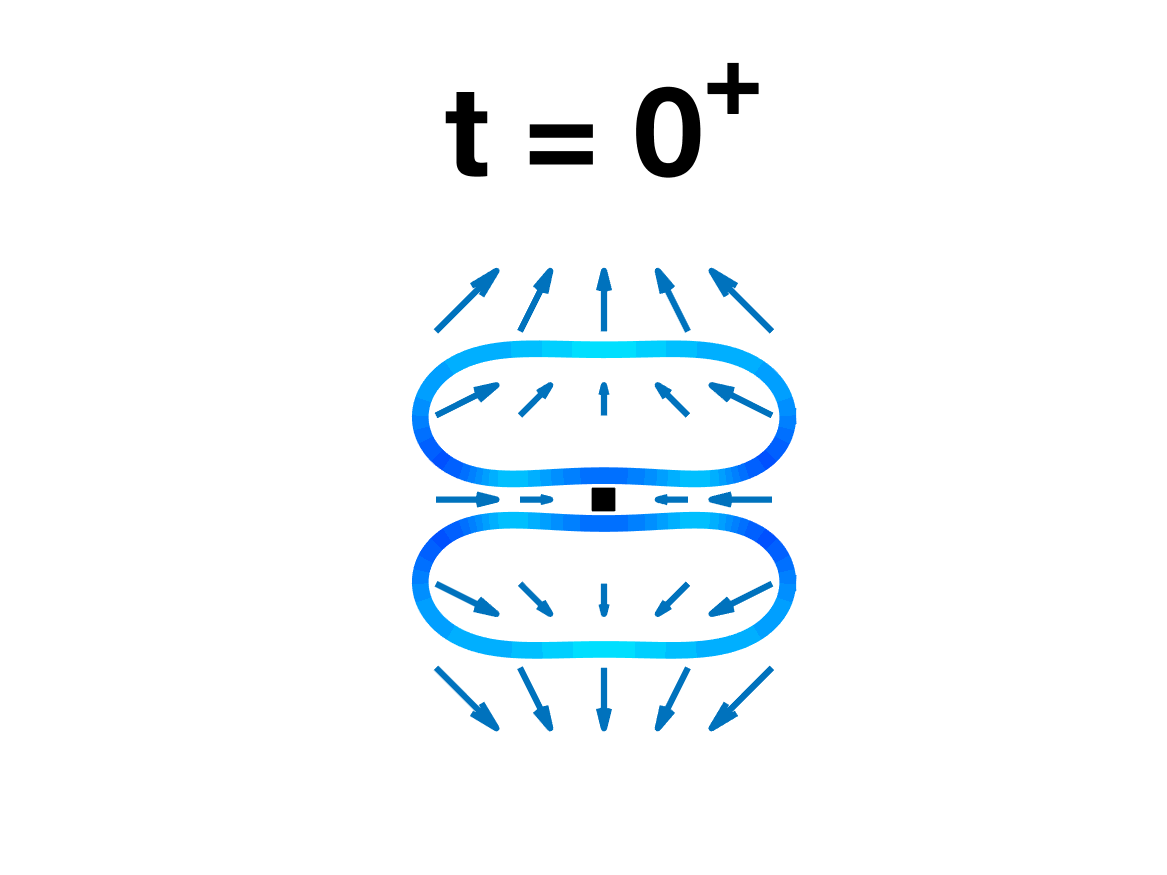}
  \includegraphics[height = 0.18\textwidth,trim={4cm 1cm 4cm 1cm},clip]
    {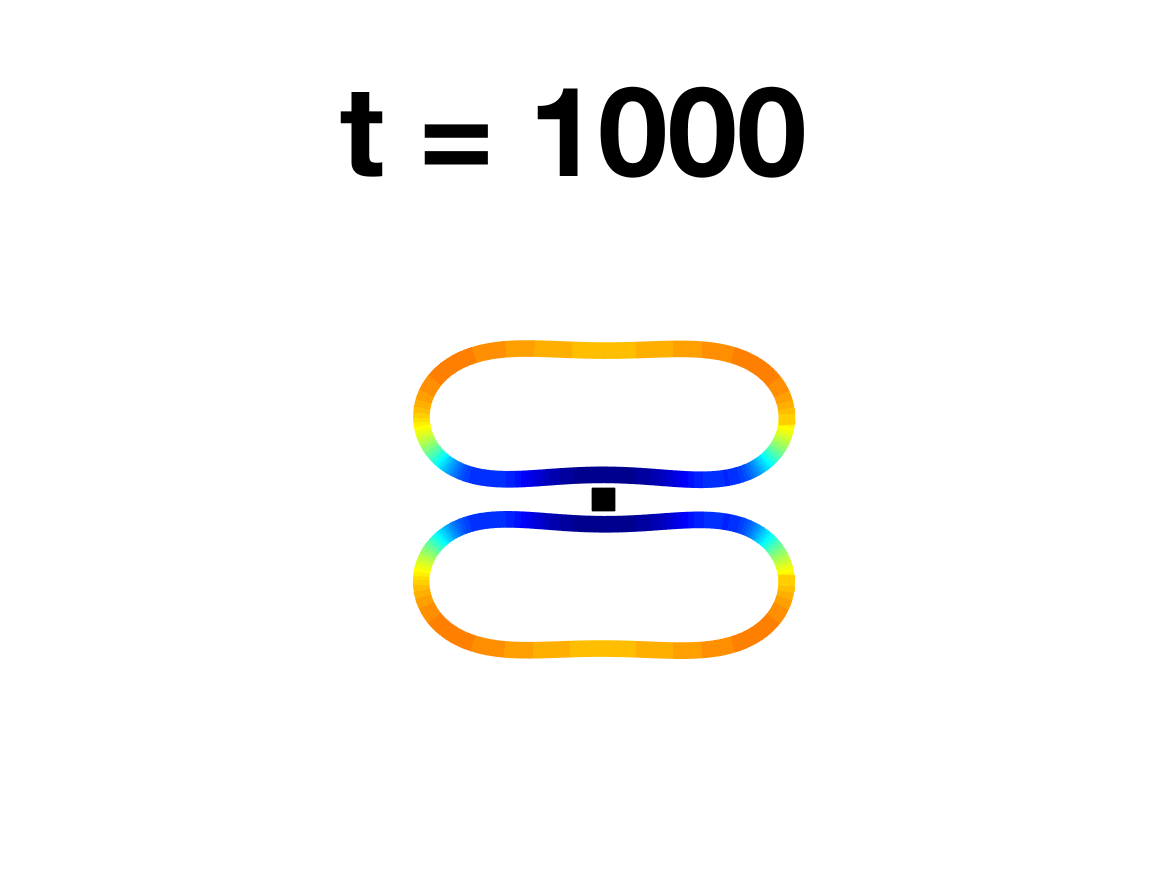}
  \includegraphics[height = 0.18\textwidth,trim={4cm 1cm 4cm 1cm},clip]
    {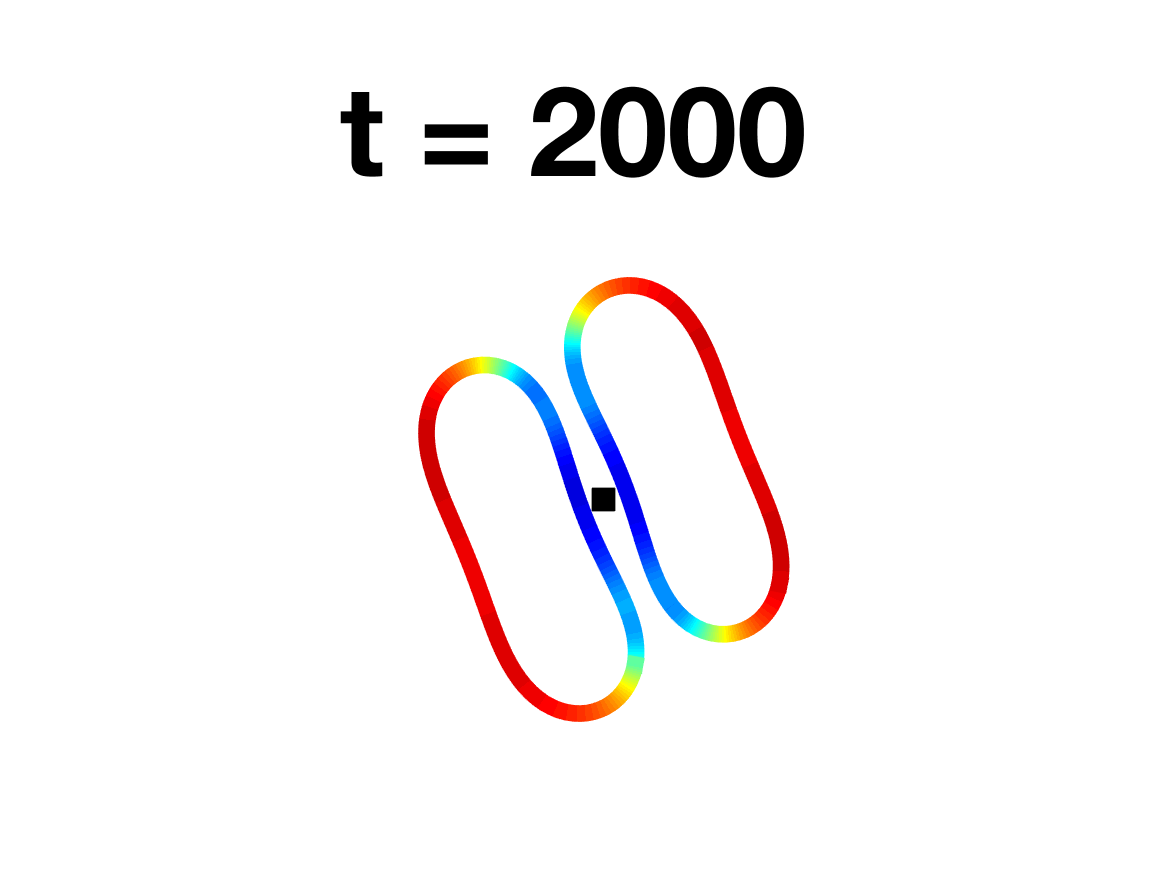}
  \includegraphics[height = 0.18\textwidth,trim={4cm 1cm 4cm 1cm},clip]
    {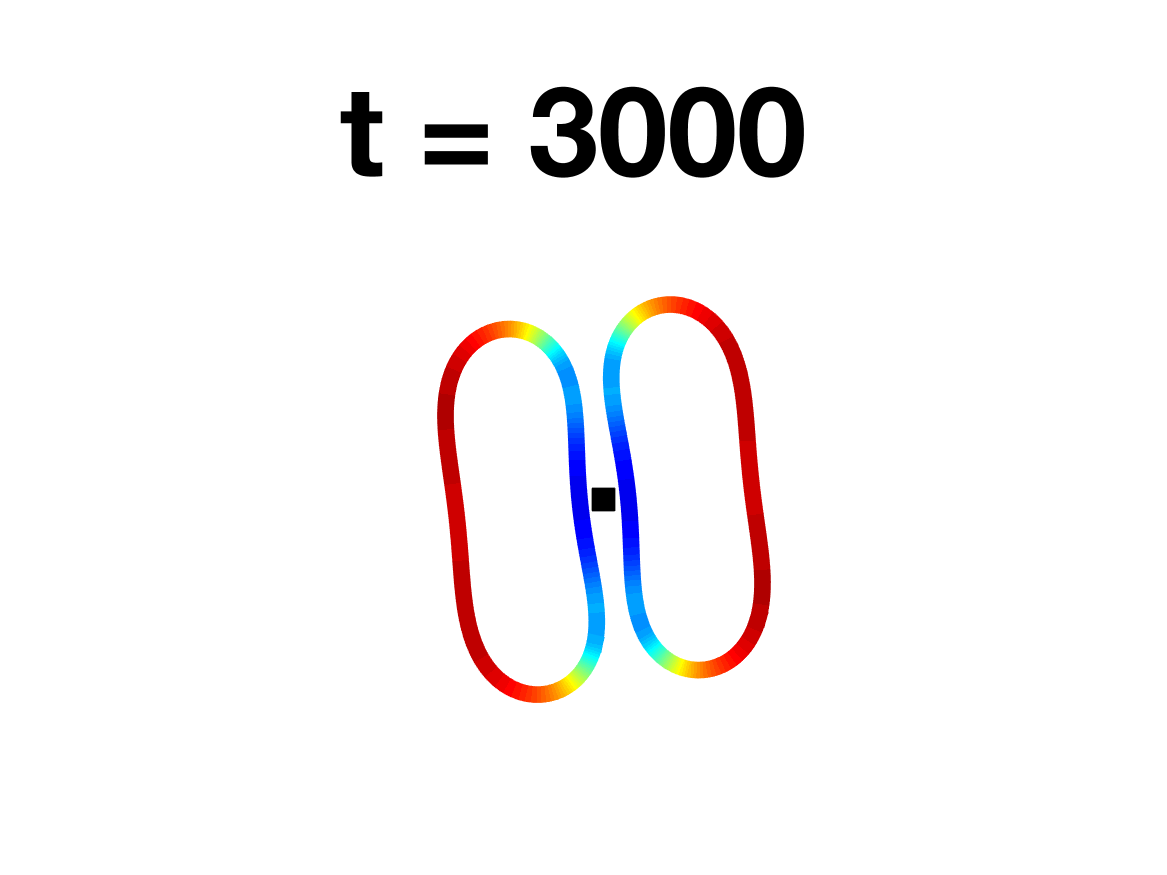}
  \includegraphics[height = 0.18\textwidth,trim={4cm 1cm 3cm 1cm},clip]
    {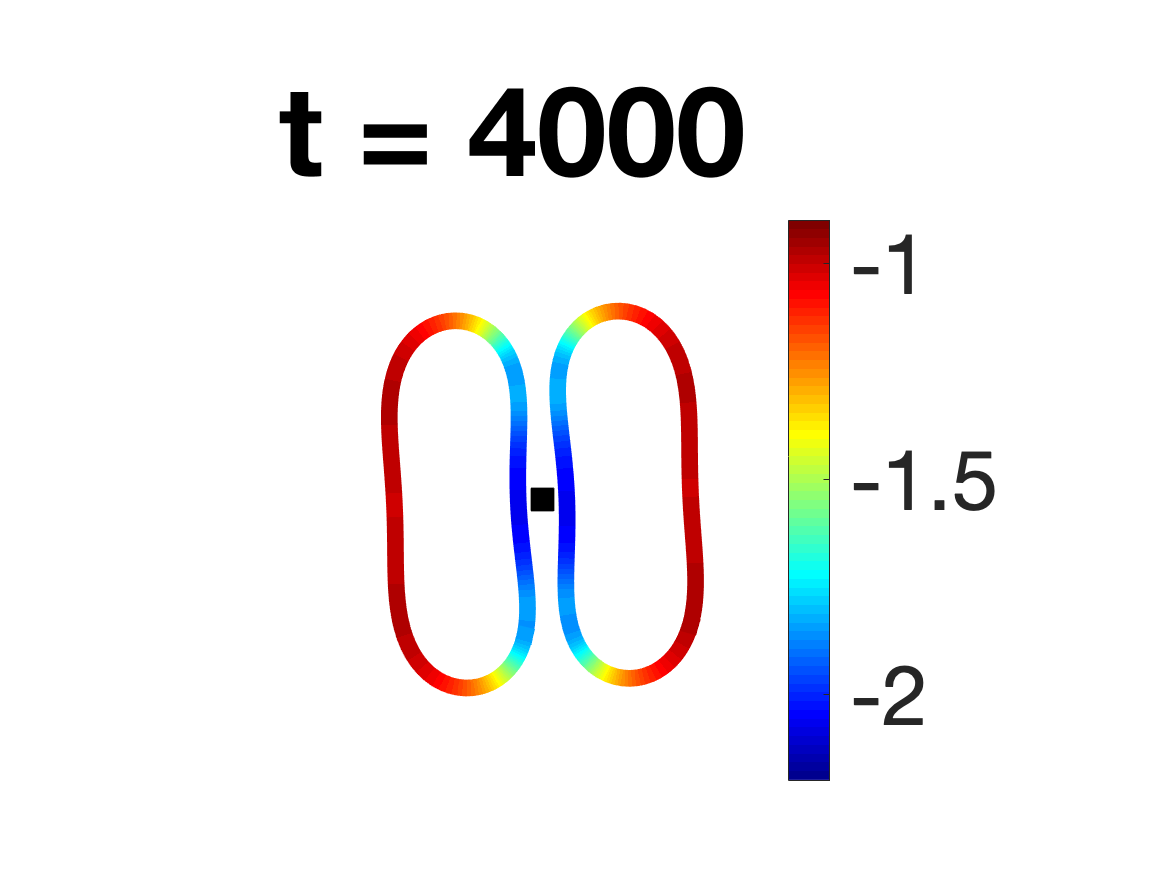}
    \\
  \includegraphics[height = 0.18\textwidth,trim={4cm 1cm 4cm 1cm},clip]
    {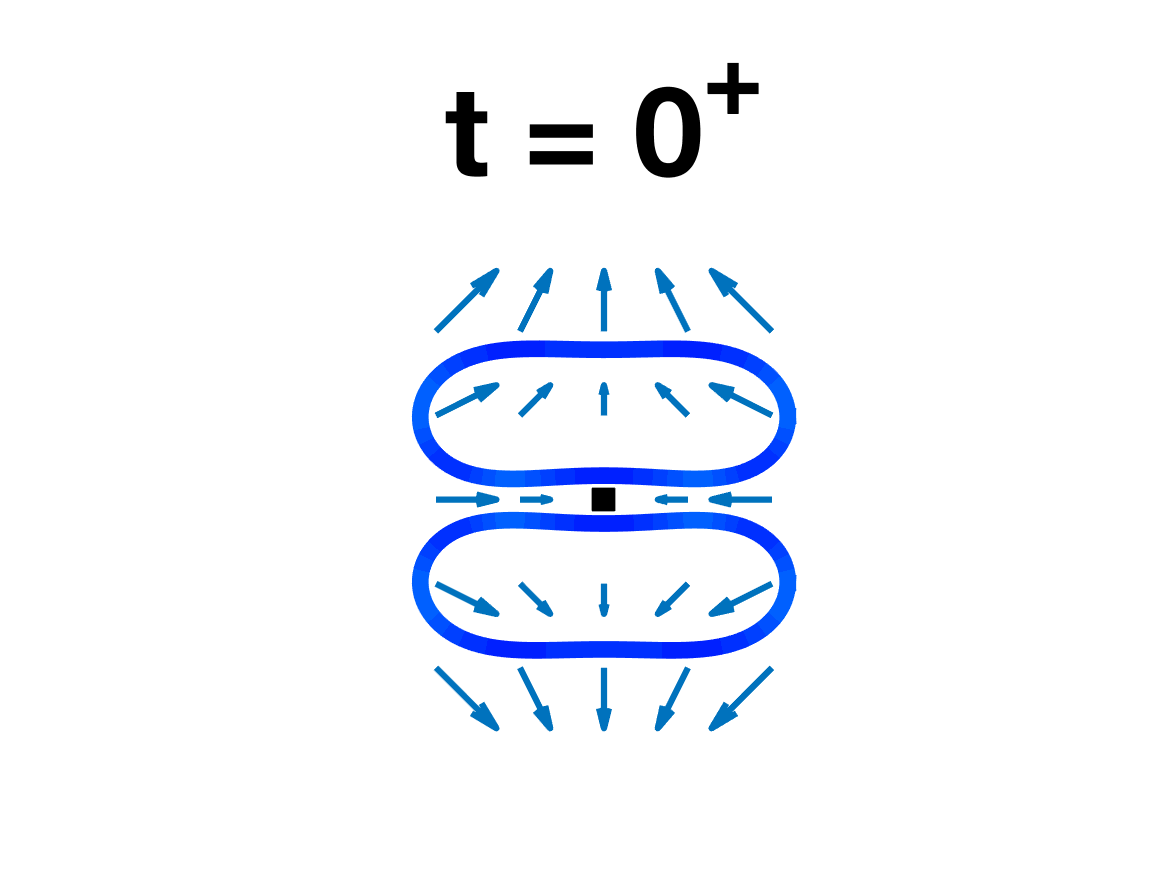}
  \includegraphics[height = 0.18\textwidth,trim={4cm 1cm 4cm 1cm},clip]
    {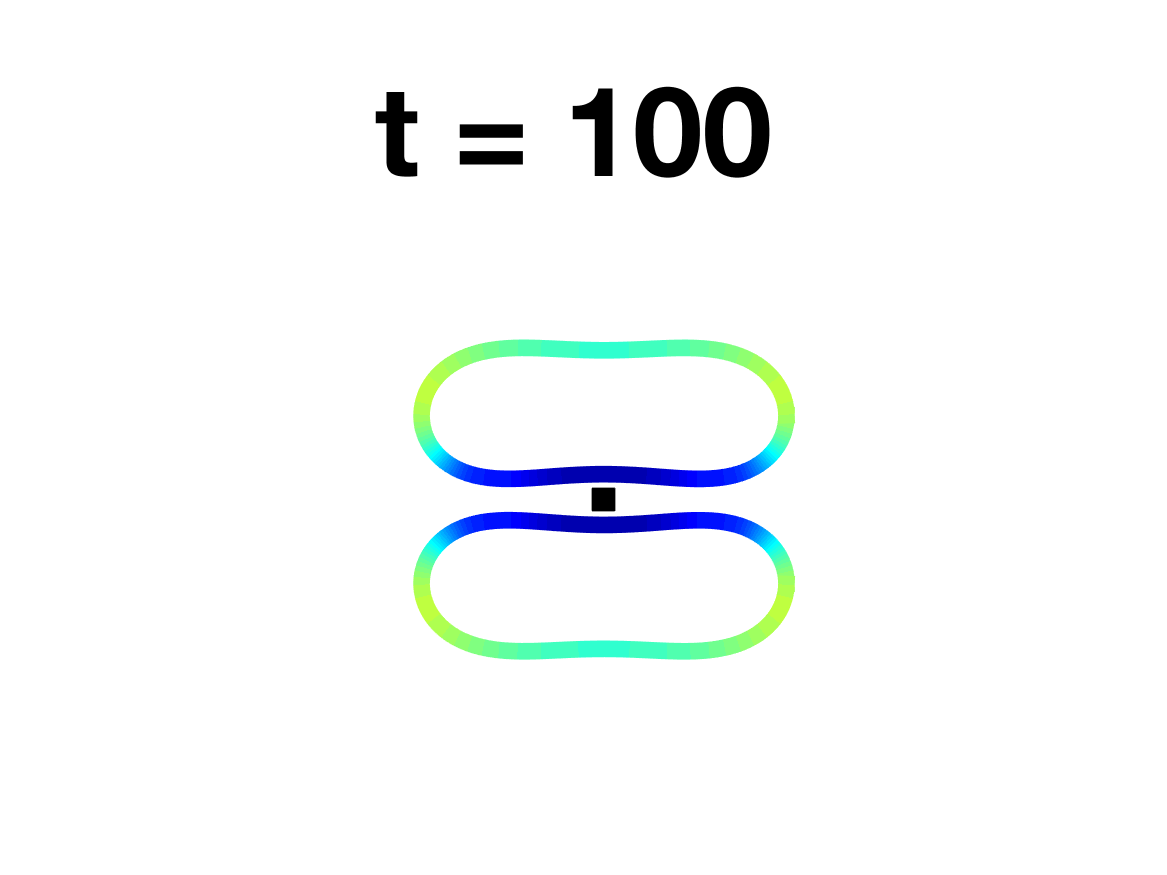}
  \includegraphics[height = 0.18\textwidth,trim={4cm 1cm 4cm 1cm},clip]
    {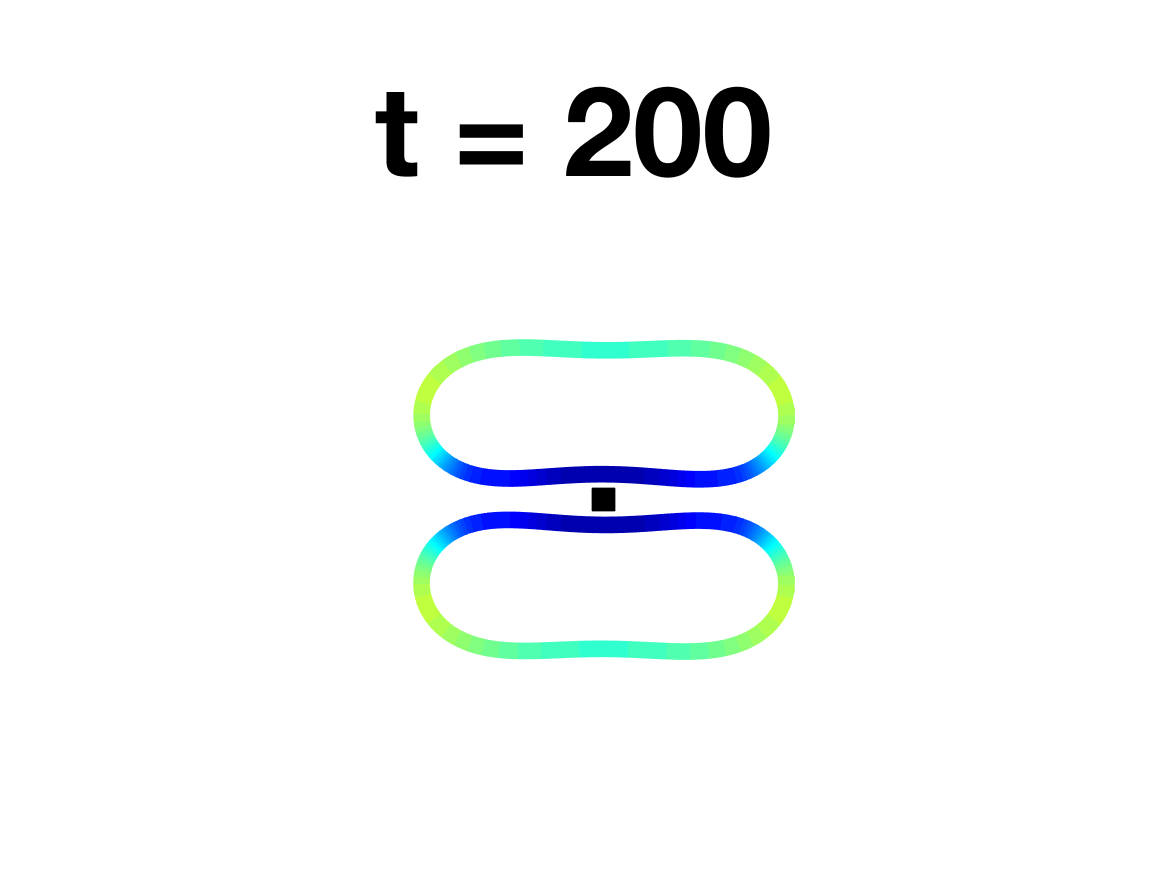}
  \includegraphics[height = 0.18\textwidth,trim={4cm 1cm 4cm 1cm},clip]
    {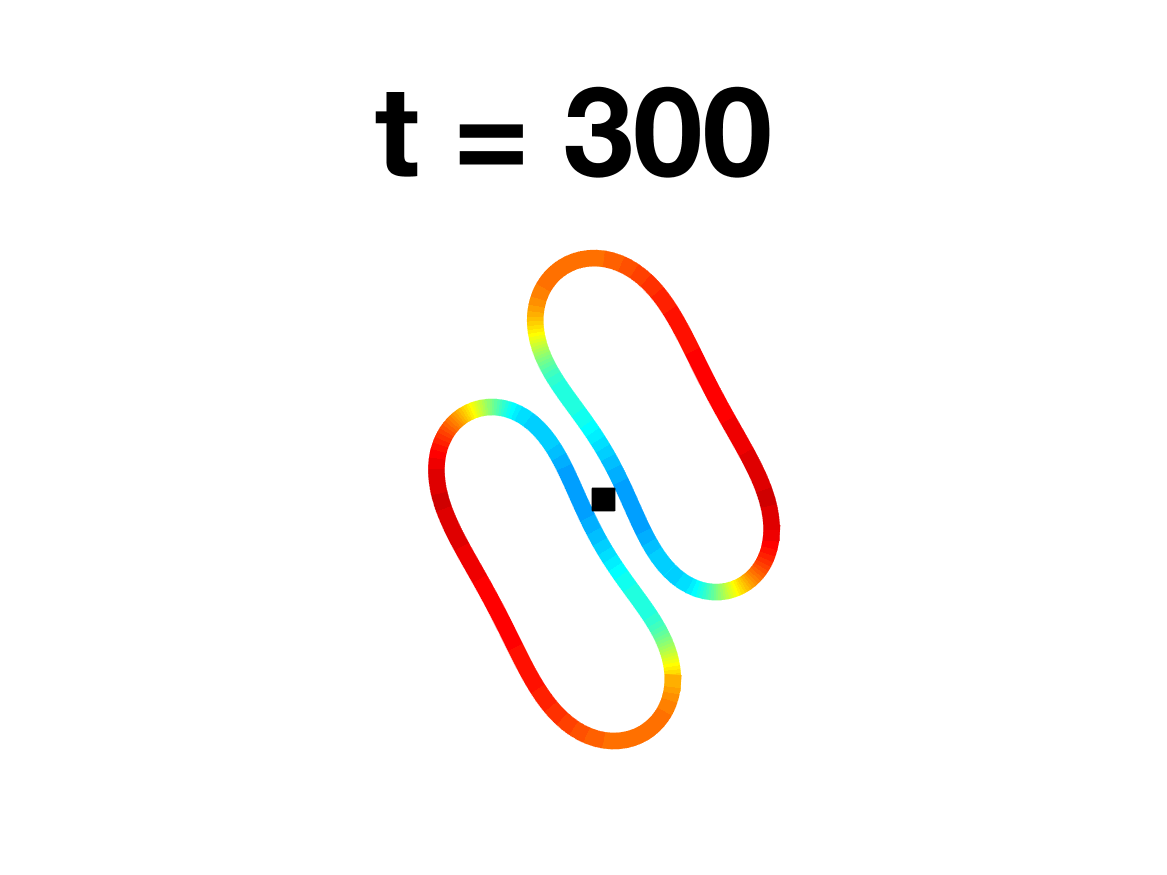}
  \includegraphics[height = 0.18\textwidth,trim={4cm 1cm 3cm 1cm},clip]
    {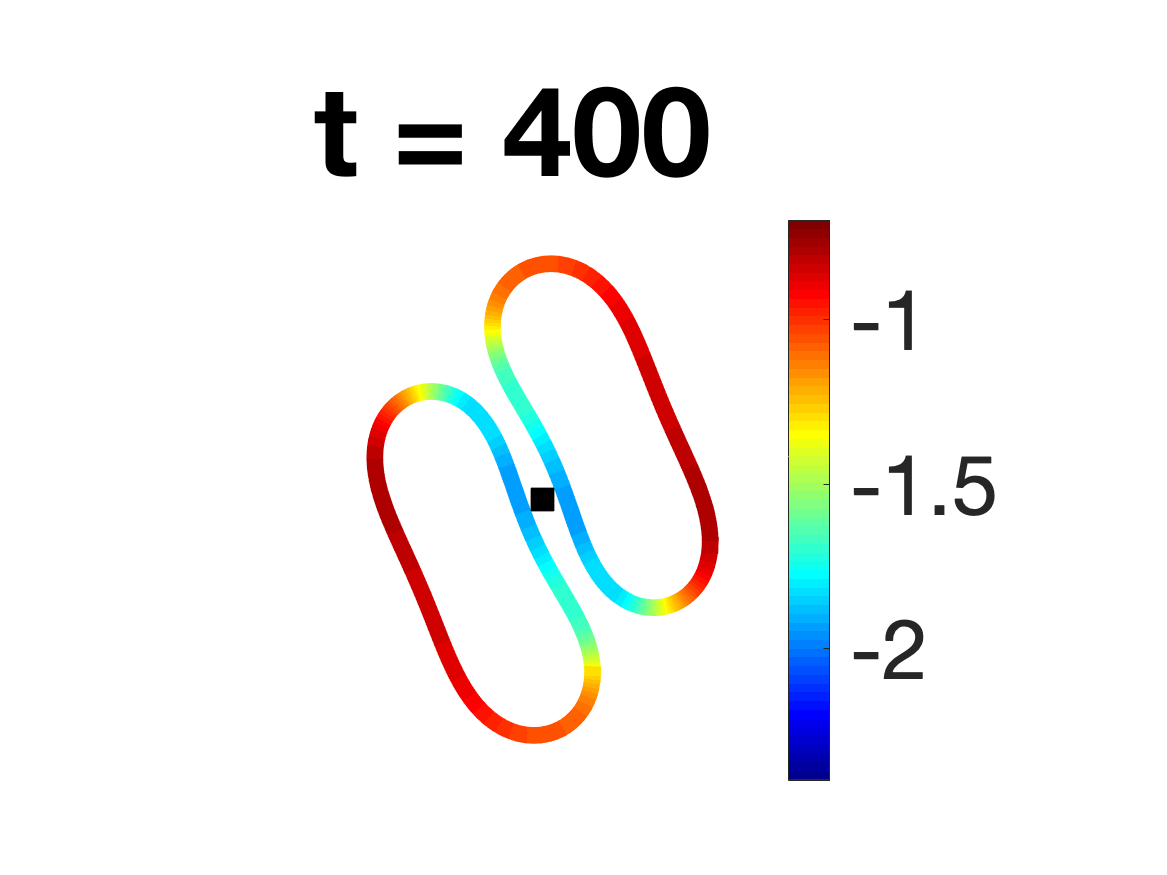}
    \\
  \includegraphics[height = 0.18\textwidth,trim={4cm 1cm 4cm 1cm},clip]
    {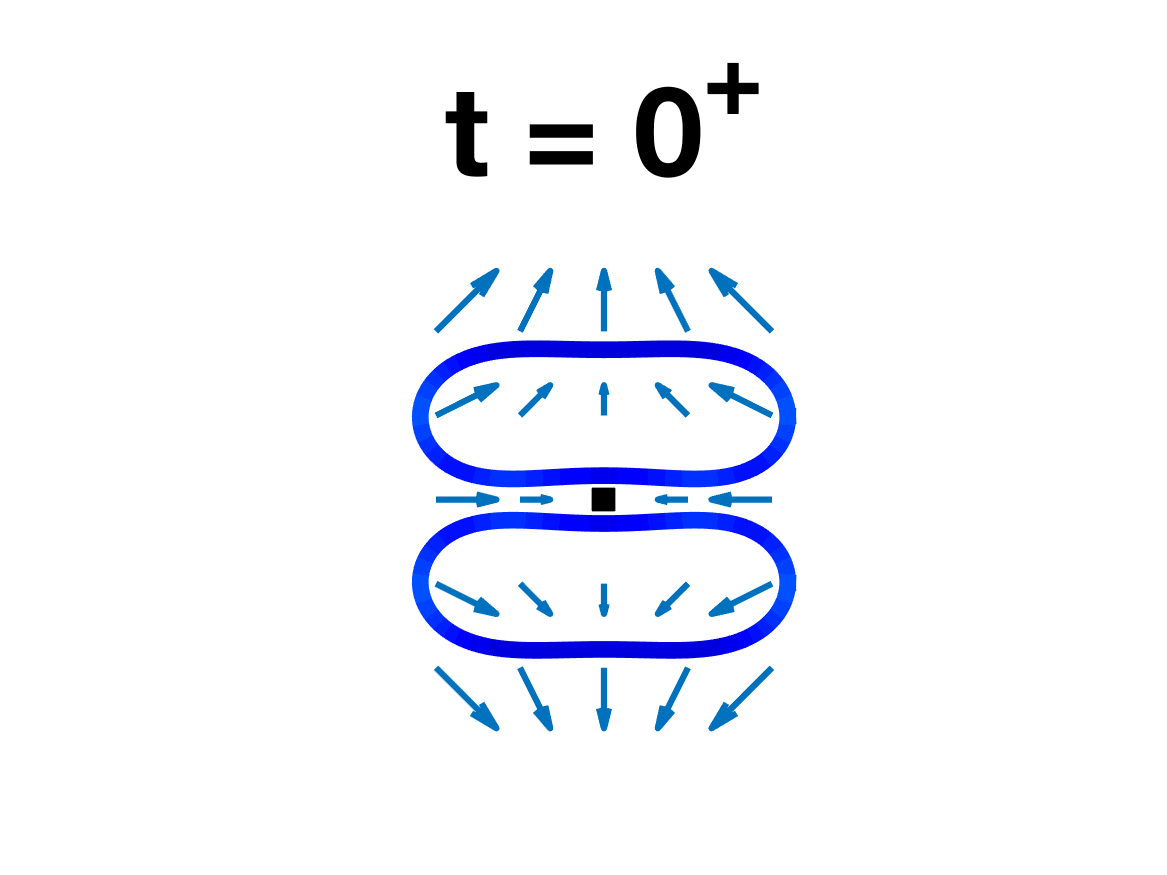}
  \includegraphics[height = 0.18\textwidth,trim={4cm 1cm 4cm 1cm},clip]
    {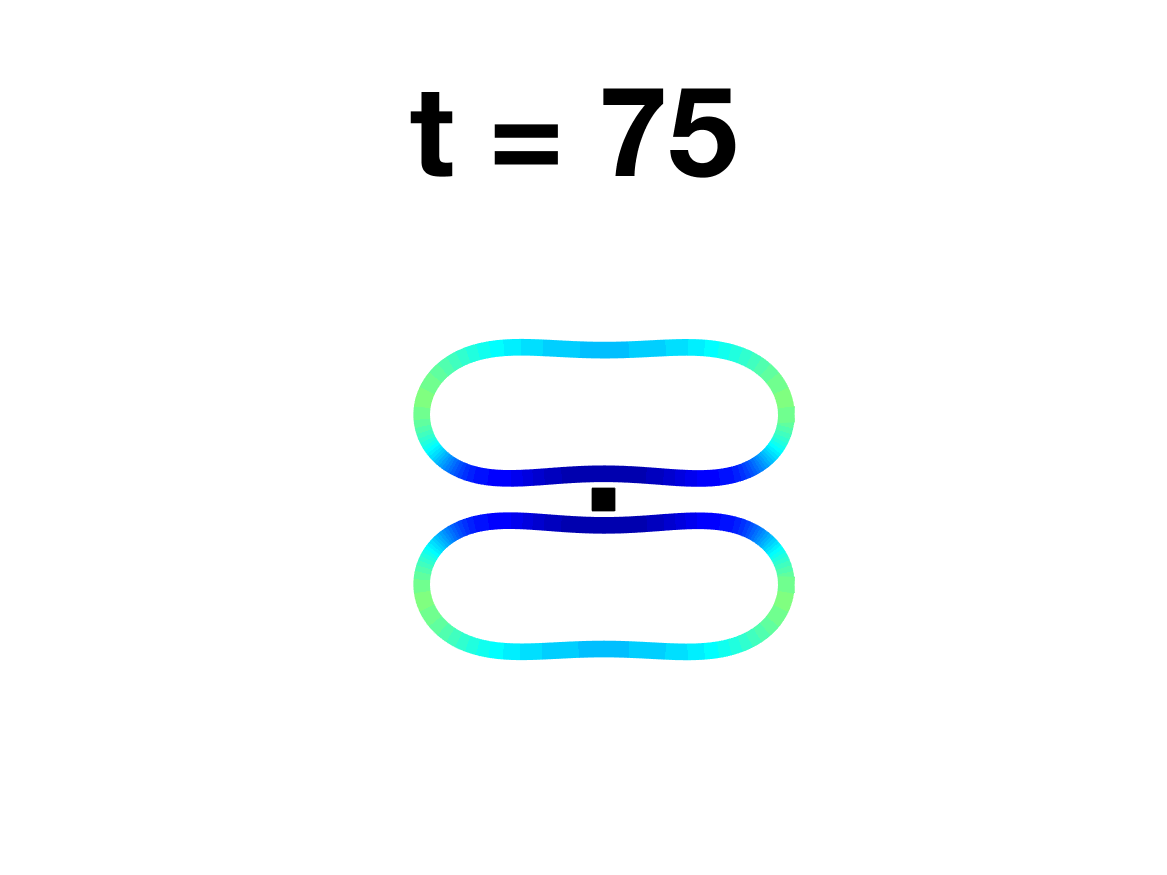}
  \includegraphics[height = 0.18\textwidth,trim={4cm 1cm 4cm 1cm},clip]
    {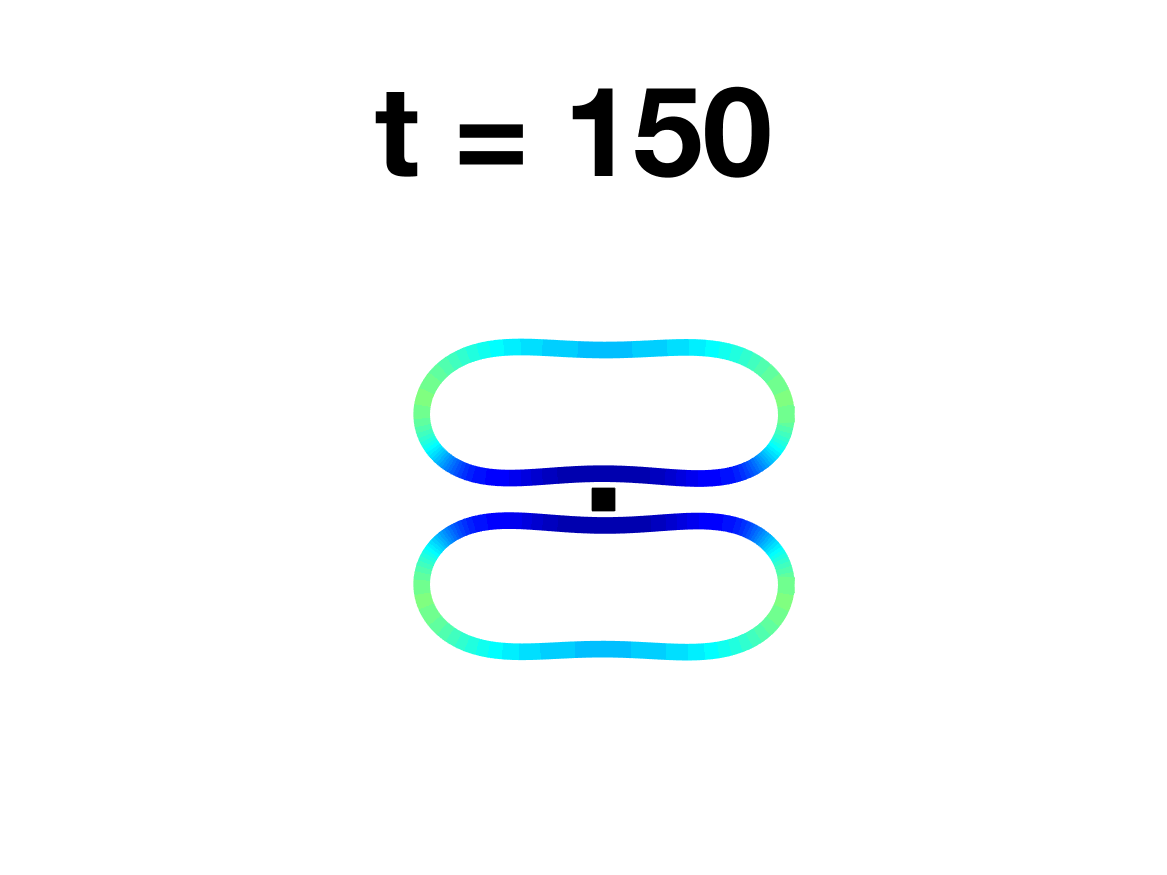}
  \includegraphics[height = 0.18\textwidth,trim={4cm 1cm 4cm 1cm},clip]
    {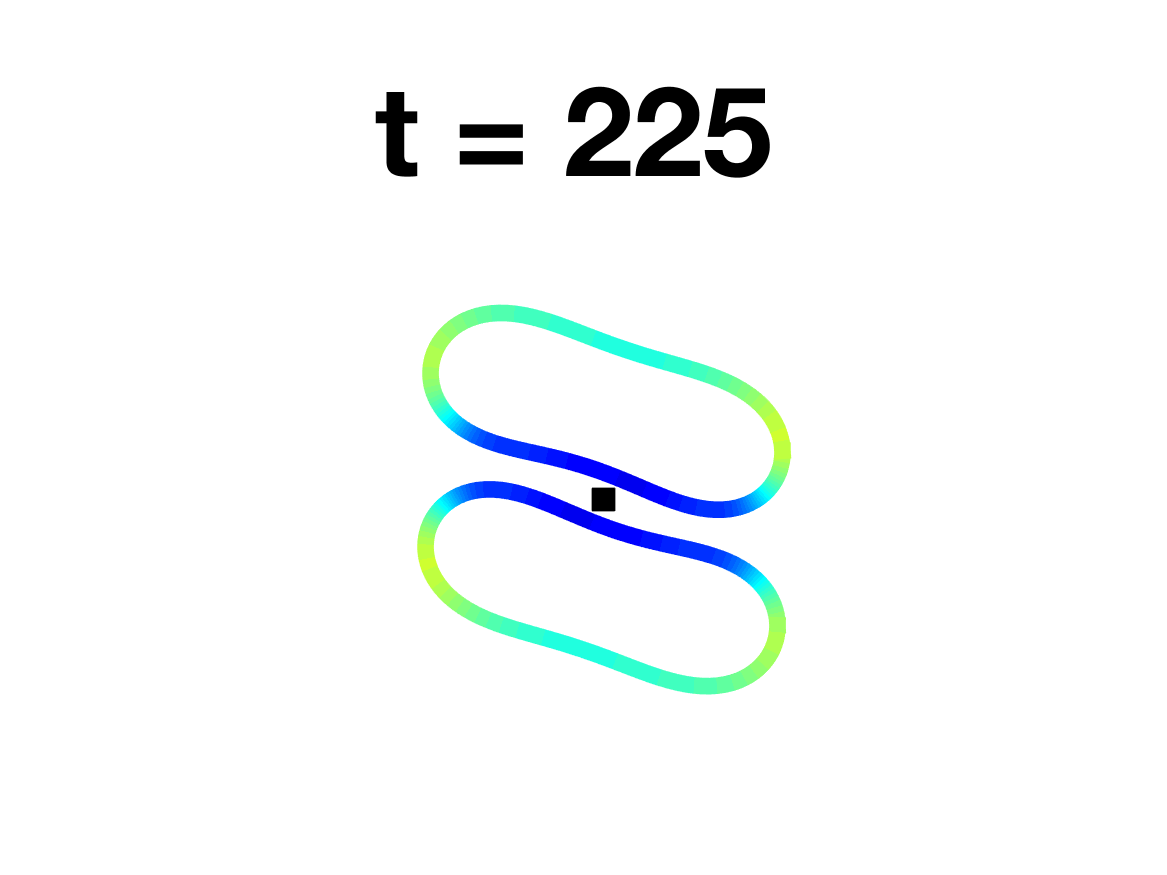}
  \includegraphics[height = 0.18\textwidth,trim={4cm 1cm 3cm 1cm},clip]
    {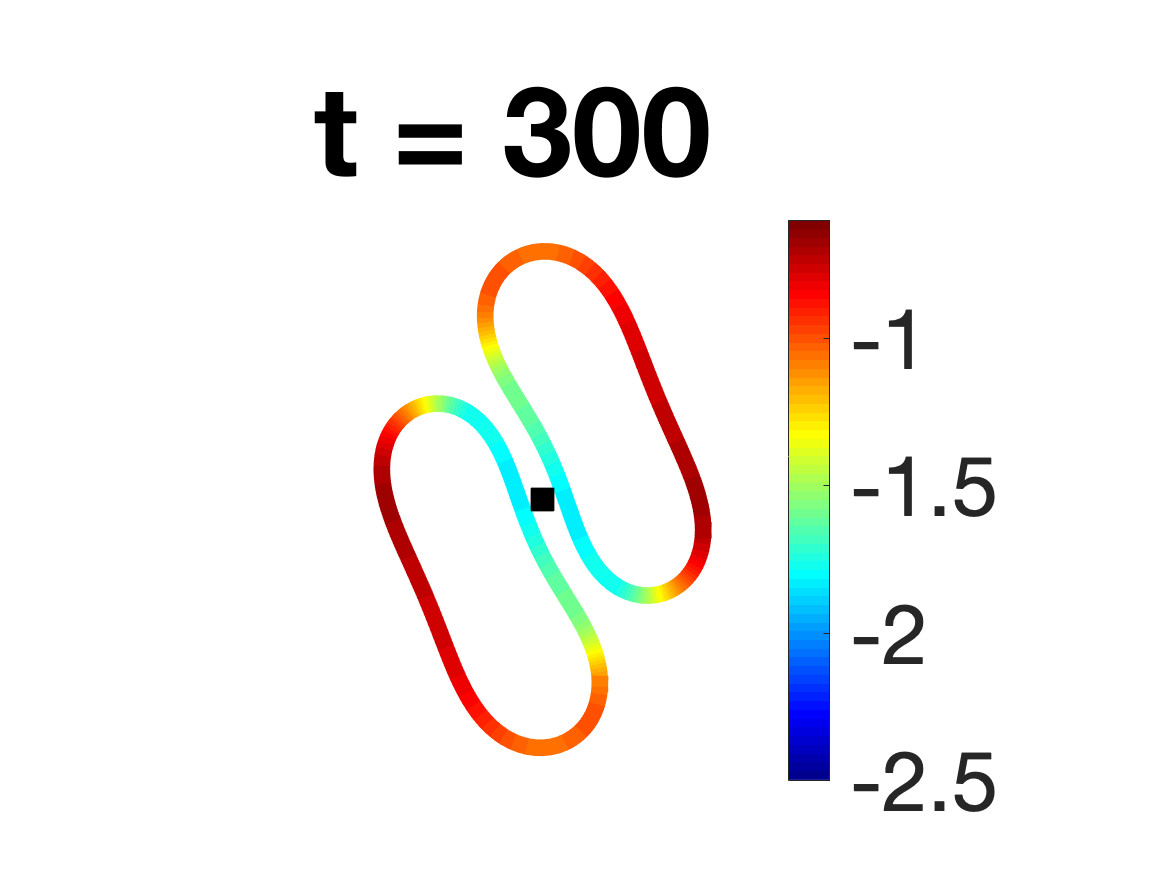}
  \caption{\label{fig:extensional1} A vesicle doublet in an extensional
  flow with an initial inclination angle $\theta=\pi$.  The reduced area
  is $\Delta A = 0.7$, the Hamaker constant is $\mathcal{H} = 0.7$, the
  separation distance is $\delta = 0.4$, and the extension rate is
  $\dot\epsilon = 0.02$ ({\em top}), $\dot\epsilon=0.07$ ({\em middle}),
  and $\dot\epsilon = 0.1$ ({\em bottom}).  The center of the vesicle,
  denoted with a black square, is the stagnation point.  The color
  coding is the tension.}
  \end{figure}

In the following numerical experiments we initially place a vesicle
doublet with $\theta=\pi$, so that the diverging flow may be strong
enough to pull vesicles away from the doublet.  At low extension rates,
we expect the doublet to stay bound at the fluid trap stagnation point.
On the other hand, the vesicle doublet may become unstable and
eventually separate at higher extension rates.  Thus we expect there to
exist a critical extension rate $\dot\epsilon_c$ above which the vesicle
doublet cannot stay bound under a given adhesion potential.  Therefore,
the dependence of the critical extension rate $\dot\epsilon_c$ on the
adhesion potential and the mechanical vesicle properties provides a
means to probe the physics of membrane adhesion.

We consider vesicle doublets with reduced areas 0.70, 0.75, 0.80, 0.85,
0.90, and 0.95, all with a length of $2\pi$, and we vary the extension
rate ($\dot\epsilon$) between $10^{-2}$ and $10^{-1}$.  Since the
stagnation point can be controlled in an experimental
setting~\cite{BentleyLeal1986_JFMa, Johnson-Chavarria2011_EMJ}, we mimic
the active control of the microfluidic experiments by moving the center
of the doublet at each time step so that the stagnation point occurs
exactly in the middle of the doublet.  With this adjustment, the
vesicles either remain as a doublet in the fluid trap centered around
the stagnation point, or the doublet is broken and the vesicles separate
from one another.  

\begin{figure}
  \includegraphics[height = 0.18\textwidth,trim={4cm 1cm 4cm 1cm},clip]
    {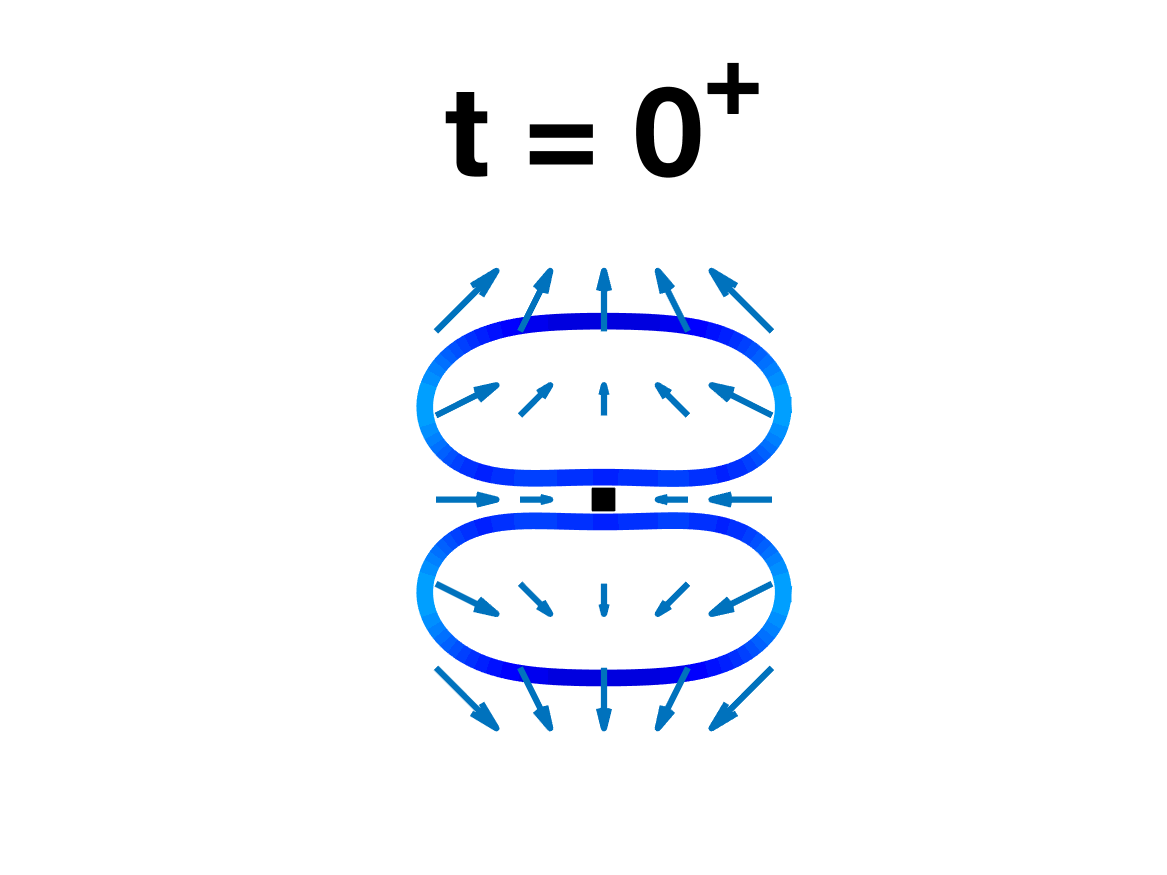}
  \includegraphics[height = 0.18\textwidth,trim={4cm 1cm 4cm 1cm},clip]
    {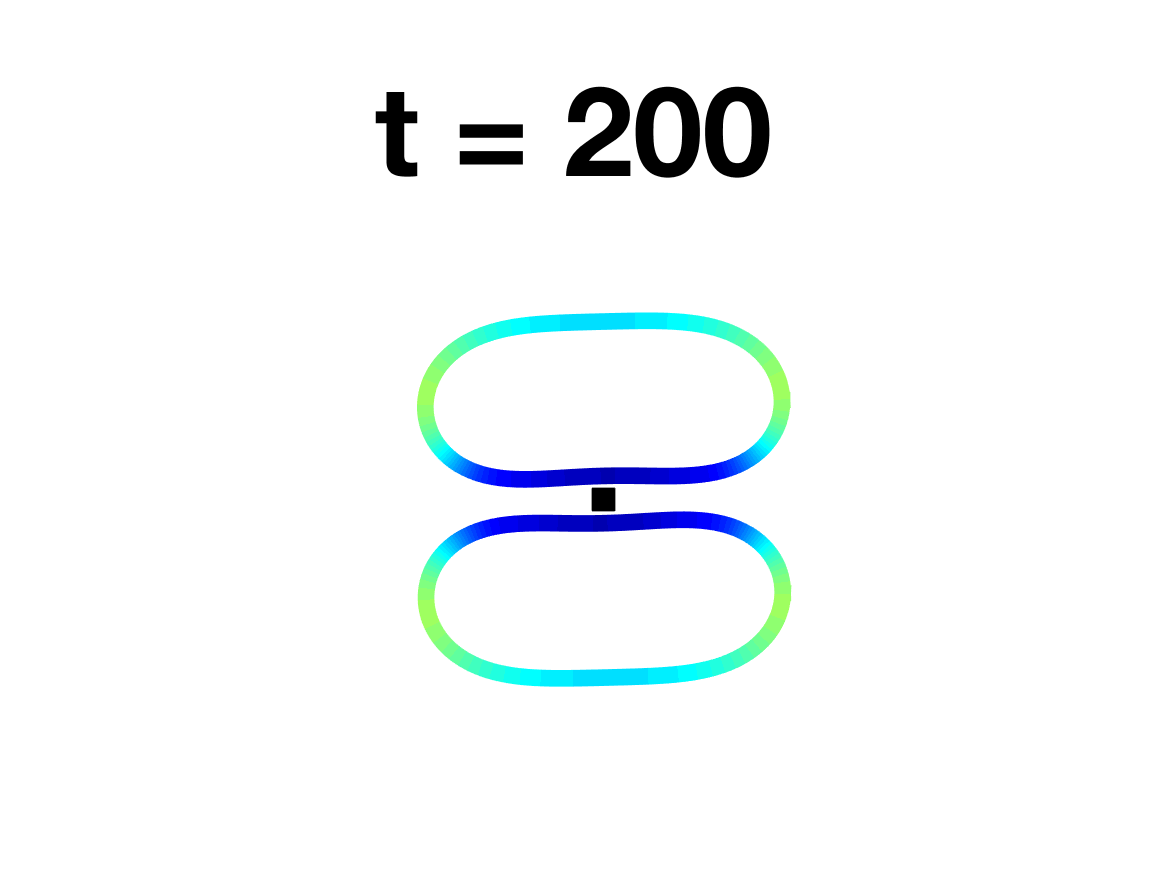}
  \includegraphics[height = 0.18\textwidth,trim={4cm 1cm 4cm 1cm},clip]
    {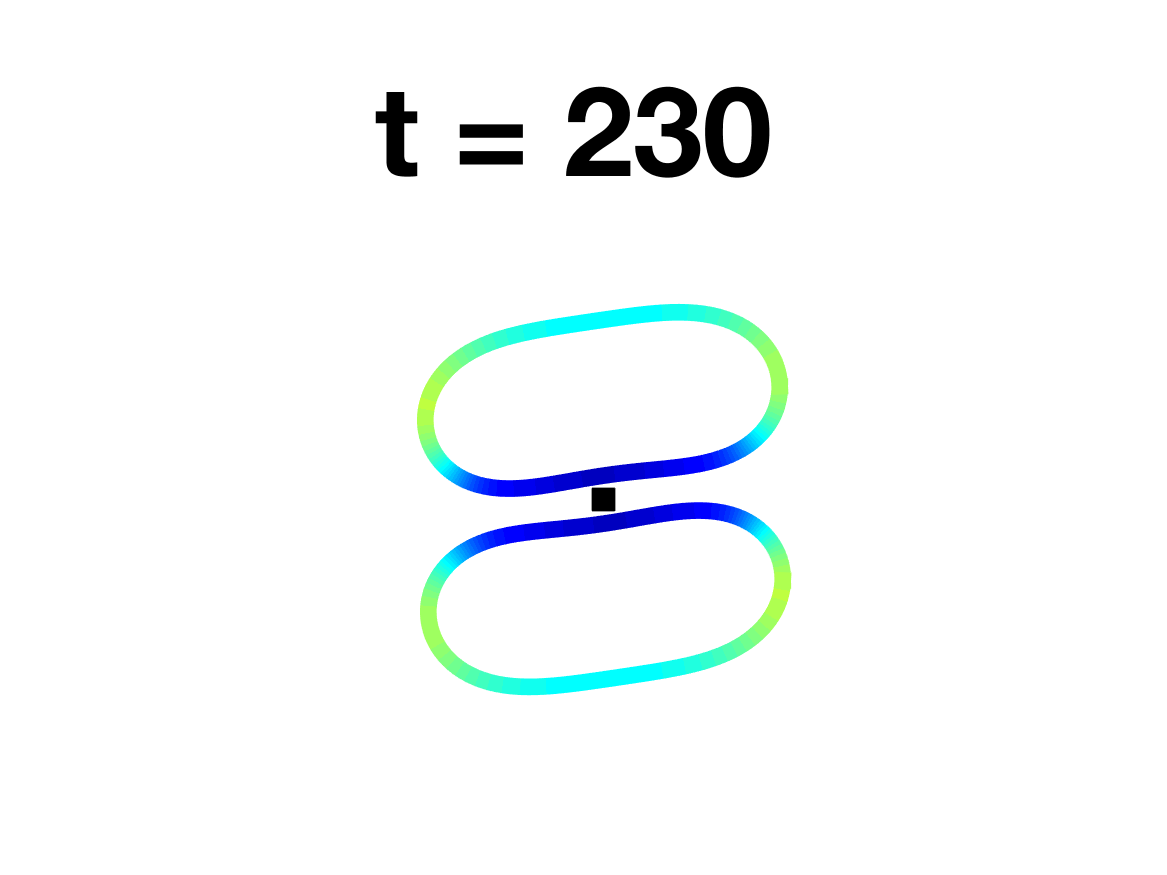}
  \includegraphics[height = 0.18\textwidth,trim={4cm 1cm 4cm 1cm},clip]
    {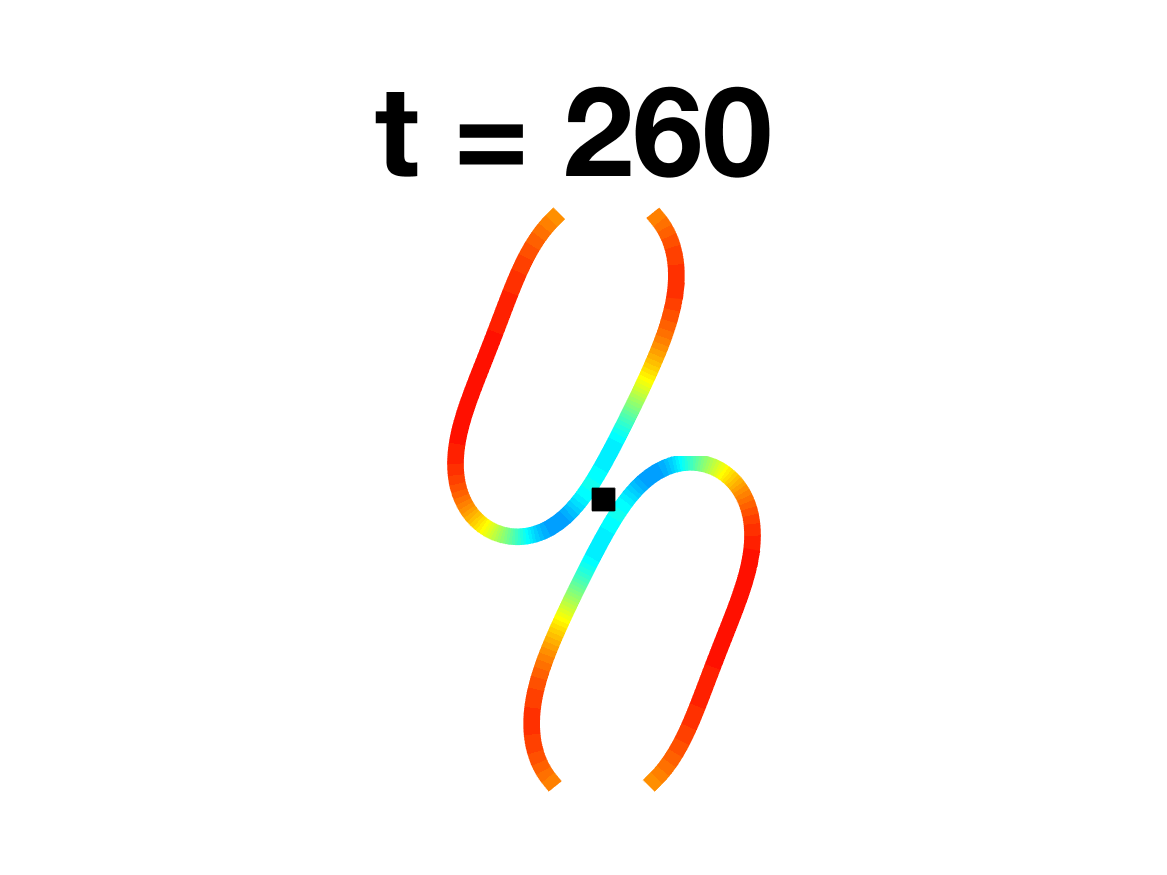}
  \includegraphics[height = 0.18\textwidth,trim={4cm 1cm 3cm 1cm},clip]
    {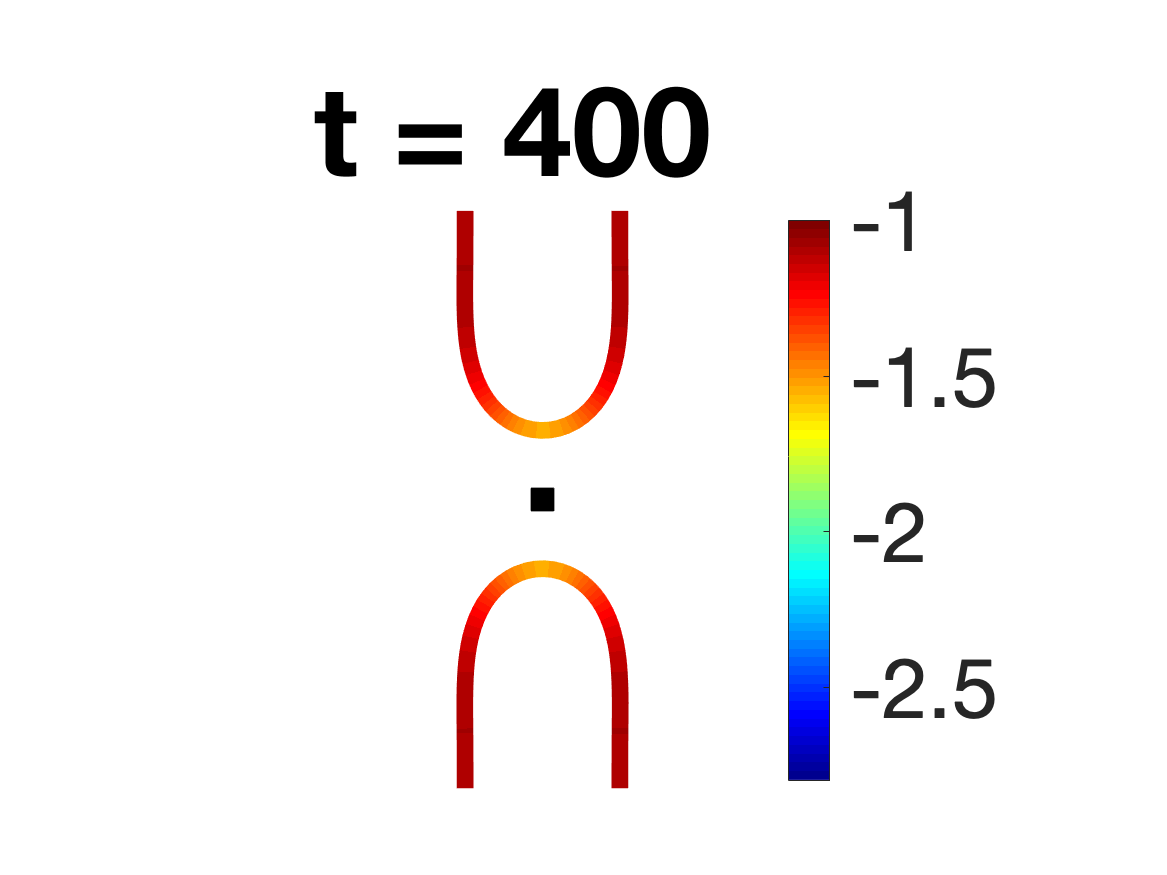}
  \caption{\label{fig:extensional3} A vesicle doublet in an extensional
  flow with an initial inclination angle $\theta=\pi$.  The reduced area
  is $\Delta A = 0.8$, the Hamaker constant is $\mathcal{H} = 0.7$, the
  separation distance is $\delta = 0.4$, and the extension rate is
  $\dot\epsilon = 0.1$.  The center of the vesicle, denoted with a black
  square, is a stagnation point.  The color coding is the tension.}
\end{figure}

Figure~\ref{fig:extensional1} shows snapshots from simulating a vesicle
doublet of reduced area $\Delta A = 0.7$ at three different extension
rates. A general feature of the doublet dynamics is that the doublet
rotates from $\theta=\pi$ configuration towards the more stable
$\theta=\pi/2$ configuration.  We also observe that with moderate
extension rates, the doublet falls short of aligning their long axis
with the diverging direction.  Moreover, the final inclination angle is
closest to the stable $\theta=\pi/2$ for the smallest extension rates.
While this doublet remains bound for all the considered extension rates,
a doublet with a reduced area of $\Delta A = 0.8$ is split with a
critical extension rate $\dot\epsilon_c = 0.1$
(Figure~\ref{fig:extensional3}).  

The equilibrium configuration for low extension rate $\dot\epsilon=0.02$
(top row in Figure~\ref{fig:extensional1}) is almost identical to the
equilibrium configuration for $\Delta y=0$ (top row of
Figure~\ref{fig:extensionalNonsymmetric}).  The rotation of the vesicle
doublet from $\theta=\pi$ to $\theta=\pi/2$ can be understood as the
transition from an unstable configuration $\theta=\pi$ to a stable
configuration $\theta=\pi/2$.  For higher extension rates (second and
third rows in Figure~\ref{fig:extensional1}) the partially rotated
vesicle pairs at equilibrium are also very similar to the case of
$\Delta y=0.02$ (second row of
Figure~\ref{fig:extensionalNonsymmetric}).  Such asymmetric vesicle
configuration is reminiscent of the asymmetric deformation of a single
vesicle in an extensional flow~\cite{KantslerSegreSteinberg2008_PRL,
Narsimhan2014_JFM, DahlNarsimhanGouveia2016_SoftMatt}.  To gain more
insight into such nonlinear dynamics, in the following we explore more
physical characterization of the transition from $\theta=\pi$ to
$\theta=\pi/2$ at different reduced area and extension rate.

\begin{figure}[htp]
  \begin{tabular}{@{}p{0.3\linewidth}@{\quad}p{0.3\linewidth}@{\quad}p{0.3\linewidth}@{}}
  \subfigimg[width=\linewidth]{(a)}{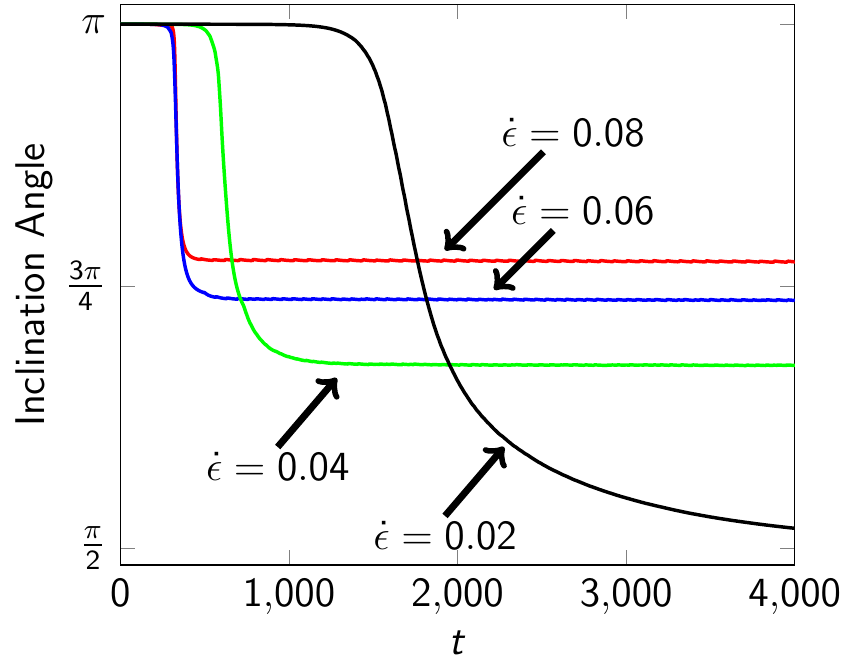} &
  \subfigimg[width=\linewidth]{(b)}{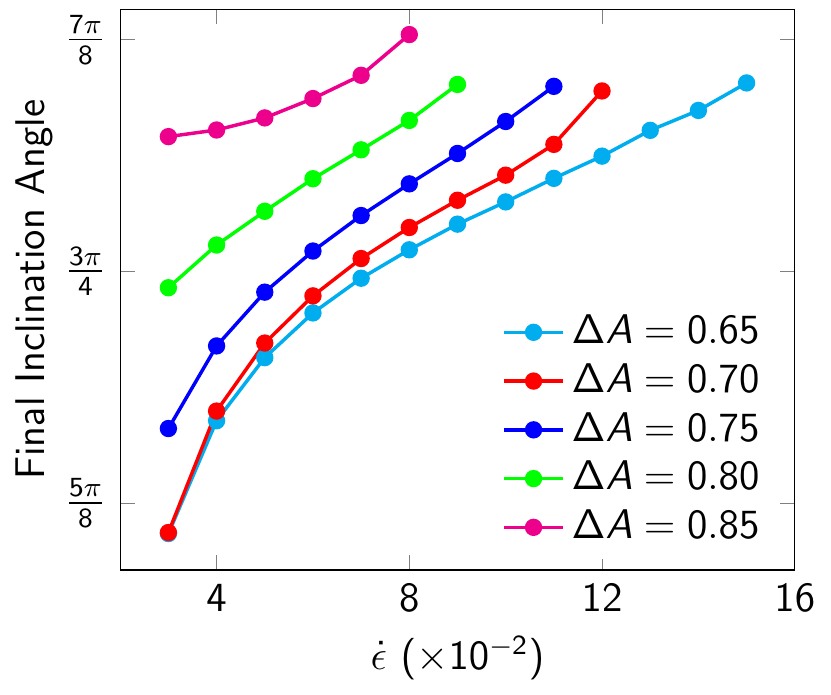} &
  \subfigimg[width=\linewidth]{(c)}{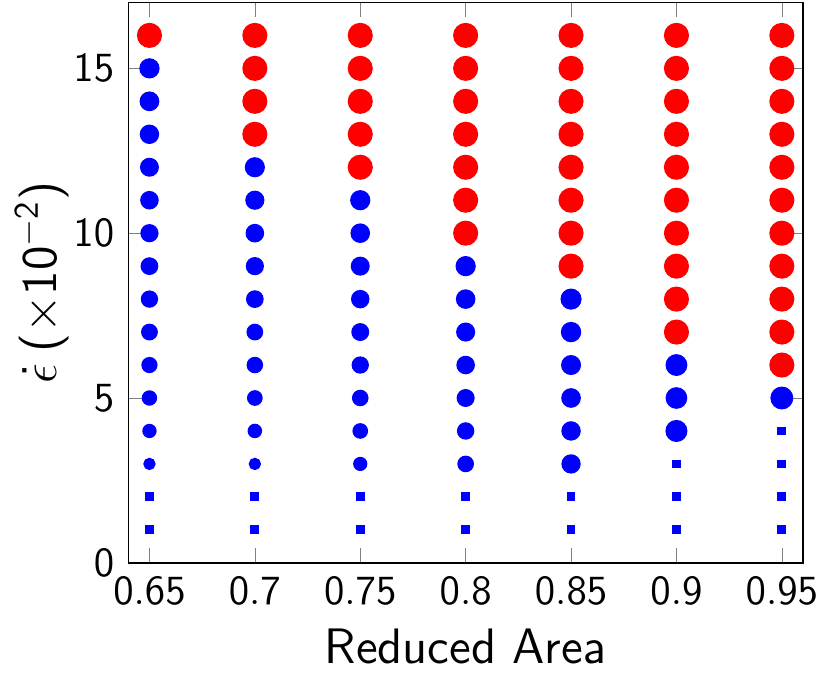}
  \end{tabular}
  \caption{\label{fig:extensionalInclinationAngle} The long time
  behavior of two vesicles in a fluid trap with Hamaker constant
  $\mathcal{H} = 0.7$ and separation distance $\delta = 0.4$. (a) The
  inclination angle of a doublet formed by vesicles of reduced area
  $\Delta A = 0.7$ in an extensional flow with an extension rate
  $\dot\epsilon$.  For all extension rates, the vesicle's orientation
  angle tends to a value in $(\frac{\pi}{2},\pi)$. (b) The final
  inclination angle of the doublets for a variety of reduced areas.  At
  larger extension rates, the doublet is broken by the flow.  (c) A
  phase diagram of the long-time behavior of a fluid trap formed by two
  adhering vesicles in an extensional flow.  Reduced areas and extension
  rates that do not form a doublet are marked in red, and those that do
  form a doublet are marked in blue.  For the blue marks, the size is
  proportional to the final inclination angle.  Squares indicate that
  the doublet had not reached an equilibrium inclination angle in the
  given time horizon.}
\end{figure}

In Figure~\ref{fig:extensionalInclinationAngle}(a), we plot the
inclination angle of the doublet as a function of time for four
different extension rates, and the final inclination angle for all
doublets that reach an equilibrium state are in
Figure~\ref{fig:extensionalInclinationAngle}(b).  We observe that not
only do smaller extension rates result in smaller inclination angles,
but smaller reduced areas also result in smaller inclination angles.  We
summarize the final inclination angle of the doublet in
Figure~\ref{fig:extensionalInclinationAngle}(c).  The size of the round
dots are scaled to the final inclination angle as defined in
Figure~\ref{fig:InclinationAngle}.  At smaller extension rates, the
vesicles come closer to aligning their long axis with the diverging
direction.  When the doublet is broken at reduced area $\Delta A =
0.80$, the extension rate is sufficiently large to align the long axis
of the vesicle with the diverging direction and then the vesicles
separate (Figure~\ref{fig:extensional3}).  For very low extension rates,
the time horizon is insufficient for the doublet to reach an equilibrium
state, and these simulations are marked with a square.
Figure~\ref{fig:extensionalInclinationAngle}(c) also summarizes the
reduced areas and extension rates that result in a bound vesicle doublet
in a fluid trap: Parameter values with a blue mark result in a fluid
trap and parameter values with a red mark result in vesicle separation.

\section{Adhesion of two vesicles in a planar shear flow}
\label{sec:sflow} 
We consider two vesicles under a planar shear flow $\uu =
\dot\gamma(y,0)$, where $\dot\gamma$ is the shear rate.  To be
consistent with simulations in previous sections, the vesicles have a
length of $2\pi$ for all reduced areas.  The initial placement of the
vesicles is chosen so that the flow dives them towards one another in a
head-on collision.  In the absence of adhesion,  two vesicles deform
significantly as they collide head-on, deflect to opposite sides of the
$y$-axis as they pass one another, and then separate.  This head-on
collision process of two vesicles or capsules in a planar shear flow is
found to be similar between two-~\cite{Breyiannis-Pozrikidis:2000,
RahimianVeerapaneniBiros2010_JCP} and
three-dimensions~\cite{LacMorelBarthes-Biesel2007_JFM,
LacBarthes-Biesel2008_PoF, OmoriIshikawaImaiYamaguchi2013_JBiomechics,
RahimianVeerapaneniZorinBiros2015_JCP}.  However, in the presence of
adhesion, the vesicles can form a doublet for certain values of the
separation distance $\delta$, Hamaker constant $\mathcal{H}$, shear rate
$\dot\gamma$, and reduced area $\Delta A$.  If a doublet does form, then
it undergoes a periodic motion with a period that depends on the same
parameters.

\begin{figure}[htp]
  \includegraphics[width=0.19\textwidth]{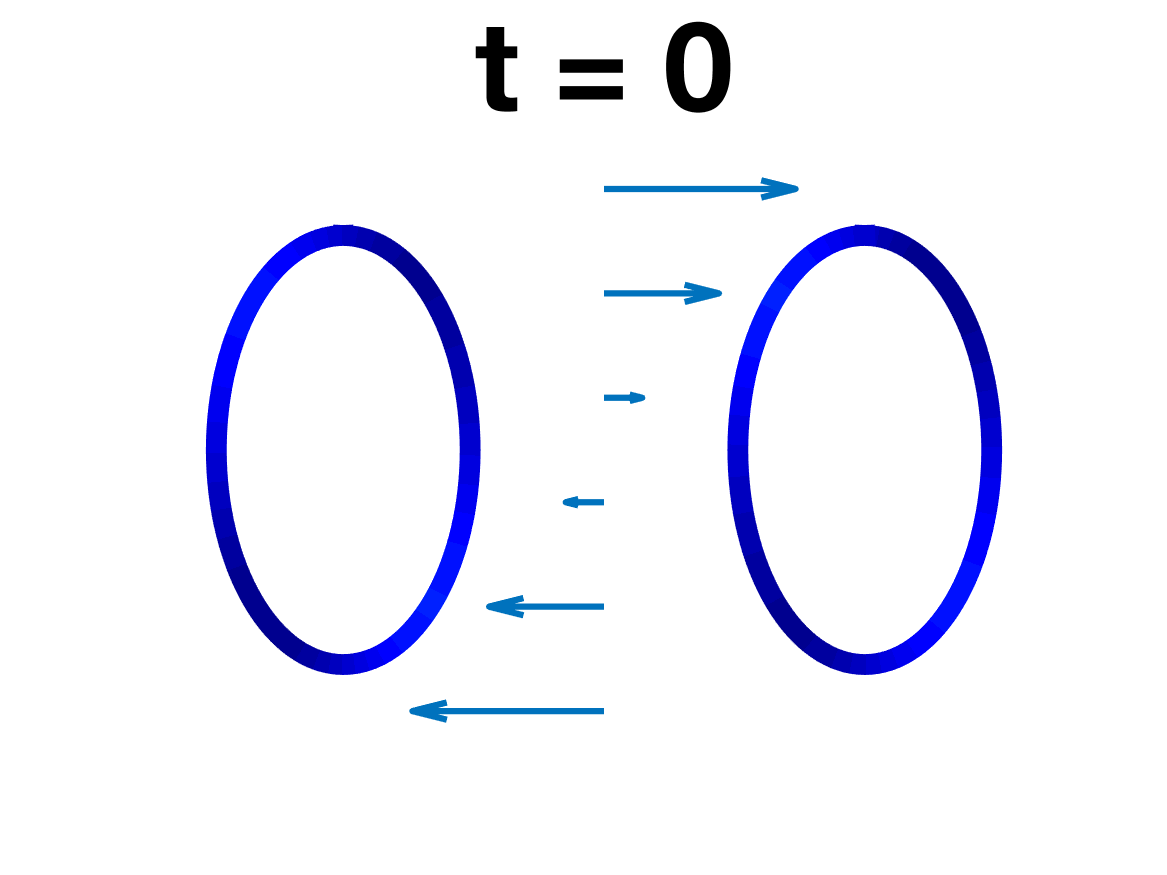}
  \includegraphics[width=0.19\textwidth]{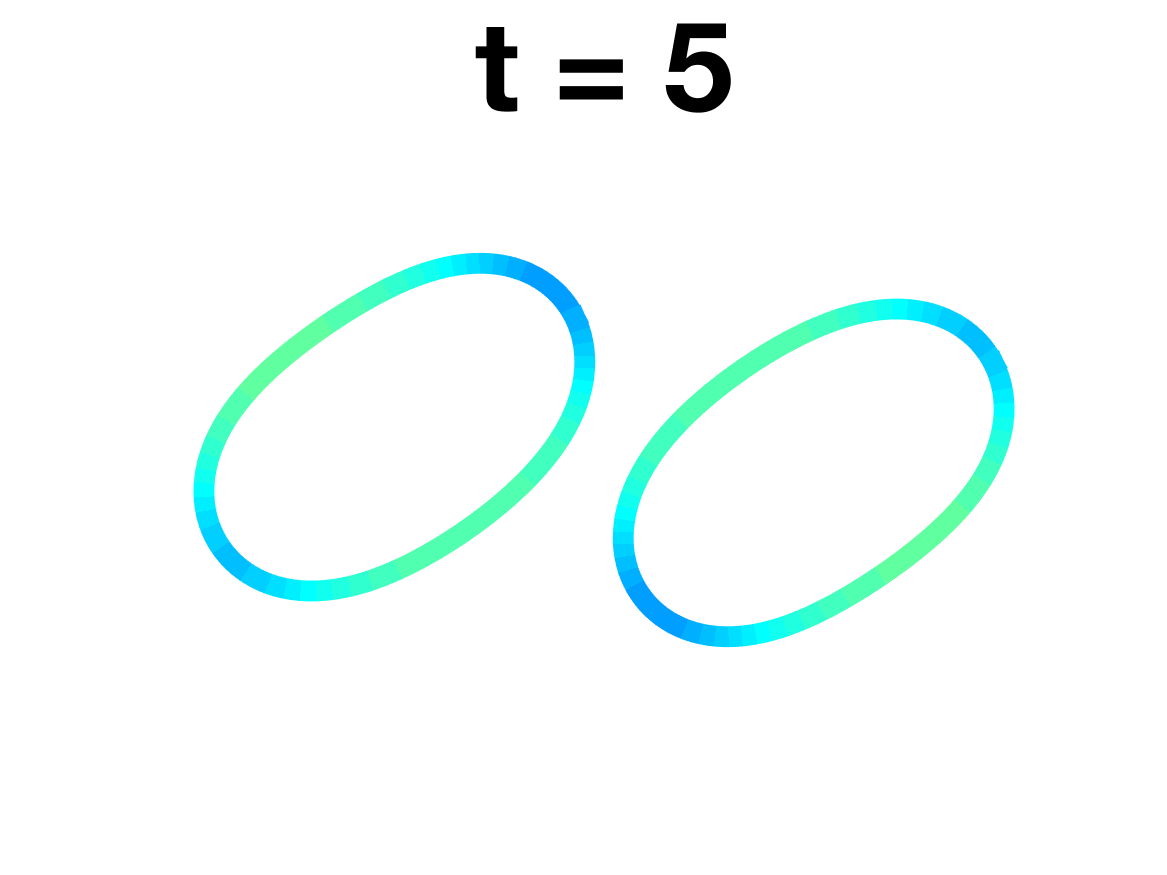}
  \includegraphics[width=0.19\textwidth]{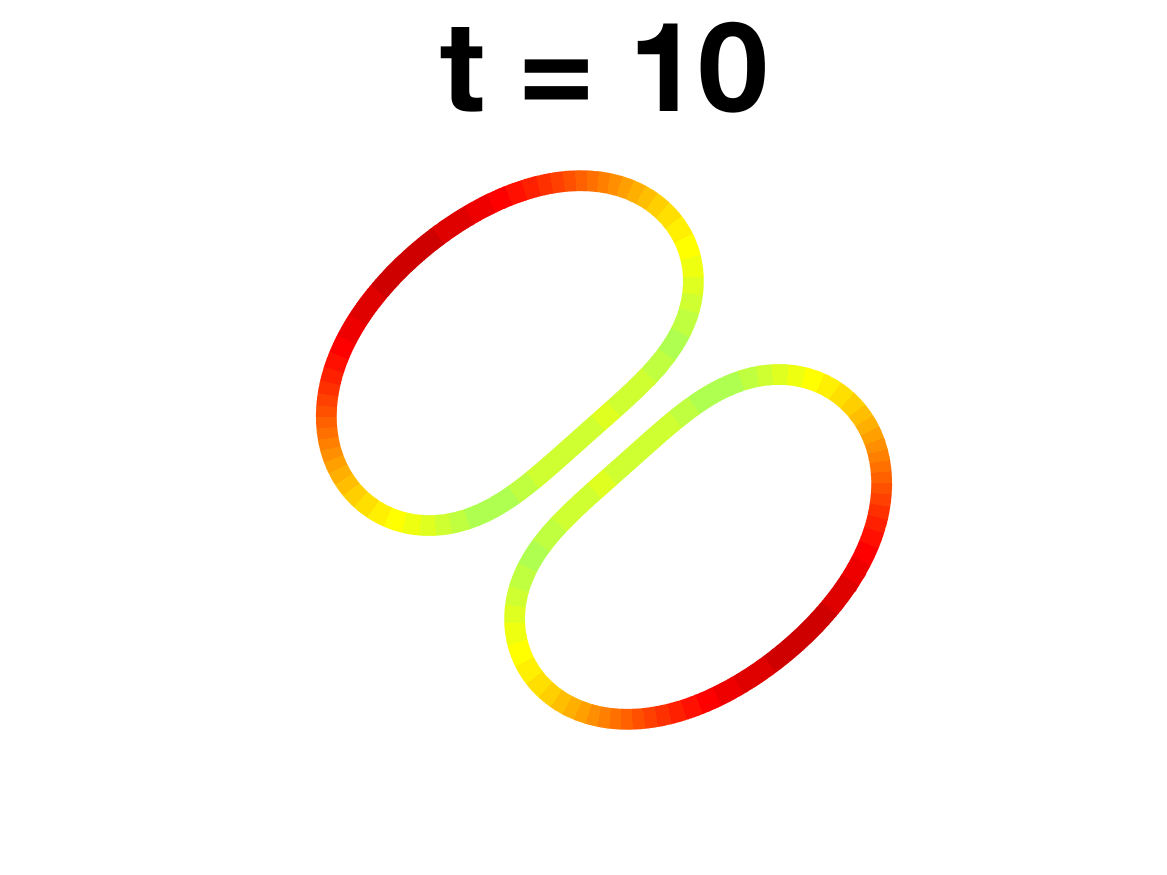}
  \includegraphics[width=0.19\textwidth]{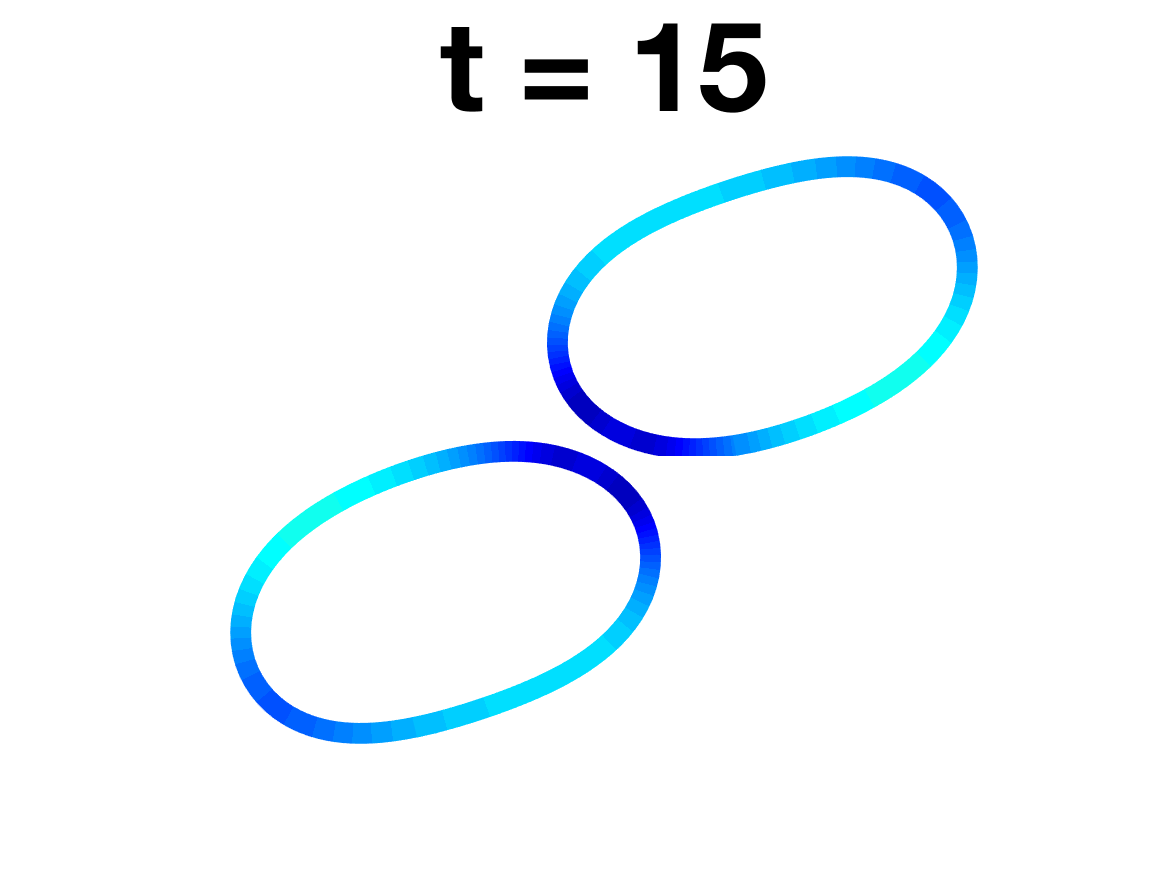}
  \includegraphics[width=0.19\textwidth]{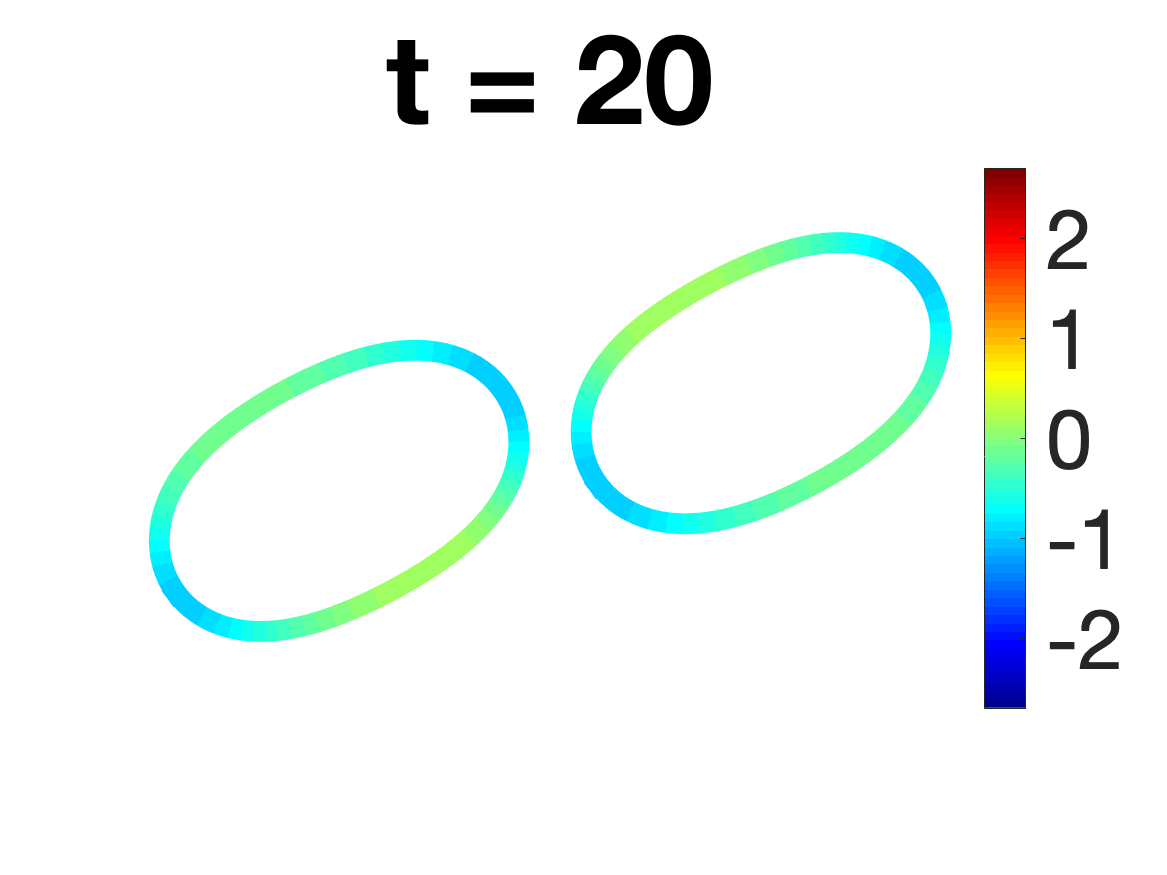}
  \includegraphics[width=0.19\textwidth]{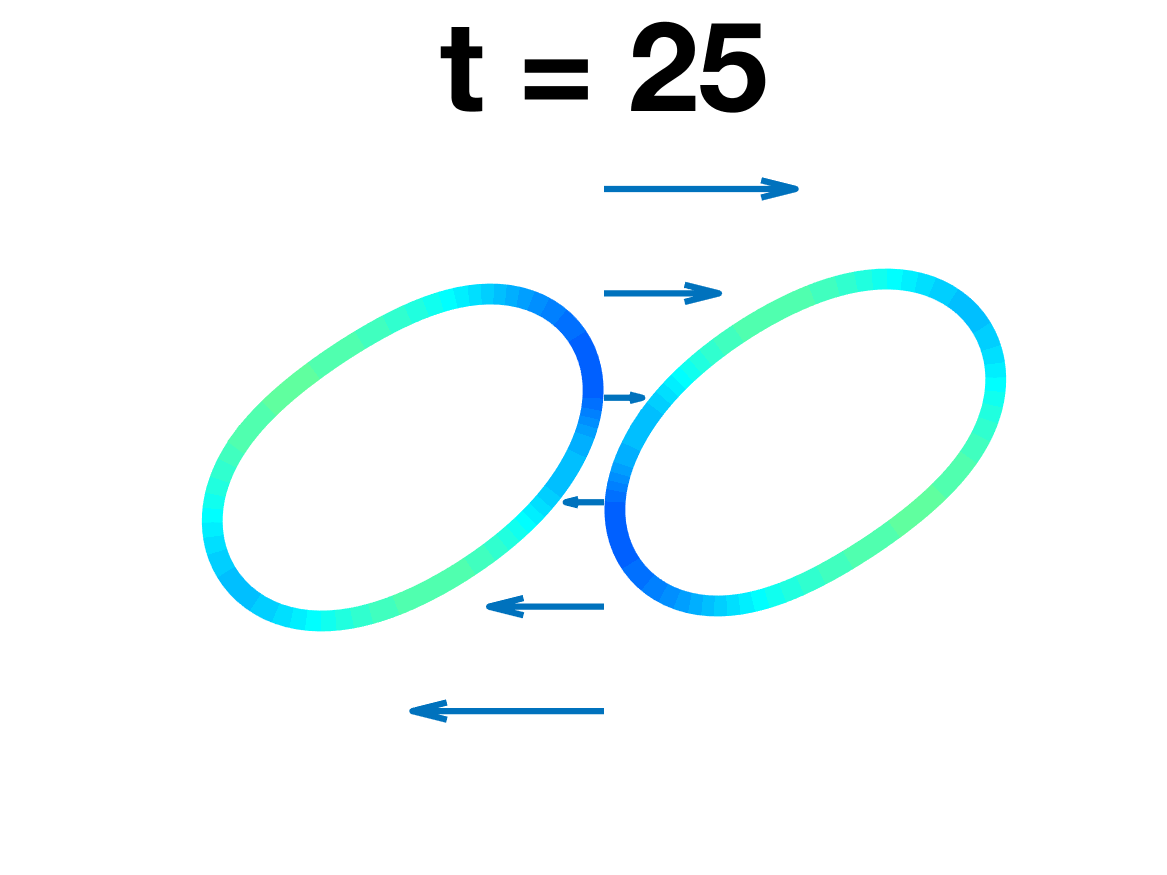}
  \includegraphics[width=0.19\textwidth]{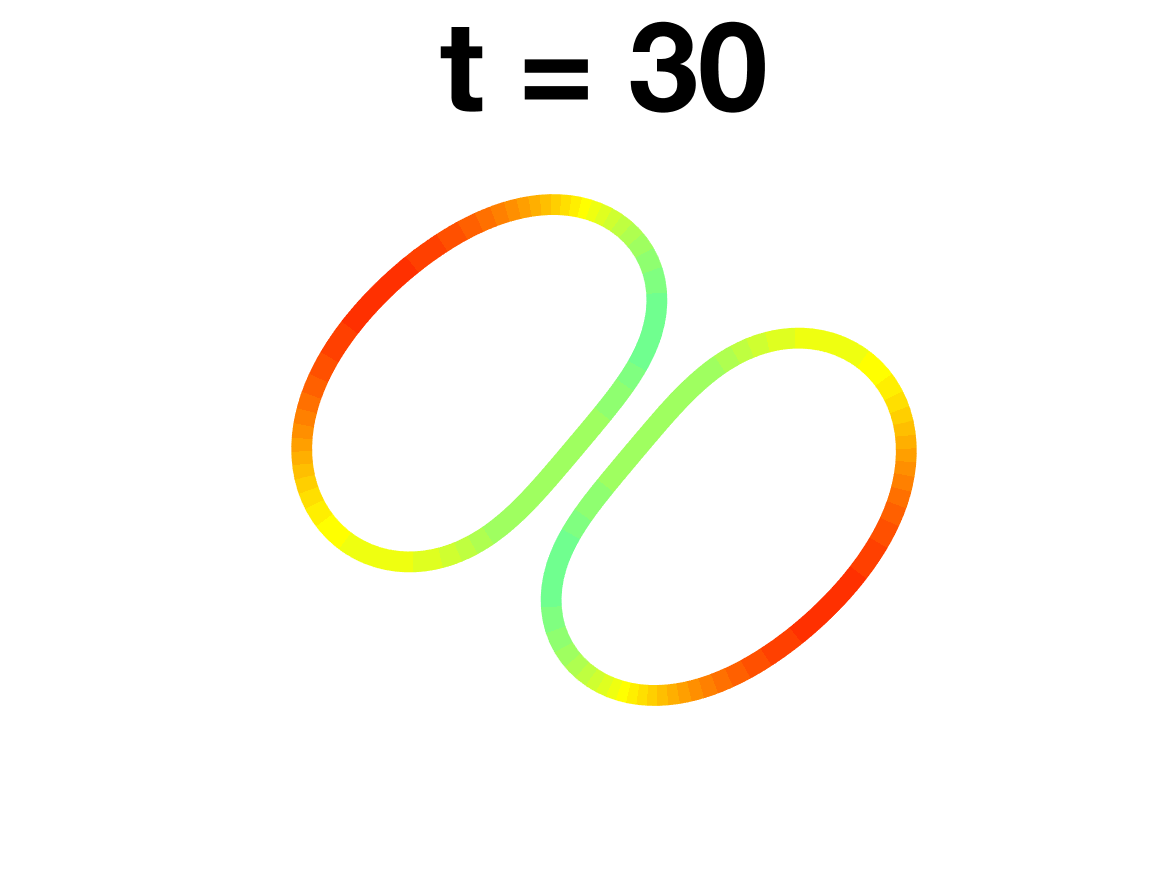}
  \includegraphics[width=0.19\textwidth]{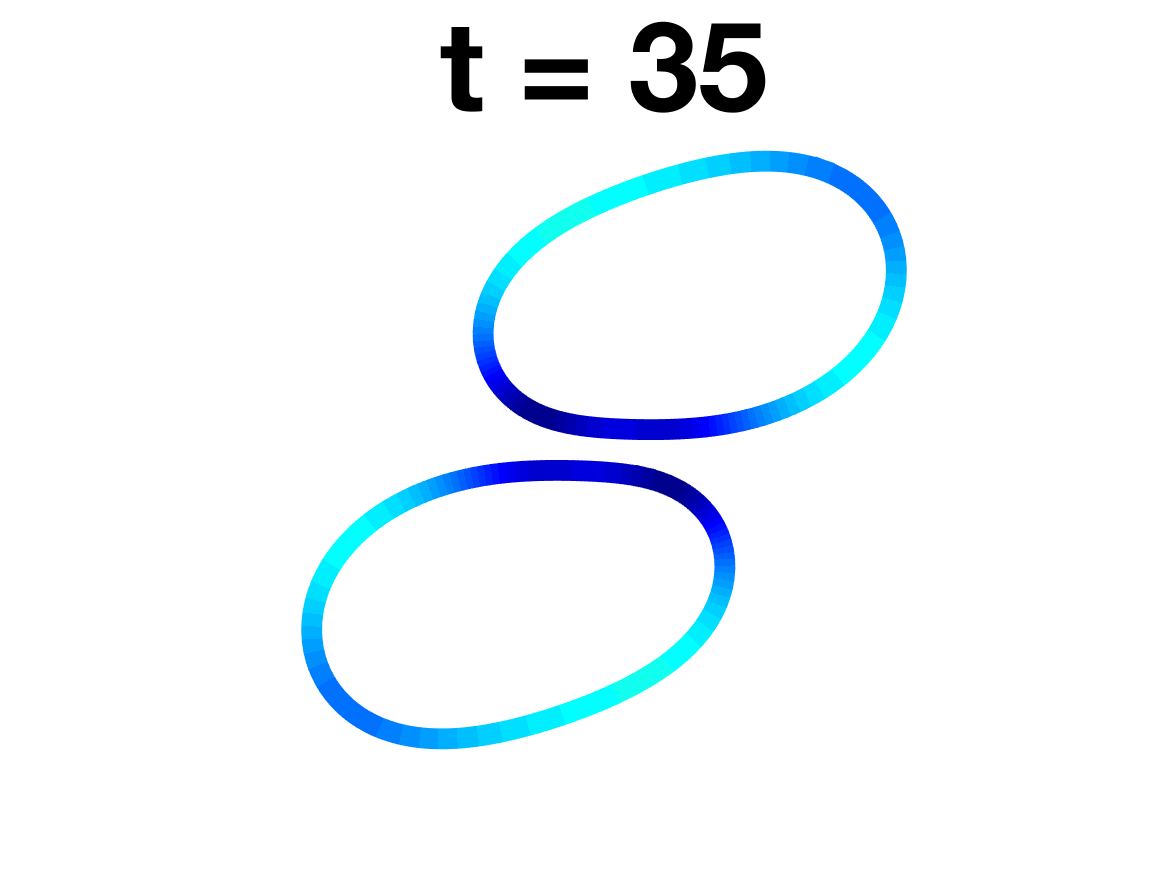}
  \includegraphics[width=0.19\textwidth]{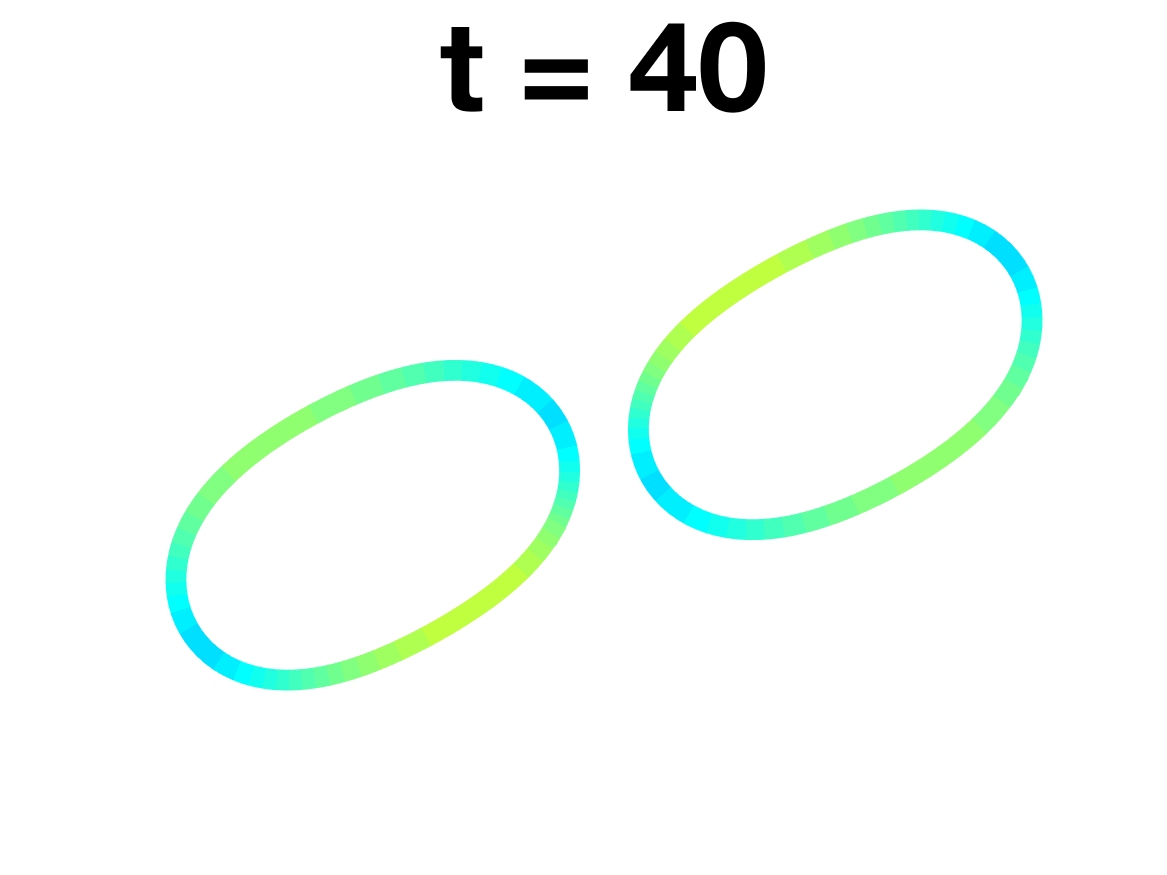}
  \includegraphics[width=0.19\textwidth]{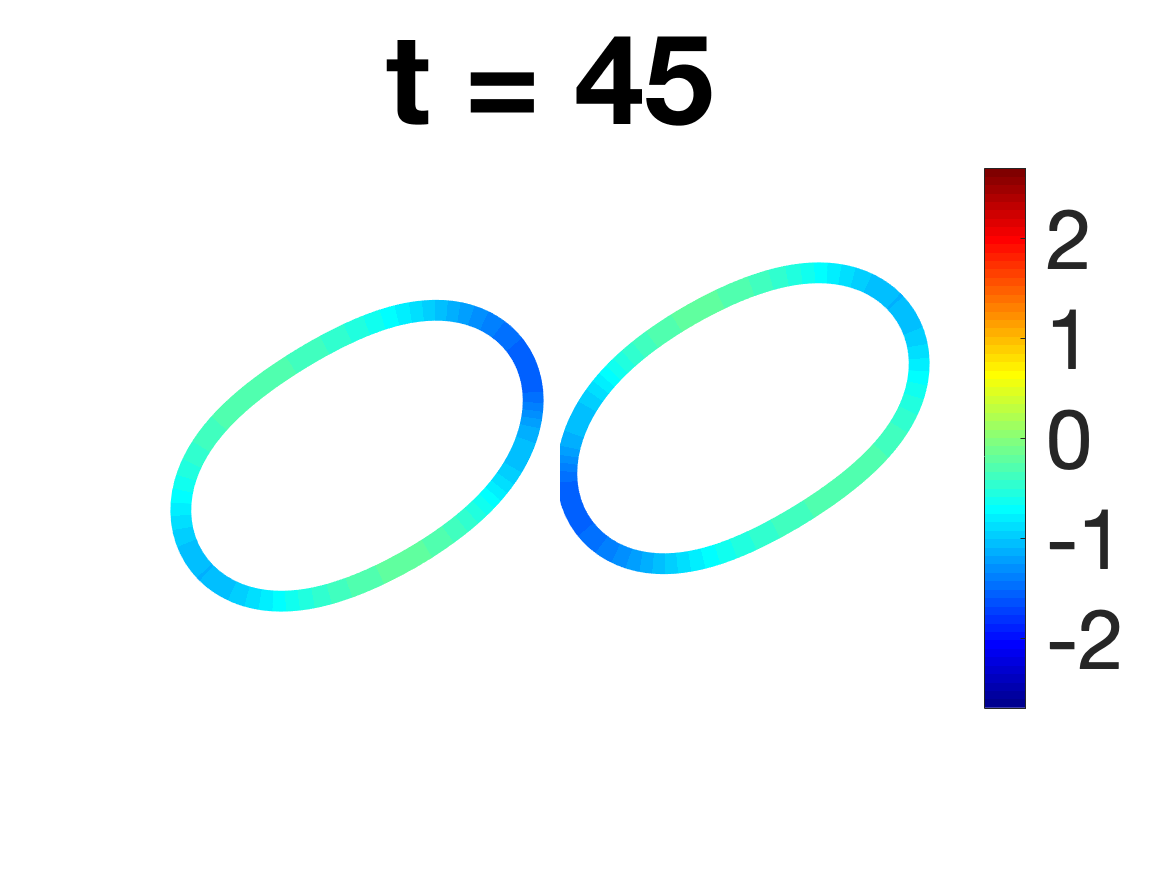}
  \caption{\label{fig:doublet090-weakAdhesion} The formation of a weakly
  adhered doublet in a shear flow.  The reduced area is $\Delta A =
  0.9$, the Hamaker constant is $\mathcal{H}=0.7$, the separation
  distance is $\delta = 0.4$, and the shear rate is $\dot\gamma=0.5$.
  The interaction between the two vesicles is a sliding motion, and the
  dynamics has a period of about 42.  The color coding is the tension.}
\end{figure}

\begin{figure}[htp]
  \includegraphics[width=0.19\textwidth]{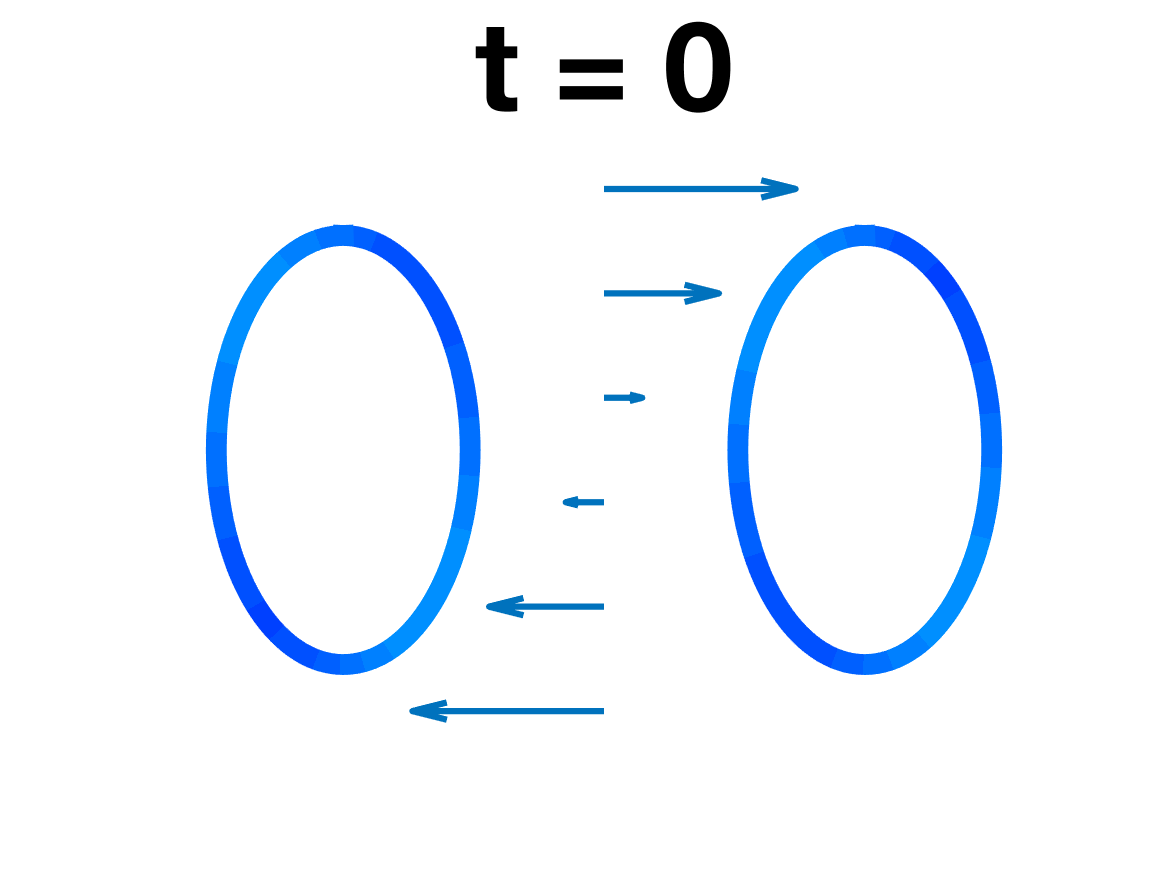}
  \includegraphics[width=0.19\textwidth]{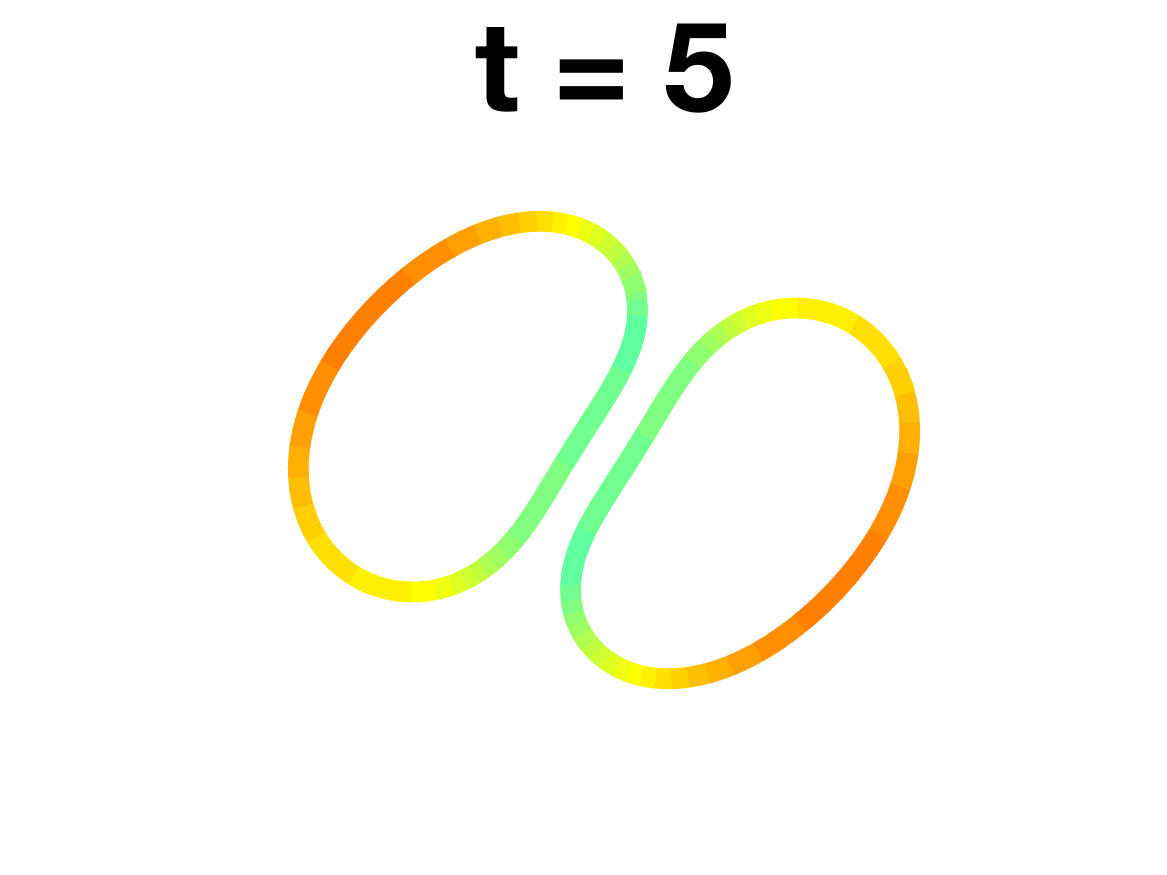}
  \includegraphics[width=0.19\textwidth]{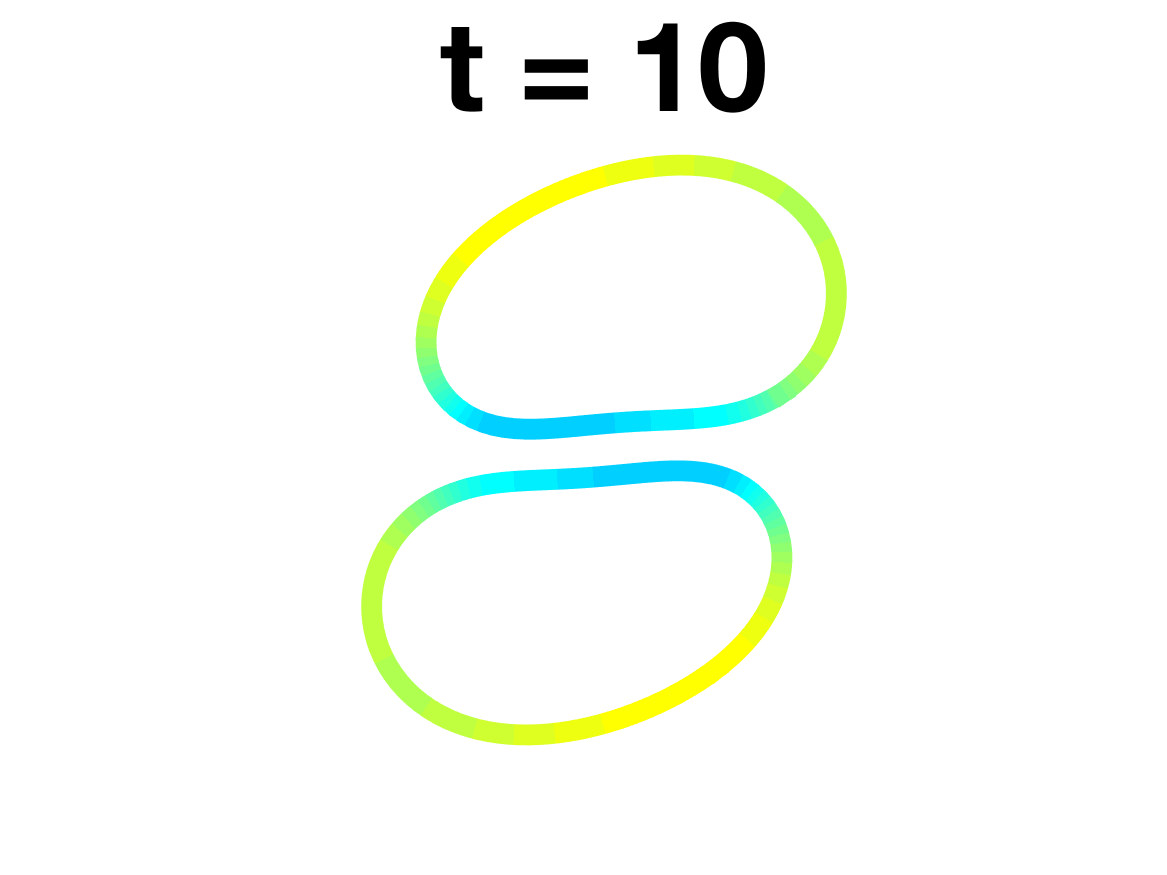}
  \includegraphics[width=0.19\textwidth]{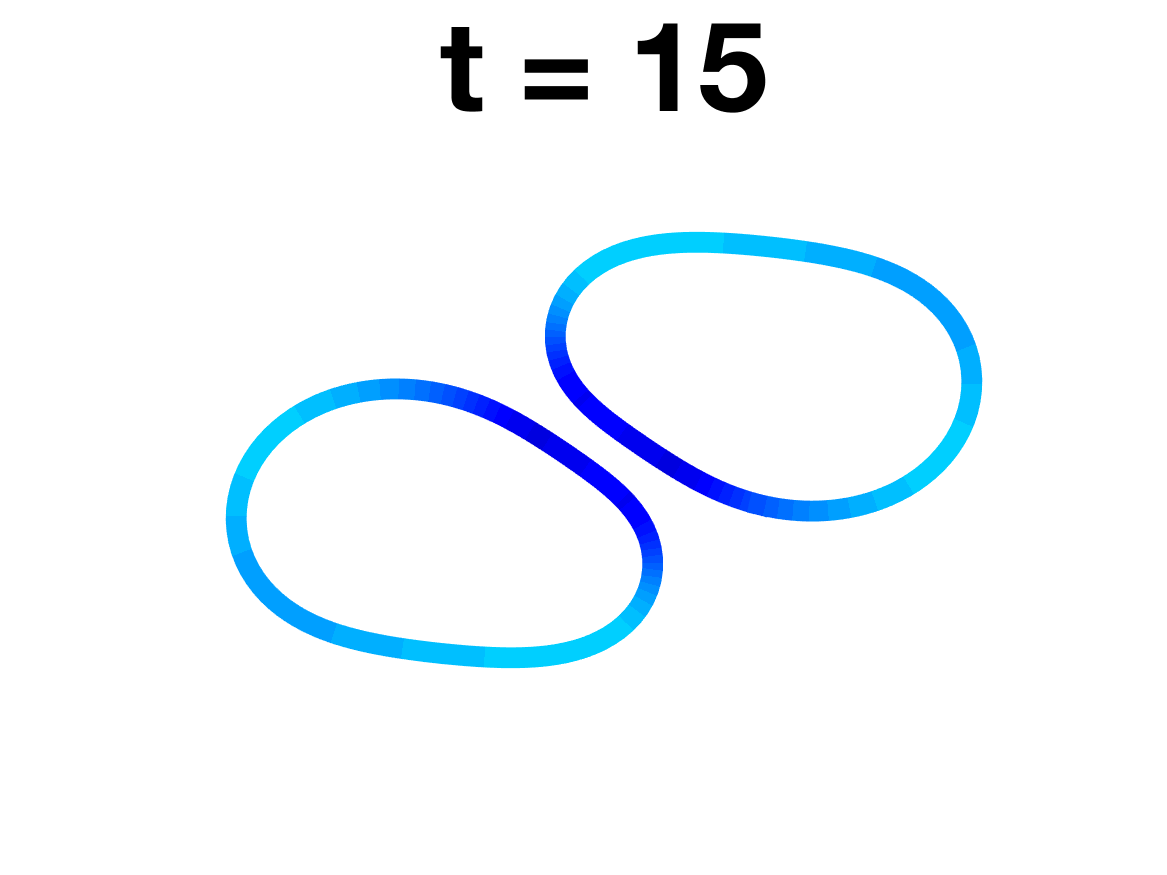}
  \includegraphics[width=0.19\textwidth]{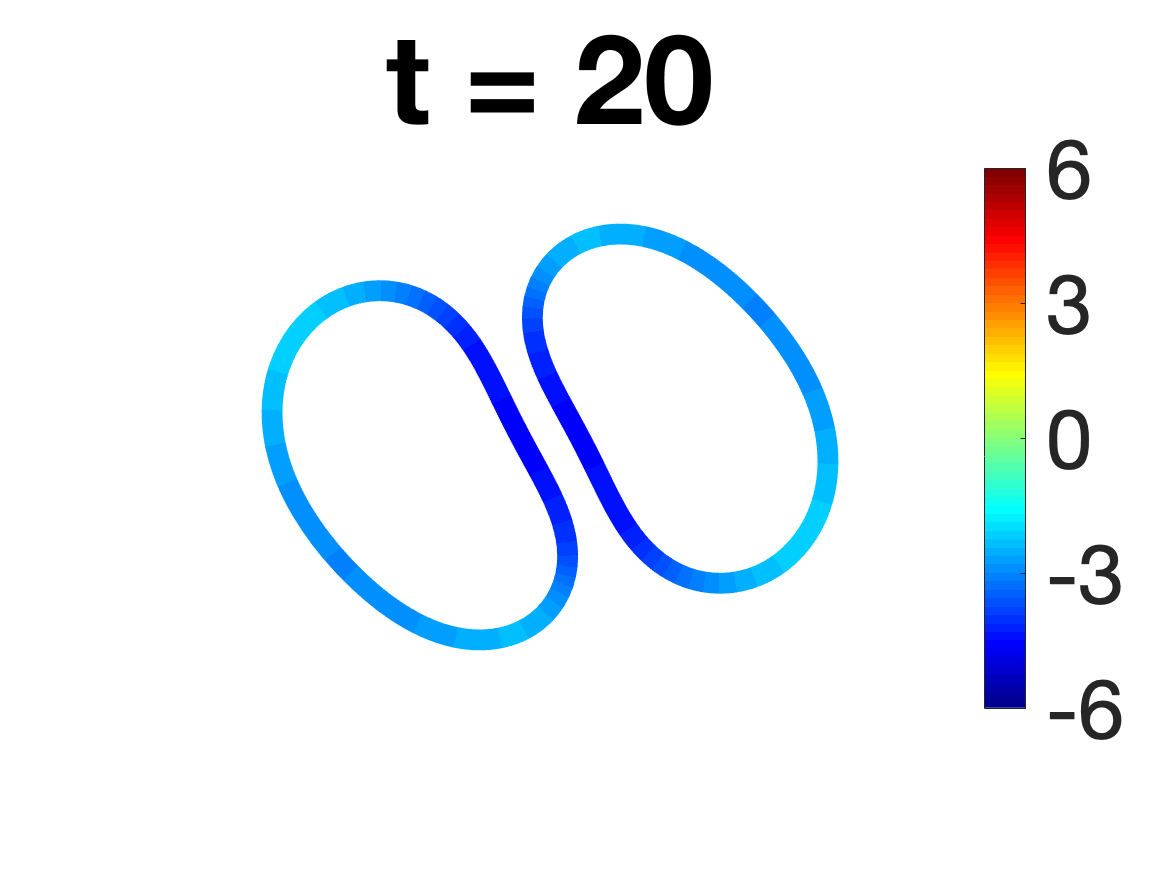}
  \includegraphics[width=0.19\textwidth]{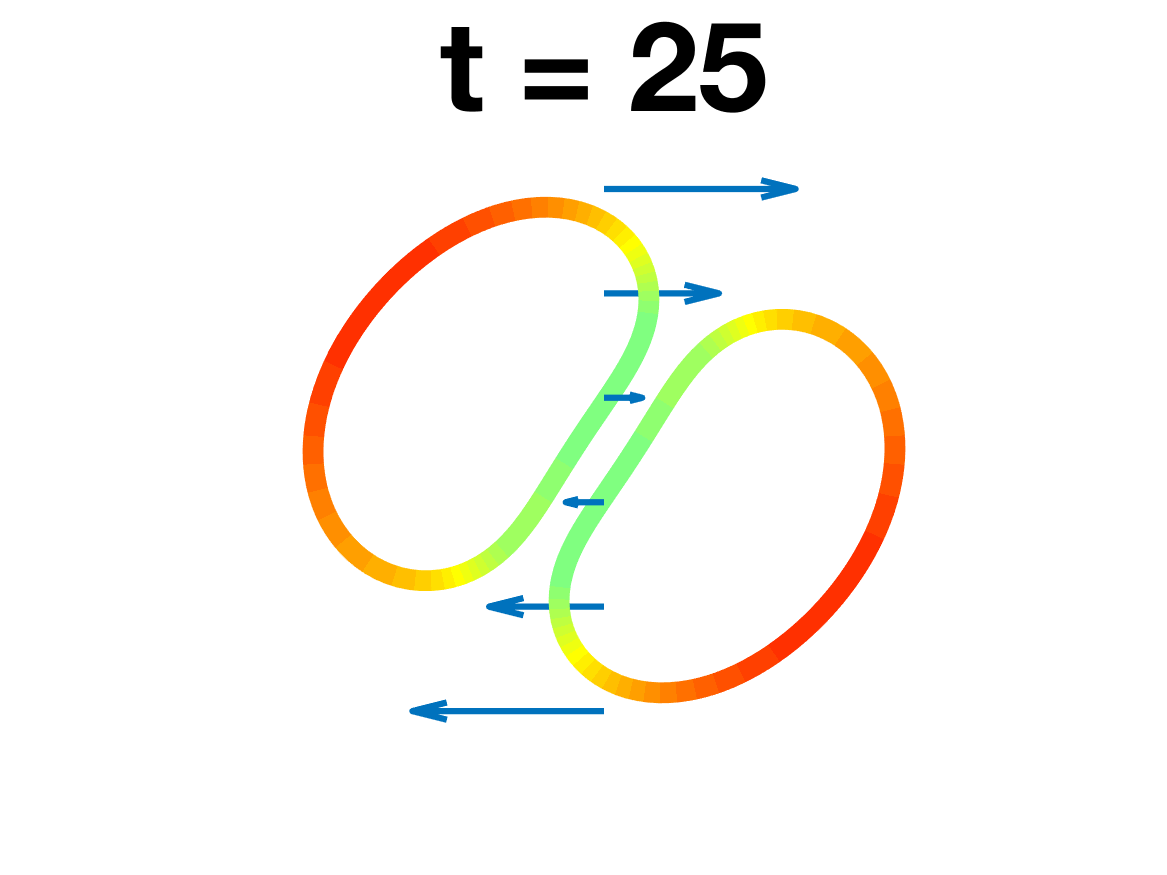}
  \includegraphics[width=0.19\textwidth]{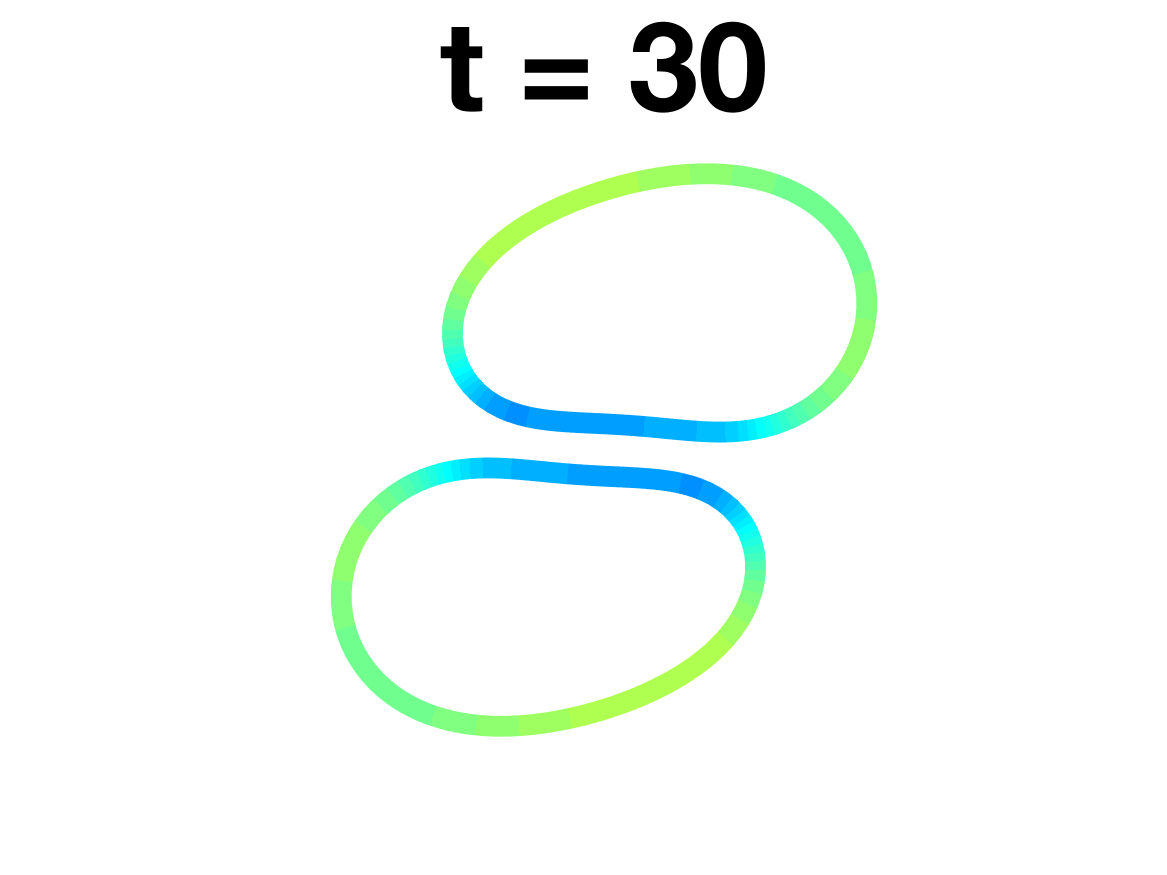}
  \includegraphics[width=0.19\textwidth]{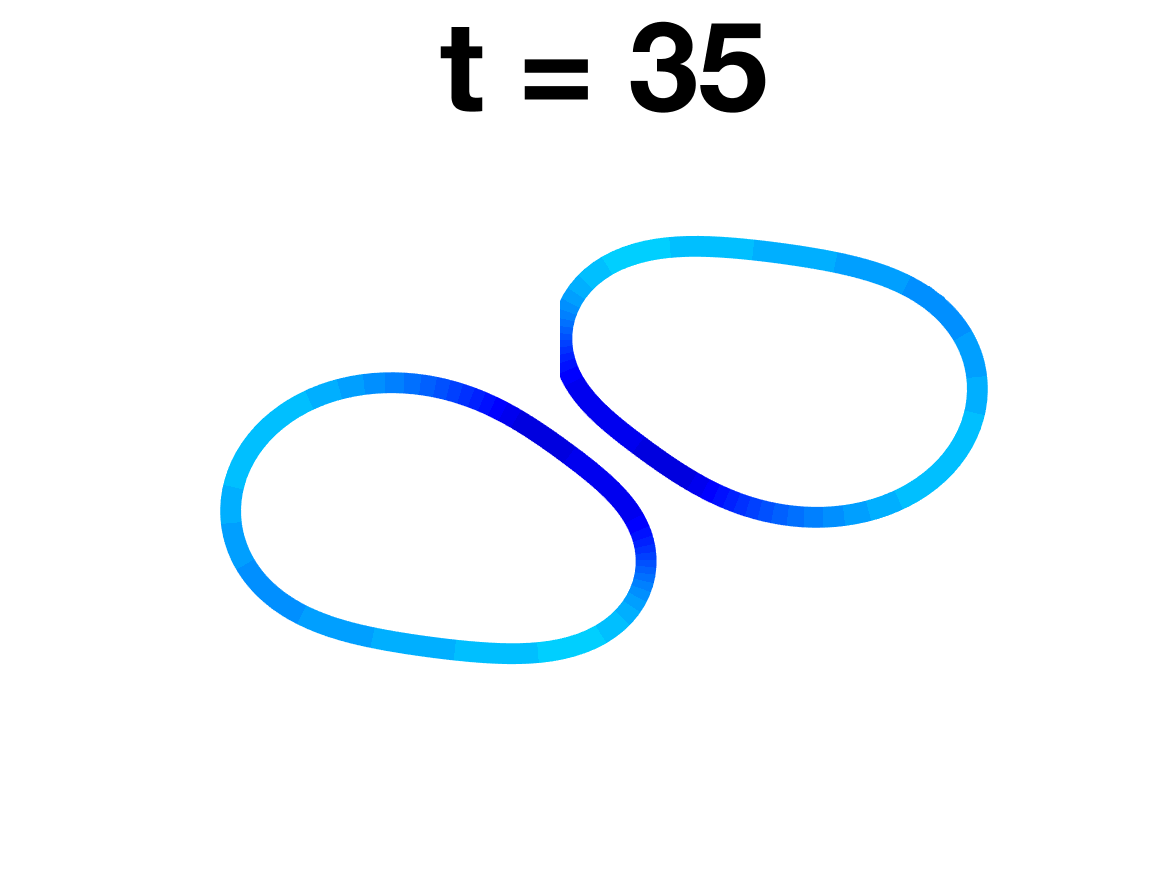}
  \includegraphics[width=0.19\textwidth]{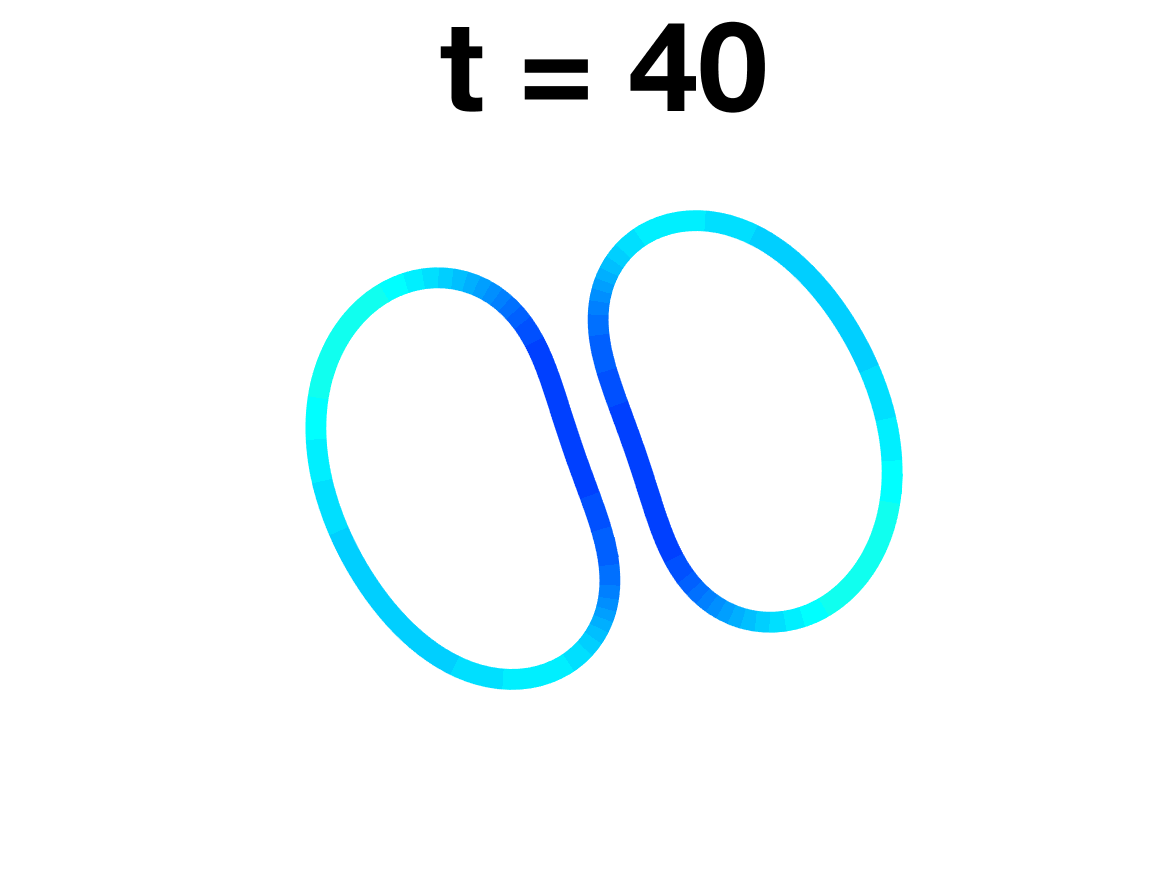}
  \includegraphics[width=0.19\textwidth]{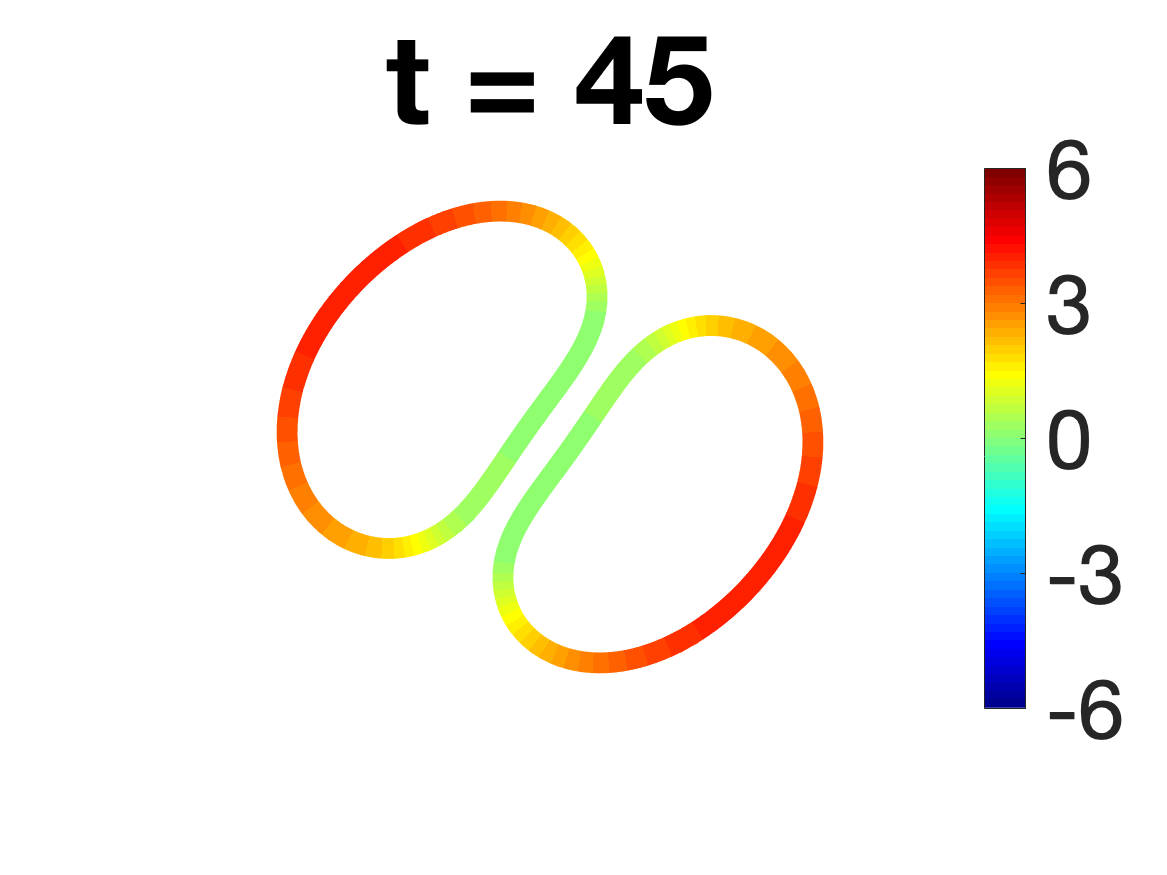}
  \caption{\label{fig:doublet090-strongAdhesion} The formation of a
  strongly adhered doublet in a shear flow.  The reduced area is $\Delta
  A = 0.9$, the Hamaker constant is $\mathcal{H}=2.1$, the separation
  distance is $\delta = 0.4$, and the shear rate is $\dot\gamma=0.5$.
  Since the contact region is unchanged throughout the simulation, the
  dynamics of the doublet resembles a rigid body motion with a period of
  about 42.  The color coding is the tension.}
\end{figure}

Figures~\ref{fig:doublet090-weakAdhesion}
and~\ref{fig:doublet090-strongAdhesion} show snapshots of two vesicles
that have formed a doublet and the color coding is the tension along the
vesicles.  Each of the vesicles have reduced area $\Delta A = 0.9$, shear
rate $\dot\gamma = 0.5$, separation distance $\delta = 0.4$, and Hamaker
constant $\mathcal{H} = 0.7$ (Figure~\ref{fig:doublet090-weakAdhesion})
and $\mathcal{H} = 2.1$ (Figure~\ref{fig:doublet090-strongAdhesion}).
Similar to the quiescent example, the membrane tension is negative in
the contact region indicating that the membrane is being compressed when
the adhesive force is strongest.  The two values for the Hamaker
constant are chosen since the oscillation period of the vesicle doublet
dynamics is both $t^* \approx 42$.  For both cases the vesicles in the
doublet move in tandem with a tank-reading motion that is much slower
than the tank-treading motion that occurs in the absence of adhesion.
However, the dynamics of a single period for these two Hamaker constants
differ significantly.  For $\mathcal{H} = 0.7$ the individual vesicles
undergo a sliding motion as they pass one another
(Figure~\ref{fig:doublet090-weakAdhesion}).  However, for $\mathcal{H} =
2.1$, once the contact region is formed, the doublet maintains the same
structure and undergoes a tumbling dynamic similar to a Jeffery orbit.

\begin{figure}
  \begin{tabular}{@{}p{0.45\linewidth}@{\quad}p{0.45\linewidth}@{}}
  \subfigimg[width=\linewidth]{(a)}{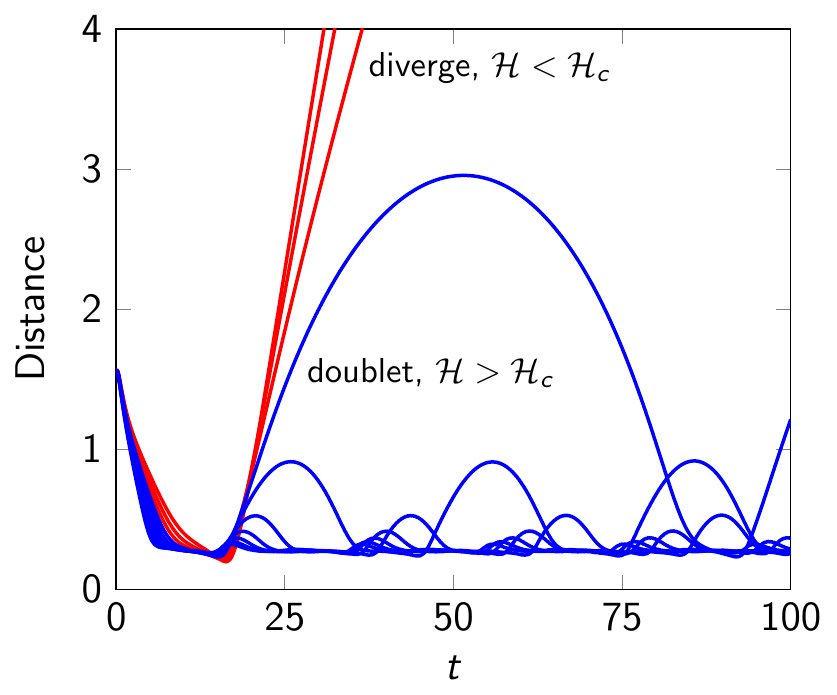} &
  \subfigimg[width=\linewidth]{(b)}{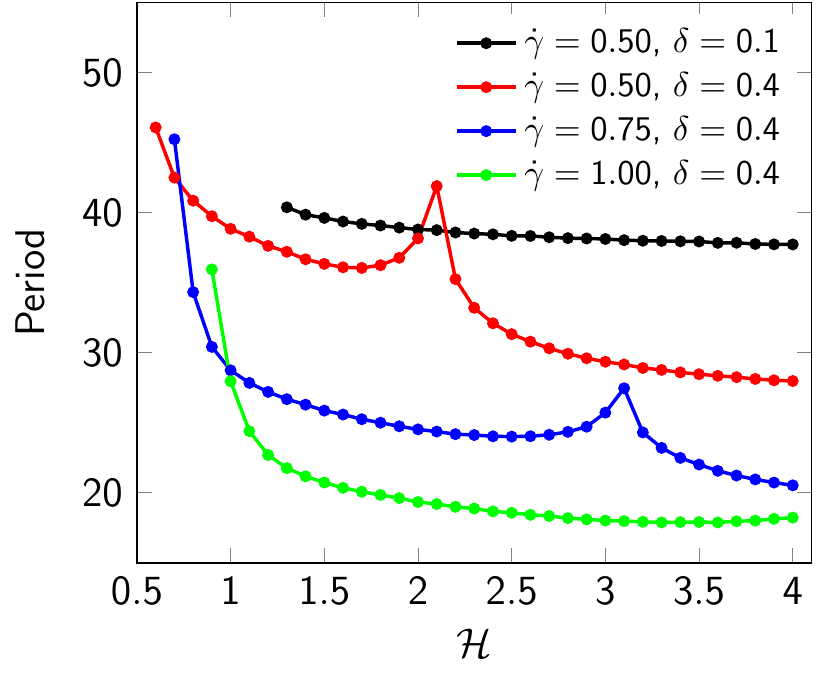}
  \end{tabular}
  \caption{\label{fig:sflow_distance} (a) The distance between a pair of
  vesicles in a planar shear flow with shear rate $\dot\gamma=0.5$, and
  separation distance $\delta = 0.4$.  The different lines correspond to
  a linear spacing of Hamaker constants ranging from $\mathcal{H}=0.1$
  and $\mathcal{H}=1.0$.  (b) The time required for a doublet to
  complete a single rotation.  The reduced area is $\Delta A = 0.90$ and
  the separation distance is $\delta = 0.4$.  The different lines
  correspond to different shear rates and separation distances.}
\end{figure}

To further understand the formation of the doublet, we compute the
minimum distance between two vesicles with reduced area $\Delta
A = 0.9$, separation distance $\delta = 0.5$, and shear rate $\dot\gamma
= 0.5$.  Figure~\ref{fig:sflow_distance}(a) shows this distance as a
function of time for varying Hamaker constants.  For Hamaker constants
less than a critical value, $\mathcal{H}_c$, the adhesive force is not
strong enough to form a doublet, and the distance between the vesicles
grows with time (red curves).  However, when $\mathcal{H} >
\mathcal{H}_c$, the adhesive force is sufficiently strong to bind the
vesicles into a doublet, and then the doublet undergoes a periodic
motion (blue curves).  In Figure~\ref{fig:sflow_distance}(b), we plot
the time required for the doublet to make a complete revolution for
three different shear rates and two different separation distances.
Initially, as the shear rate increases, the period decreases.  However,
the transition from the weakly adhesive doublet
(Figure~\ref{fig:doublet090-weakAdhesion}) to the stronger adhesive
doublet (Figure~\ref{fig:doublet090-strongAdhesion}) results in an
increase in the period.  Once the Hamaker constant is sufficiently
strong that the contact region remains fixed throughout the dynamics,
the period of the doublet again decreases as the shear rate increases.

Especially at the large Hamaker constants, the dynamics of the doublet
resemble those of a Jeffery orbit.  An ellipse with aspect ratio
$\lambda$ undergoes a periodic tumbling motion with
period~\cite{jef1922}
\begin{align}
  p = \frac{2\pi}{\dot\gamma}\left(\lambda + \lambda^{-1} \right).
  \label{eqn:JefferyPeriod}
\end{align}
Instead of measuring the aspect ratio of a doublet in a flow with shear
rate $\dot\gamma$, we use the computed period to define an effective
aspect ratio $\lambda_{\mathrm{eff}}$ that
satisfies~\eqref{eqn:JefferyPeriod}.  Therefore, the effective aspect
ratio is the aspect ratio of an ellipse in a shear flow that is
undergoing a Jeffery orbit with period $p$.  For the Hamaker constants
$\mathcal{H} = 1.5$ and $\mathcal{H} = 3.5$, the shear rates, periods,
and effective aspect ratio are summarized in
Table~\ref{tbl:aspectRatio}, which illustrates that the effective aspect
ratio of the doublet depends on both the Hamaker constant and the shear
rate.
\begin{table}[htp]
\begin{minipage}{0.45\textwidth}
  $\mathcal{H} = 1.5$ \\
  \begin{tabular}{>{\centering\arraybackslash}p{2cm}
>{\centering\arraybackslash}p{2cm}
>{\centering\arraybackslash}p{2cm}}
  $\dot\gamma$ & $p$ & $\lambda_{\mathrm{eff}}$ \\
  \hline
  0.50 & 36.34 & 2.49 \\
  0.75 & 25.86 & 2.72 \\
  1.00 & 20.73 & 2.96 
\end{tabular}
\end{minipage}
\begin{minipage}{0.45\textwidth}
  $\mathcal{H} = 3.5$ \\
  \begin{tabular}{>{\centering\arraybackslash}p{2cm}
>{\centering\arraybackslash}p{2cm}
>{\centering\arraybackslash}p{2cm}}
  $\dot\gamma$ & $p$ & $\lambda_{\mathrm{eff}}$ \\
  \hline
  0.50 & 28.46 & 1.66 \\
  0.75 & 22.01 & 2.17 \\
  1.00 & 17.91 & 2.44 
\end{tabular}
\end{minipage}
\caption{\label{tbl:aspectRatio} The shear rate, period, and effective
aspect ratio of a vesicle doublet with reduced area $\Delta A = 0.9$,
separation distance $\delta = 0.4$, and Hamaker constants $\mathcal{H} =
1.5$ and $\mathcal{H} = 3.5$.  The effective aspect ratios depend on
both the Hamaker constant and the shear rate, indicating that for the
reported Hamaker constants, the dynamics of the doublet differ from that
of a Jeffery orbit.}
\end{table}
Furthermore the oscillatory dynamics in
Figure~\ref{fig:doublet090-weakAdhesion} and
Figure~\ref{fig:doublet090-strongAdhesion} illustrate that for similar
oscillation periods, the actual aspect ratio of the vesicle doublet may
be a dynamical variable as well.  This indicates that the vesicle
doublet's oscillatory dynamics in a shear flow is more complicated than
the Jeffery orbit of a rigid body.

\begin{figure}
  \begin{tabular}{@{}p{0.45\linewidth}@{\quad}p{0.45\linewidth}@{}}
  \subfigimg[width=\linewidth]{(a)}{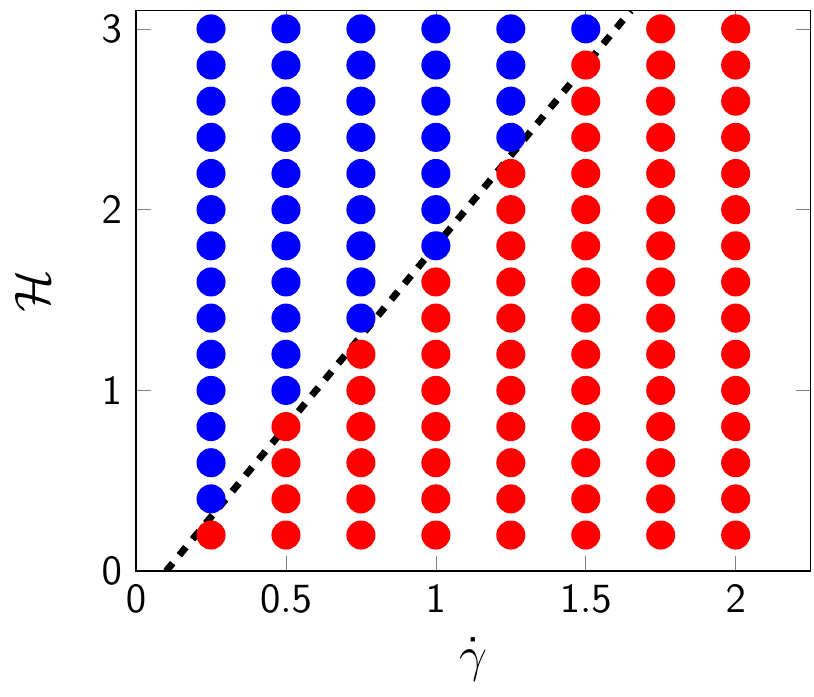} &
  \subfigimg[width=\linewidth]{(b)}{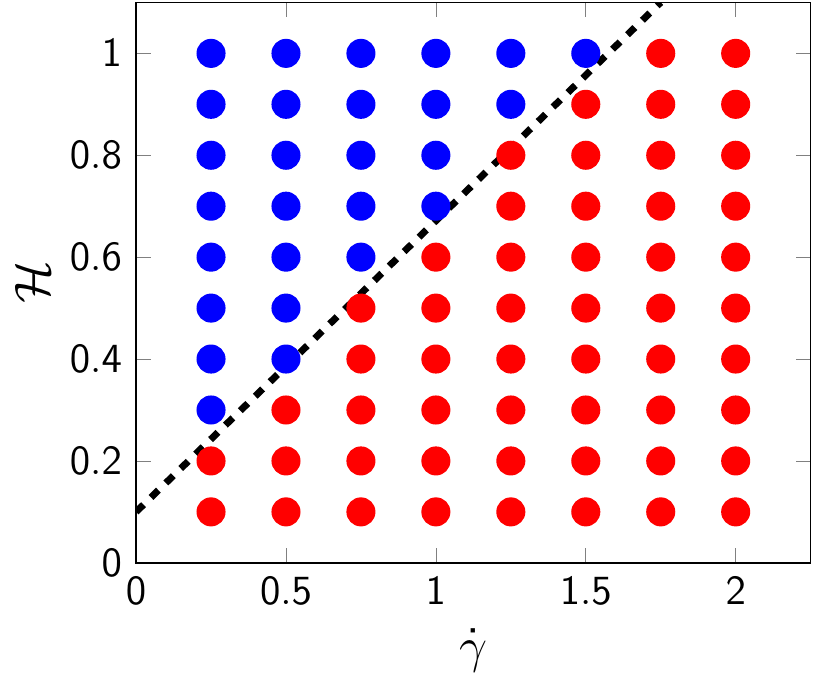}
  \end{tabular}
  \caption{\label{fig:sflow_phase_diagram} The phase diagram of vesicle
  hydrodynamics in a shear flow with $\delta = 0.1$ (a) and $\delta =
  0.4$ (b) with reduced area $\Delta A = 0.9$. The critical Hamaker
  constant $\mathcal{H}$ for binding of two vesicles in a shear flow
  depends on the shear rate $\dot\gamma$.  For a doublet to form, lower
  separation distances require a stronger Hamaker constants at a given
  shear rate.  The stability limit appears to be linear, indicating that
  there is potentially a stability criteria that depends on a
  dimensionless number involving the ratio of the Hamaker constant and
  shear rate.}
\end{figure}

Next, for several shear rates, we determine the critical Hamaker
constant, $\mathcal{H}_c$, that determines if the vesicles form a
doublet or separate.  Figure~\ref{fig:sflow_phase_diagram} indicates
whether a doublet with $\Delta A = 0.9$ is formed (blue) or not (red)
for two different separation distances.  The plot includes a line of
best fit between the stable and unstable regions.  Since the transition
is linear for both separation distances, this indicates that a
dimensionless number involving the ratio of the Hamaker constant and
shear rate could be used to determine if a doublet forms or not.  We
note that the critical Hamaker constant also depends on the reduced
area.  For smaller reduced areas, we expect a smaller contact region to
form and this will lower the Hamaker constant, and this behavior will be
investigated in future work.

Finally, we investigate the rheology of a suspension of a doublet by
computing the effective viscosity of a doublet and compare it to the
effective viscosity of a single tank-treading vesicle. The effective
viscosity is defined as the viscosity of a homogeneous Newtonian fluid
with the same energy dissipation per macroscopic element of fluid.  In a
simple shear flow, the intrinsic viscosity, $[\mu]$ is
\begin{align*}
  [\mu]:= \frac{\mu_{\mathrm{eff}} - \mu_0}{\phi \mu_0} = 
  \frac{1}{\dot\gamma \mu_0 (T_e - T_i)} \int_{T_i}^{T_e} 
  \langle \tau_{12} \rangle dt,
\end{align*}
where
\begin{align*}
  \langle \tau \rangle = \frac{1}{|\omega|} \int_{\gamma}
    \xxi \otimes \xx ds,
\end{align*}
$\phi$ is the area fraction of vesicles, $\tau$ is the stress due to the
vesicles, $\langle \cdot \rangle$ is the spatial average, $|\omega|$ is
the total area of both vesicles, and $\xxi$ is the traction as defined
in equation~\eqref{eqn:traction}.  

\begin{figure}[h]
  \begin{tabular}{@{}p{0.45\linewidth}@{\quad}p{0.45\linewidth}@{}}
  \subfigimg[width=\linewidth]{(a)}{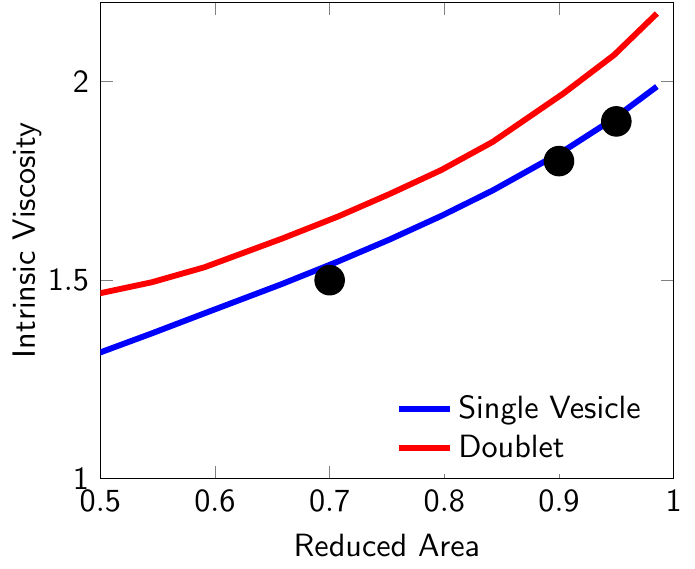} &
  \subfigimg[width=\linewidth]{(b)}{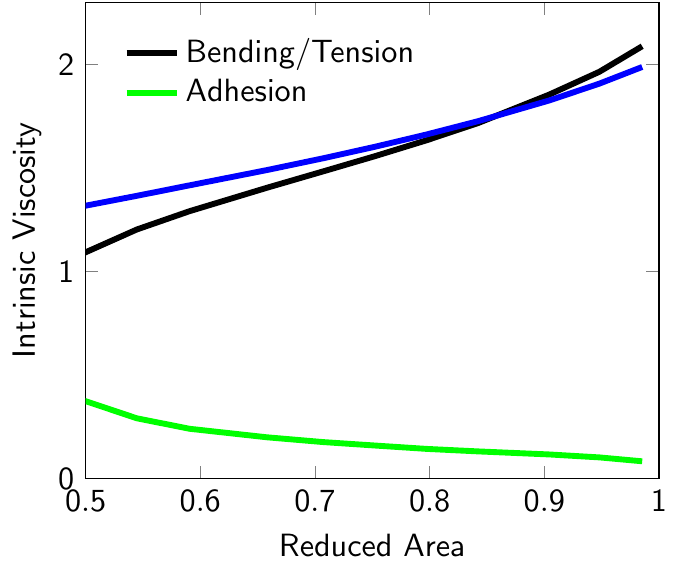}
  \end{tabular}
  \caption{\label{fig:shearIntrinsicViscosity} (a) The intrinsic
  viscosity of a tank-treading vesicle (blue) and a doublet (red).  The
  shear rate is $\dot\gamma = 0.5$, the Hamaker constant is $\mathcal{H}
  = 0.7$, and the separation distance is $\delta = 0.4$.  The black
  marks denote intrinsic viscosity values computed by Ghigliotti {\em et
  al.}~\cite{GhigliottiBibenMisbah2010_JFM} (cf.~Figure 5).  (b) The
  decomposition of the intrinsic viscosity into the contributions for
  the bending and tension (black) of the vesicles, and the contribution
  from the vesicle adhesion (green).  Also included is the intrinsic
  viscosity of a single tank-treading vesicle (blue).}
\end{figure}
In Figure~\ref{fig:shearIntrinsicViscosity}(a), we compare the intrinsic
viscosity of a single tank-treading vesicle to a doublet with Hamaker
constant $\mathcal{H} = 0.7$, separation distance $\delta = 0.4$, and
shear rate $\dot\gamma = 0.5$.  To validate our simulations, we
superimpose (black marks) the intrinsic viscosity calculated by
Ghigliotti {\em et al.}~\cite{GhigliottiBibenMisbah2010_JFM}.  The
presence of the doublet slightly increases the intrinsic viscosity at
all the reduced areas.  We also compute the intrinsic viscosity of a
doublet with $\Delta A = 0.9$, $\dot\gamma = 0.5$, and $\mathcal{H} =
2.1$.  Again, this doublet has a similar period to the case $\mathcal{H}
= 0.7$. At this larger Hamaker constant, the effective viscosity
increases slightly from $1.97$ to $2.06$.  To further characterize the
effect of adhesion, in Figure~\ref{fig:shearIntrinsicViscosity}(b) we
decompose the intrinsic viscosity into the contributions from the
bending and tension (black), and the contribution from the adhesion
(green).  We also superimpose the intrinsic viscosity of a single
tank-treading vesicle (blue) to demonstrate that the bending and tension
of the doublet behave similarly, but not identically, to a dilute
suspensions of non-adhering tank-treading vesicles.  We see that the
effect of the adhesion on the intrinsic viscosity is largest for
vesicles with small reduced areas.  We also decompose the intrinsic
viscosity contributions of the doublet with $\mathcal{H} = 2.1$ and
$\Delta A = 0.9$.  When compared with the smaller Hamaker constant, the
intrinsic viscosity contribution due to bending and tension is reduced
from $1.85$ to $1.71$, and the contribution due to adhesion is increased
from $0.12$ to $0.36$.  Therefore, as the Hamaker constant increases,
the increase in the intrinsic viscosity is due to the adhesive force
rather than the bending and tension.

\section{Conclusions\label{sec:conclusions}}
In this work we use a boundary integral formulation with adaptive
time-stepping to simulate hydrodynamics of two vesicles with adhesive
interactions.  In a quiescent flow, two vesicles that are initially
sufficiently far apart move towards each other under a long-range
attraction.  We use a lubrication theory to estimate the time required
to reach the separation distance $\delta$, and the theoretical scaling
is in good agreement with numerical results.  Once two membranes are
within separation distance $\delta$, the adhesive force turns repulsive
and the membranes flatten to form a contact region.  Our simulations
show that the membranes in the contact region are actually curved with
end points at the shortest distance, consistent with results for
moderate adhesion strength in~\cite{FlormannAouane2017_SciReports}.
Once a vesicle doublet forms, we examine the dependence of membrane
bending and adhesion energies on the reduced area, the Hamaker constant,
and the separation distance.

Next we conduct numerical simulations of two vesicles in a planar
extensional flow.  With initial configurations where two vesicles
collide around the stagnation point (one vesicle at the stagnation point
and the other vesicle placed close to the converging axis), the two
vesicles form a doublet and stay around the flow center only when two
vesicles collide head-on with no displacement $\Delta y=0$.  For $\Delta
y > 0$ we find that the vesicles rotate and move away from the
stagnation point, and eventually move farther from each other due to the
strong diverging flow in the far-field.

This inspires us to conduct a numerical experiment where a vesicle
doublet is placed at the center of a fluid trap, which can be actively
controlled in microfluidic channel so that fluid trap center is
effectively the stagnation point of an extensional flow characterized by
a extensional flow rate.  At low flow rate, we find the doublet to
rotate nearly ninety degrees to align with the flow such that the long
axis of the doublet is parallel to the divergent axis and the convergent
stream is pushing the two vesicles together. As the flow rate increases,
the vesicle doublet rotates less, and when the flow rate exceeds the
critical value, the diverging flow breaks the doublet structure by
pulling the vesicles apart.  These results indicate that it is possible
to use the fluid trap to separate a vesicle doublet under adhesion, and
thus provide a means to probe the adhesion strength between membranes.
For a pair of $\mu m$-sized vesicles with a bending modulus of $\sim
10^{-19}$ $J$, an extensional flow rate of $0.5$ $s^{-1}$ is expected to
separate a vesicle doublet with reduced area of $\Delta A = 0.8$ and a
Hamaker constant $\mathcal{H}=0.7$, which corresponds to $\sim 1$ $\mu
J/m^2$. These conditions are quite realizable in microfluidic
experiments, and we hope that our simulations will inspire microfluidic
experiments to be designed to measure membrane-membrane adhesion.

We also examine how adhesive interaction may dynamically lead to the
formation of a vesicle doublet.  We simulate two vesicles approaching
each other in a planar shear flow, and examine how their adhesive
interactions lead to doublet formation.  Once a doublet forms, the two
vesicle membranes rotate around each other as they deform dynamically.
The usual tank-treading motion of a vesicle under shear flow is not
observed in each of the two vesicles. We compute the effective shear
viscosity of a dilute suspension of vesicle doublets, and found it to be
more than twice the effective shear viscosity of a dilute suspension of
single vesicles.  Furthermore, we find that the membrane adhesion
contribution to the shear viscosity increases with decreasing reduced
area while the bending/tension contribution increases with the reduced
area.

In our formulation we did not include any electrostatic interactions
between membranes under adhesion. When the electrostatic interaction is
important, an electro-osmotic pressure in the thin film is found to be
responsible for the observed membrane
undulation~\cite{SteinkuhlerAgudo-Canalejo2016_BJ}.  In addition,
simulations presented in this work are for identical vesicles (same
reduced area, length, and bending modulus) in the doublet.  Flormann
{\em et al.}~demonstrated that asymmetric vesicle reduced area may lead
to various equilibrium doublet shape such as male-female, asymmetric
S-shape, and parachute shape.  It is also possible that viscosity
contrast may also lead to different equilibrium doublet shapes.  Future
work includes three dimensional simulations, dispersive vesicle
properties (such as reduced area and bending modulus), viscosity
contrast, and effects of confinement on adhesive interactions.  Another
future direction is to consider the clustering and packing of an
unbounded suspension of vesicles~\cite{FlormannAouane2017_SciReports}.

Finally, it is not clear how thermal fluctuations may affect the
hydrodynamics of vesicles under adhesion. For example, does the
fluctuating hydrodynamics in the thin film between two vesicles enhance
adhesion to keep vesicles bound under linear flows as speculated in
cells~\cite{FenzBihrSchmidt2017_NaturePhys}?  Recently Liu {\em et
al.}~\cite{LiuChuNewbyRead2018_bioRxiv} used immersed boundary
simulations to show that, at a separation distance of tens of
nanometers, the thin film between the two membranes facilitates the
coupling between membranes via strong hydrodynamic interactions. In
particular, they demonstrate numerically that the fluctuation in one
membrane is highly correlated to the other membrane without any physical
contact. We are actively pursuing this direction with hydrodynamic
modeling and simulations of adhesive membranes with thermal
fluctuations.

\acknowledgments

BQ acknowledges support from Florida State University startup funds and
Simons Foundation Mathematics and Physical Sciences-Collaboration Grants
for Mathematicians 527139.  SV acknowledges support from NSF under
grants DMS-1719834 and DMS-1454010.  YNY acknowledges support from
NSF-DMS 1614863 and NSF-DMS 1412789.  Both SV and YNY were also supported
by the Flatiron Institute, a division of Simons Foundation.

BQ and YNY contributed equally to model development and simulations. BQ
developed the code. All authors contributed equally to writing the
manuscript.

\begin{appendices}
\section{Integral Equation Formulation}
\label{sec:AppendixB}
Using potential theory and following~\cite{Veerapaneni2009_JCP}, we recast the governing equations~\eqref{eq:gov} as integro-differential equations for the evolution of membrane positions: 
\begin{align*}
  &\dot{\xx} = \uu_{\infty}(\xx) + \SS[\xxi](\xx), \\
  &\xx_s \cdot \dot{\xx}_s = 0,
 \end{align*}
where the single-layer potential $\SS[\cdot]$ is defined by
\begin{align*}
  \SS[\xxi](\xx) &= \frac{1}{4\pi\mu} \int_\gamma \left(
    -\log \|\xx - \yy\| + \frac{(\xx - \yy) \otimes (\xx - \yy)}{\|\xx - \yy\|^2} \right) 
    \xxi(\yy) ds_\yy, 
\end{align*}
and the membrane force $\xxi$ is a sum of the bending, tension and, adhesion forces: 
\begin{align}
\xxi = -\kappa_b \xx_{ssss} + (\sigma \xx_s)_s + \AA\xx.
\label{eqn:traction}
\end{align}
Defining the bending operator as $\BB[\ff](\xx) = -\kappa_b \ff_{ssss}$,
the tension operator $\TT[\sigma](\xx) = (\sigma \xx_s)_s$, and using
IMEX-Euler, the no-slip boundary results in the time stepping method
\begin{align*}
  \xx^{N+1} - \Delta t \SS^N \BB^N \xx^{N+1} - 
    \Delta t \SS^N \TT^N \sigma^{N+1} = \xx^N + 
    \Delta t \SS^N \AA^N \xx^N,
\end{align*}
and the inextensibility constraint that is discretized as
\begin{align*}
  \xx_s^{N} \cdot \xx_{s}^{N+1} = 1.
\end{align*}
We discretize the vesicles at a set of collocation points, compute the
bending and tension terms with Fourier differentiation, and apply Alpert
quadrature~\cite{alp1999} to the weakly-singular single-layer potential
$\SS$.  The source and target points of the adhesion force never
coincide since they are always on different vesicles, so the adhesion
force~\eqref{eqn:adhesionForce} is computed with the spectrally accurate
trapezoid rule~\cite{tre-wei2014}.  

The dynamics of a doublet undergoes many different time scales over time
horizons that are sufficiently large to characterize the formation of a
doublet and its rheological properties.  Therefore, time adaptivity is
crucial so that a user-specified tolerance is achieved without using a
guess-and-check procedure to find an appropriately small fixed time step
size.  To control the error and achieve second-order accuracy in time,
we use a time adaptive spectral deferred correction method that
applies IMEX-Euler twice per time step~\cite{quaife2016adaptive}.

\section{Adhesion Force}
\label{sec:appendixA}
Consider a suspension of two vesicles $\gamma_1$ and $\gamma_2$
parameterized as $\xx_1(s)$ and $\xx_2(s)$, respectively, with $s \in
[0,1]$.  Here $s$ is the arclength, and we have assumed, without loss of
generality, that both vesicles have length one.  We use the  L.-J.~type
potential
\begin{align*}
  \phi(z) = \mathcal{H} \left[ 
    \left(\frac{\delta}{z}\right)^m - \frac{m}{n}\left(\frac{\delta}{z}\right)^n \right],
\end{align*}
where $z$ is the distance between two points on a pair of vesicles.
Then, we define the total adhesive energy on $\gamma_1$ to be
\begin{align*}
  U_1 = \int_{\gamma_1} \int_{\gamma_2} \phi(\|\xx_1 - \xx_2\|) 
    ds_{\xx_2} ds_{\xx_1}.
\end{align*}
Perturbing $\xx_1$ to $\tilde{\xx}_1 = \xx_1 +  \delta \xx_1$ results
in a new vesicle $\tilde{\gamma}_1$, and the perturbed adhesive energy is
\begin{align*}
  \widetilde{U}_1 = \int_{\tilde{\gamma}_1} \int_{\gamma_2}
  \phi(\|\tilde{\xx}_1 - \xx_2\|) ds_{\xx_2} ds_{\tilde{\xx}_1},
\end{align*}
and the change in the energy is
\begin{align*}
  \delta U_1 = \int_{\tilde{\gamma}_1} \int_{\gamma_2}
  \phi(\|\tilde{\xx}_1 - \xx_2\|) ds_{\xx_2} ds_{\tilde{\xx}_1} - 
  \int_{\gamma_1} \int_{\gamma_2} \phi(\|\xx_1 - \xx_2\|) 
  ds_{\xx_2} ds_{\xx_1}.
\end{align*}

We now decompose the perturbation into normal and tangential components
as $\delta \xx_1 = \epsilon \yy(s) = \epsilon(u\tt + v\nn)$. The
perturbed arclength term, to leading order, is
\begin{align*}
  \|\tilde{\xx}'_1\| \approx 1 + \epsilon(u_s + \nu\kappa),
\end{align*}
where $\kappa$ is the curvature.  To leading order, inextensible
perturbations satisfy $u_s + \kappa v = 0$, so the arclength term of
$\gamma_1$ and $\tilde{\gamma}_1$ are identical to leading order.
Therefore,
\begin{align*}
  \delta U_1 &= \int_{\gamma_1} \int_{\gamma_2} \left(
  \phi(\|\xx_1 + \epsilon \yy  - \xx_2\|) - \phi(\|\xx_1 - \xx_2\|)
  \right) ds_{\xx_2} ds_{\xx_1} \\
  &\approx \epsilon \int_{\gamma_1} \int_{\gamma_2}
  \nabla \phi (\|\xx_1 - \xx_2\|) \cdot \yy ds_{\xx_2} ds_{\xx_1}
\end{align*}
and the adhesive force applied by vesicle 2 on vesicle 1 is
\begin{align*}
 \int_{\gamma_2}\nabla \phi(\|\xx_1 - \xx_2\|)ds_{\xx_2} = 
  -m \mathcal{H}\delta^{n} \int_{\gamma_k} 
  \frac{\xx - \yy}{\|\xx - \yy\|^{m+2}} 
  \left(\delta^{m-n} - \|\xx - \yy\|^{m-n} \right) ds_\yy.
  \end{align*}
When $(m,n) = (4,2)$, the above expression becomes
\begin{align*}
  \int_{\gamma_2}\nabla \phi(\|\xx_1 - \xx_2\|)ds_{\xx_2} = 
  -4 \mathcal{H} \delta^2 \int_{\gamma_2}
  \frac{\xx_1 - \xx_2}{\|\xx_1 - \xx_2\|^6} 
  \left(\delta^2 - \|\xx_1 - \xx_2\|^2 \right) ds_{\xx_2}.
\end{align*}
A similar expression holds for the adhesive force applied by vesicle 1
on vesicle 2, and equation~\eqref{eqn:adhesionForce} gives the adhesive
force for a suspension of $M$ vesicles.

\end{appendices}

\bibliography{refs}

\begin{thebibliography}{86}%
\makeatletter
\providecommand \@ifxundefined [1]{%
 \@ifx{#1\undefined}
}%
\providecommand \@ifnum [1]{%
 \ifnum #1\expandafter \@firstoftwo
 \else \expandafter \@secondoftwo
 \fi
}%
\providecommand \@ifx [1]{%
 \ifx #1\expandafter \@firstoftwo
 \else \expandafter \@secondoftwo
 \fi
}%
\providecommand \natexlab [1]{#1}%
\providecommand \enquote  [1]{``#1''}%
\providecommand \bibnamefont  [1]{#1}%
\providecommand \bibfnamefont [1]{#1}%
\providecommand \citenamefont [1]{#1}%
\providecommand \href@noop [0]{\@secondoftwo}%
\providecommand \href [0]{\begingroup \@sanitize@url \@href}%
\providecommand \@href[1]{\@@startlink{#1}\@@href}%
\providecommand \@@href[1]{\endgroup#1\@@endlink}%
\providecommand \@sanitize@url [0]{\catcode `\\12\catcode `\$12\catcode
  `\&12\catcode `\#12\catcode `\^12\catcode `\_12\catcode `\%12\relax}%
\providecommand \@@startlink[1]{}%
\providecommand \@@endlink[0]{}%
\providecommand \url  [0]{\begingroup\@sanitize@url \@url }%
\providecommand \@url [1]{\endgroup\@href {#1}{\urlprefix }}%
\providecommand \urlprefix  [0]{URL }%
\providecommand \Eprint [0]{\href }%
\providecommand \doibase [0]{http://dx.doi.org/}%
\providecommand \selectlanguage [0]{\@gobble}%
\providecommand \bibinfo  [0]{\@secondoftwo}%
\providecommand \bibfield  [0]{\@secondoftwo}%
\providecommand \translation [1]{[#1]}%
\providecommand \BibitemOpen [0]{}%
\providecommand \bibitemStop [0]{}%
\providecommand \bibitemNoStop [0]{.\EOS\space}%
\providecommand \EOS [0]{\spacefactor3000\relax}%
\providecommand \BibitemShut  [1]{\csname bibitem#1\endcsname}%
\let\auto@bib@innerbib\@empty
\bibitem [{\citenamefont {Sackmann}(1996)}]{sackmann1996}%
  \BibitemOpen
  \bibfield  {author} {\bibinfo {author} {\bibfnamefont {E.}~\bibnamefont
  {Sackmann}},\ }\bibfield  {title} {\enquote {\bibinfo {title} {{Supported
  Membranes: Scientific and Practical Applications}},}\ }\href@noop {}
  {\bibfield  {journal} {\bibinfo  {journal} {Science}\ }\textbf {\bibinfo
  {volume} {271}},\ \bibinfo {pages} {43--48} (\bibinfo {year}
  {1996})}\BibitemShut {NoStop}%
\bibitem [{\citenamefont {Fenz}\ and\ \citenamefont
  {Sengupta}(2012)}]{FenzSengupta2012_IntegrBiol}%
  \BibitemOpen
  \bibfield  {author} {\bibinfo {author} {\bibfnamefont {S.~F.}\ \bibnamefont
  {Fenz}}\ and\ \bibinfo {author} {\bibfnamefont {K.}~\bibnamefont
  {Sengupta}},\ }\bibfield  {title} {\enquote {\bibinfo {title} {Giant vesicles
  as cell models},}\ }\href@noop {} {\bibfield  {journal} {\bibinfo  {journal}
  {Integr. Biol.}\ }\textbf {\bibinfo {volume} {4}},\ \bibinfo {pages}
  {982--995} (\bibinfo {year} {2012})}\BibitemShut {NoStop}%
\bibitem [{\citenamefont {Barthes-Biesel}(2016)}]{Barthes-Biesel2016_ARFM}%
  \BibitemOpen
  \bibfield  {author} {\bibinfo {author} {\bibfnamefont {D.}~\bibnamefont
  {Barthes-Biesel}},\ }\bibfield  {title} {\enquote {\bibinfo {title} {Motion
  and deformation of elastic capsules and vesicles in flow},}\ }\href@noop {}
  {\bibfield  {journal} {\bibinfo  {journal} {Annu. Rev. Fluid Mech.}\ }\textbf
  {\bibinfo {volume} {48}},\ \bibinfo {pages} {23--52} (\bibinfo {year}
  {2016})}\BibitemShut {NoStop}%
\bibitem [{\citenamefont
  {Dobereiner}(2000)}]{Dobereiner2000_CurrentOpinionCIS}%
  \BibitemOpen
  \bibfield  {author} {\bibinfo {author} {\bibfnamefont {H.-G.}\ \bibnamefont
  {Dobereiner}},\ }\bibfield  {title} {\enquote {\bibinfo {title} {Properties
  of giant vesicles},}\ }\href@noop {} {\bibfield  {journal} {\bibinfo
  {journal} {Current Opinion Colloid Int. Sci.}\ }\textbf {\bibinfo {volume}
  {5}},\ \bibinfo {pages} {256--263} (\bibinfo {year} {2000})}\BibitemShut
  {NoStop}%
\bibitem [{\citenamefont {Evans}\ \emph {et~al.}(2013)\citenamefont {Evans},
  \citenamefont {Rawicz},\ and\ \citenamefont
  {Smith}}]{EvansRawiczSmith2013_FaradayDiscussions}%
  \BibitemOpen
  \bibfield  {author} {\bibinfo {author} {\bibfnamefont {E.}~\bibnamefont
  {Evans}}, \bibinfo {author} {\bibfnamefont {W.}~\bibnamefont {Rawicz}}, \
  and\ \bibinfo {author} {\bibfnamefont {B.~A.}\ \bibnamefont {Smith}},\
  }\bibfield  {title} {\enquote {\bibinfo {title} {Concluding remarks back to
  the future: mechanics and thermodynamics of lipid biomembrane},}\ }\href@noop
  {} {\bibfield  {journal} {\bibinfo  {journal} {Faraday Discussions}\ }\textbf
  {\bibinfo {volume} {161}},\ \bibinfo {pages} {591--611} (\bibinfo {year}
  {2013})}\BibitemShut {NoStop}%
\bibitem [{\citenamefont {Sugiyama}\ and\ \citenamefont
  {Toyota}(2018)}]{SugiyamaToyota2018_Life}%
  \BibitemOpen
  \bibfield  {author} {\bibinfo {author} {\bibfnamefont {H.}~\bibnamefont
  {Sugiyama}}\ and\ \bibinfo {author} {\bibfnamefont {T.}~\bibnamefont
  {Toyota}},\ }\bibfield  {title} {\enquote {\bibinfo {title} {Toward
  experimental evolution with giant vesicles},}\ }\href@noop {} {\bibfield
  {journal} {\bibinfo  {journal} {Life}\ }\textbf {\bibinfo {volume} {8}},\
  \bibinfo {pages} {53} (\bibinfo {year} {2018})}\BibitemShut {NoStop}%
\bibitem [{\citenamefont {Barthes-Biesel}\ and\ \citenamefont
  {Rallison}(1981)}]{Barthes-BieselRallison1981_JFM}%
  \BibitemOpen
  \bibfield  {author} {\bibinfo {author} {\bibfnamefont {D.}~\bibnamefont
  {Barthes-Biesel}}\ and\ \bibinfo {author} {\bibfnamefont {J.~M.}\
  \bibnamefont {Rallison}},\ }\bibfield  {title} {\enquote {\bibinfo {title}
  {The time-dependent deformation of a capsule freely suspended in a linear
  shear flow},}\ }\href@noop {} {\bibfield  {journal} {\bibinfo  {journal}
  {Journal of Fluid Mechanics}\ }\textbf {\bibinfo {volume} {113}},\ \bibinfo
  {pages} {251--267} (\bibinfo {year} {1981})}\BibitemShut {NoStop}%
\bibitem [{\citenamefont {Misbah}(2006)}]{Misbah2006_PRL}%
  \BibitemOpen
  \bibfield  {author} {\bibinfo {author} {\bibfnamefont {C.}~\bibnamefont
  {Misbah}},\ }\bibfield  {title} {\enquote {\bibinfo {title} {Vascillating
  breathing and tumbling of vesicles under shear flow},}\ }\href@noop {}
  {\bibfield  {journal} {\bibinfo  {journal} {Physical Review Letters}\
  }\textbf {\bibinfo {volume} {96}},\ \bibinfo {pages} {028104} (\bibinfo
  {year} {2006})}\BibitemShut {NoStop}%
\bibitem [{\citenamefont {Vlahovska}\ and\ \citenamefont
  {Gracia}(2007)}]{Vlahovska2007_PRE}%
  \BibitemOpen
  \bibfield  {author} {\bibinfo {author} {\bibfnamefont {P.~M.}\ \bibnamefont
  {Vlahovska}}\ and\ \bibinfo {author} {\bibfnamefont {R.}~\bibnamefont
  {Gracia}},\ }\bibfield  {title} {\enquote {\bibinfo {title} {Dynamics of a
  viscous vesicle in linear flows},}\ }\href@noop {} {\bibfield  {journal}
  {\bibinfo  {journal} {Physical Review E}\ }\textbf {\bibinfo {volume} {75}},\
  \bibinfo {pages} {016313} (\bibinfo {year} {2007})}\BibitemShut {NoStop}%
\bibitem [{\citenamefont {Finken}\ \emph {et~al.}(2008)\citenamefont {Finken},
  \citenamefont {Lamura}, \citenamefont {Seifert},\ and\ \citenamefont
  {Gompper}}]{Finken2008_EPL}%
  \BibitemOpen
  \bibfield  {author} {\bibinfo {author} {\bibfnamefont {R.}~\bibnamefont
  {Finken}}, \bibinfo {author} {\bibfnamefont {A.}~\bibnamefont {Lamura}},
  \bibinfo {author} {\bibfnamefont {U.}~\bibnamefont {Seifert}}, \ and\
  \bibinfo {author} {\bibfnamefont {G.}~\bibnamefont {Gompper}},\ }\bibfield
  {title} {\enquote {\bibinfo {title} {Two-dimensional fluctuating vesicles in
  linear shear flow},}\ }\href@noop {} {\bibfield  {journal} {\bibinfo
  {journal} {European Physical Journal E}\ }\textbf {\bibinfo {volume} {25}},\
  \bibinfo {pages} {309--321} (\bibinfo {year} {2008})}\BibitemShut {NoStop}%
\bibitem [{\citenamefont {Zhang}\ \emph {et~al.}(2013)\citenamefont {Zhang},
  \citenamefont {Zahn}, \citenamefont {Tan},\ and\ \citenamefont
  {Lin}}]{ZhangZahnTanLin2013_PoF}%
  \BibitemOpen
  \bibfield  {author} {\bibinfo {author} {\bibfnamefont {J.}~\bibnamefont
  {Zhang}}, \bibinfo {author} {\bibfnamefont {J.}~\bibnamefont {Zahn}},
  \bibinfo {author} {\bibfnamefont {W.}~\bibnamefont {Tan}}, \ and\ \bibinfo
  {author} {\bibfnamefont {H.}~\bibnamefont {Lin}},\ }\bibfield  {title}
  {\enquote {\bibinfo {title} {A transient solution for vesicle
  electrodeformation and relaxation},}\ }\href@noop {} {\bibfield  {journal}
  {\bibinfo  {journal} {Physics of Fluids}\ }\textbf {\bibinfo {volume} {25}},\
  \bibinfo {pages} {071903} (\bibinfo {year} {2013})}\BibitemShut {NoStop}%
\bibitem [{\citenamefont {Nganguia}\ and\ \citenamefont
  {Young}(2013)}]{Nganguia2013_PRE}%
  \BibitemOpen
  \bibfield  {author} {\bibinfo {author} {\bibfnamefont {H.}~\bibnamefont
  {Nganguia}}\ and\ \bibinfo {author} {\bibfnamefont {Y.-N.}\ \bibnamefont
  {Young}},\ }\bibfield  {title} {\enquote {\bibinfo {title} {Equilibrium
  electrodeformation of a spheroidal vesicle in an ac electric field},}\
  }\href@noop {} {\bibfield  {journal} {\bibinfo  {journal} {Physical Review
  E}\ }\textbf {\bibinfo {volume} {88}},\ \bibinfo {pages} {052718} (\bibinfo
  {year} {2013})}\BibitemShut {NoStop}%
\bibitem [{\citenamefont {Bagchi}\ \emph {et~al.}(2005)\citenamefont {Bagchi},
  \citenamefont {Johnson},\ and\ \citenamefont
  {Popel}}]{BagchiJohoson2005_JBE}%
  \BibitemOpen
  \bibfield  {author} {\bibinfo {author} {\bibfnamefont {P.}~\bibnamefont
  {Bagchi}}, \bibinfo {author} {\bibfnamefont {P.~C.}\ \bibnamefont {Johnson}},
  \ and\ \bibinfo {author} {\bibfnamefont {A.~S.}\ \bibnamefont {Popel}},\
  }\bibfield  {title} {\enquote {\bibinfo {title} {Computational fluid dynamic
  simulation of aggregation of deformable cells in a shear flow},}\ }\href@noop
  {} {\bibfield  {journal} {\bibinfo  {journal} {Journal of Biomechanical
  Engineering}\ }\textbf {\bibinfo {volume} {127}},\ \bibinfo {pages}
  {1070--1080} (\bibinfo {year} {2005})}\BibitemShut {NoStop}%
\bibitem [{\citenamefont {Biben}(2005)}]{Biben2005_EJP}%
  \BibitemOpen
  \bibfield  {author} {\bibinfo {author} {\bibfnamefont {T.}~\bibnamefont
  {Biben}},\ }\bibfield  {title} {\enquote {\bibinfo {title} {Phase-field
  models for free-boundary problems},}\ }\href@noop {} {\bibfield  {journal}
  {\bibinfo  {journal} {European Journal of Physics}\ }\textbf {\bibinfo
  {volume} {26}},\ \bibinfo {pages} {47--55} (\bibinfo {year}
  {2005})}\BibitemShut {NoStop}%
\bibitem [{\citenamefont {Veerapaneni}\ \emph {et~al.}(2009)\citenamefont
  {Veerapaneni}, \citenamefont {Gueyffier}, \citenamefont {Zorin},\ and\
  \citenamefont {Biros}}]{Veerapaneni2009_JCP}%
  \BibitemOpen
  \bibfield  {author} {\bibinfo {author} {\bibfnamefont {S.~K.}\ \bibnamefont
  {Veerapaneni}}, \bibinfo {author} {\bibfnamefont {D.}~\bibnamefont
  {Gueyffier}}, \bibinfo {author} {\bibfnamefont {D.}~\bibnamefont {Zorin}}, \
  and\ \bibinfo {author} {\bibfnamefont {G.}~\bibnamefont {Biros}},\ }\bibfield
   {title} {\enquote {\bibinfo {title} {A boundary integral method for
  simulating the dynamics of inextensible vesicles suspended in a viscous fluid
  in 2{D}},}\ }\href@noop {} {\bibfield  {journal} {\bibinfo  {journal}
  {Journal of Computational Physics}\ }\textbf {\bibinfo {volume} {228}},\
  \bibinfo {pages} {2334--2353} (\bibinfo {year} {2009})}\BibitemShut {NoStop}%
\bibitem [{\citenamefont {Seol}\ \emph {et~al.}(2016)\citenamefont {Seol},
  \citenamefont {Hu}, \citenamefont {Kim},\ and\ \citenamefont
  {Lai}}]{SeolHuKimLai2016_JCP}%
  \BibitemOpen
  \bibfield  {author} {\bibinfo {author} {\bibfnamefont {Y.}~\bibnamefont
  {Seol}}, \bibinfo {author} {\bibfnamefont {W.-F.}\ \bibnamefont {Hu}},
  \bibinfo {author} {\bibfnamefont {Y.}~\bibnamefont {Kim}}, \ and\ \bibinfo
  {author} {\bibfnamefont {M.-C.}\ \bibnamefont {Lai}},\ }\bibfield  {title}
  {\enquote {\bibinfo {title} {An immersed boundary method for simulating
  vesicle dynamics in three dimensions},}\ }\href@noop {} {\bibfield  {journal}
  {\bibinfo  {journal} {Journal of Computational Physics}\ }\textbf {\bibinfo
  {volume} {322}},\ \bibinfo {pages} {125--141} (\bibinfo {year}
  {2016})}\BibitemShut {NoStop}%
\bibitem [{\citenamefont {Veerapaneni}\ \emph {et~al.}(2011)\citenamefont
  {Veerapaneni}, \citenamefont {Young}, \citenamefont {Vlahovska},\ and\
  \citenamefont {B{\l}azdzwicz}}]{Veerapaneni2011_PRL}%
  \BibitemOpen
  \bibfield  {author} {\bibinfo {author} {\bibfnamefont {S.}~\bibnamefont
  {Veerapaneni}}, \bibinfo {author} {\bibfnamefont {Y.-N.}\ \bibnamefont
  {Young}}, \bibinfo {author} {\bibfnamefont {P.~M.}\ \bibnamefont
  {Vlahovska}}, \ and\ \bibinfo {author} {\bibfnamefont {J.}~\bibnamefont
  {B{\l}azdzwicz}},\ }\bibfield  {title} {\enquote {\bibinfo {title} {Dynamics
  of a compound vesicle in shear flow},}\ }\href@noop {} {\bibfield  {journal}
  {\bibinfo  {journal} {Physical Review Letters}\ }\textbf {\bibinfo {volume}
  {106}},\ \bibinfo {pages} {158103} (\bibinfo {year} {2011})}\BibitemShut
  {NoStop}%
\bibitem [{\citenamefont {Vitkova}\ \emph {et~al.}(2008)\citenamefont
  {Vitkova}, \citenamefont {Mader}, \citenamefont {Polack}, \citenamefont
  {Misbah},\ and\ \citenamefont {Podgorski}}]{Vitkova2008_BJ}%
  \BibitemOpen
  \bibfield  {author} {\bibinfo {author} {\bibfnamefont {V.}~\bibnamefont
  {Vitkova}}, \bibinfo {author} {\bibfnamefont {M.}~\bibnamefont {Mader}},
  \bibinfo {author} {\bibfnamefont {B.}~\bibnamefont {Polack}}, \bibinfo
  {author} {\bibfnamefont {C.}~\bibnamefont {Misbah}}, \ and\ \bibinfo {author}
  {\bibfnamefont {T.}~\bibnamefont {Podgorski}},\ }\bibfield  {title} {\enquote
  {\bibinfo {title} {Micro-macro link in rheology of erythrocyte and vesicle
  suspensions},}\ }\href@noop {} {\bibfield  {journal} {\bibinfo  {journal}
  {Biophysical Journal}\ }\textbf {\bibinfo {volume} {95}},\ \bibinfo {pages}
  {L33--L35} (\bibinfo {year} {2008})}\BibitemShut {NoStop}%
\bibitem [{\citenamefont {Ghigliotti}\ \emph {et~al.}(2010)\citenamefont
  {Ghigliotti}, \citenamefont {Biben},\ and\ \citenamefont
  {Misbah}}]{GhigliottiBibenMisbah2010_JFM}%
  \BibitemOpen
  \bibfield  {author} {\bibinfo {author} {\bibfnamefont {G.}~\bibnamefont
  {Ghigliotti}}, \bibinfo {author} {\bibfnamefont {T.}~\bibnamefont {Biben}}, \
  and\ \bibinfo {author} {\bibfnamefont {C.}~\bibnamefont {Misbah}},\
  }\bibfield  {title} {\enquote {\bibinfo {title} {Rheology of a dilute
  two-dimensional suspension of vesicles},}\ }\href@noop {} {\bibfield
  {journal} {\bibinfo  {journal} {Journal of Fluid Mechanics}\ }\textbf
  {\bibinfo {volume} {653}},\ \bibinfo {pages} {489--518} (\bibinfo {year}
  {2010})}\BibitemShut {NoStop}%
\bibitem [{\citenamefont {Deschamps}\ \emph {et~al.}(2009)\citenamefont
  {Deschamps}, \citenamefont {Kantsler}, \citenamefont {Serge},\ and\
  \citenamefont {Steinberg}}]{DeschampsKantsler2009_PNAS}%
  \BibitemOpen
  \bibfield  {author} {\bibinfo {author} {\bibfnamefont {J.}~\bibnamefont
  {Deschamps}}, \bibinfo {author} {\bibfnamefont {V.}~\bibnamefont {Kantsler}},
  \bibinfo {author} {\bibfnamefont {E.}~\bibnamefont {Serge}}, \ and\ \bibinfo
  {author} {\bibfnamefont {V.}~\bibnamefont {Steinberg}},\ }\bibfield  {title}
  {\enquote {\bibinfo {title} {Dynamics of a vesicle in general flow},}\
  }\href@noop {} {\bibfield  {journal} {\bibinfo  {journal} {Proc. Nat. Acad.
  Sci.}\ }\textbf {\bibinfo {volume} {106}},\ \bibinfo {pages} {11444--11447}
  (\bibinfo {year} {2009})}\BibitemShut {NoStop}%
\bibitem [{\citenamefont {Kantsler}\ \emph
  {et~al.}(2008{\natexlab{a}})\citenamefont {Kantsler}, \citenamefont {Segre},\
  and\ \citenamefont {Steinberg}}]{KantslerSegreSteinberg2008_EPL}%
  \BibitemOpen
  \bibfield  {author} {\bibinfo {author} {\bibfnamefont {V.}~\bibnamefont
  {Kantsler}}, \bibinfo {author} {\bibfnamefont {E.}~\bibnamefont {Segre}}, \
  and\ \bibinfo {author} {\bibfnamefont {V.}~\bibnamefont {Steinberg}},\
  }\bibfield  {title} {\enquote {\bibinfo {title} {Dynamics of interacting
  vesicles and rheology of vesicle suspension in shear flow},}\ }\href@noop {}
  {\bibfield  {journal} {\bibinfo  {journal} {Europhysics Letters}\ }\textbf
  {\bibinfo {volume} {82}},\ \bibinfo {pages} {58005} (\bibinfo {year}
  {2008}{\natexlab{a}})}\BibitemShut {NoStop}%
\bibitem [{\citenamefont {Zabusky}\ \emph {et~al.}(2011)\citenamefont
  {Zabusky}, \citenamefont {Segre}, \citenamefont {Deschamps}, \citenamefont
  {Kantsler},\ and\ \citenamefont {Steinberg}}]{ZabuskySegreDeschamps2011_PoF}%
  \BibitemOpen
  \bibfield  {author} {\bibinfo {author} {\bibfnamefont {N.}~\bibnamefont
  {Zabusky}}, \bibinfo {author} {\bibfnamefont {E.}~\bibnamefont {Segre}},
  \bibinfo {author} {\bibfnamefont {J.}~\bibnamefont {Deschamps}}, \bibinfo
  {author} {\bibfnamefont {V.}~\bibnamefont {Kantsler}}, \ and\ \bibinfo
  {author} {\bibfnamefont {V.}~\bibnamefont {Steinberg}},\ }\bibfield  {title}
  {\enquote {\bibinfo {title} {Dynamics of vesicles in shear and rotational
  flows: modal dynamics and phase diagram},}\ }\href@noop {} {\bibfield
  {journal} {\bibinfo  {journal} {Physics of Fluids}\ }\textbf {\bibinfo
  {volume} {23}},\ \bibinfo {pages} {041905} (\bibinfo {year}
  {2011})}\BibitemShut {NoStop}%
\bibitem [{\citenamefont {Kantsler}\ \emph
  {et~al.}(2008{\natexlab{b}})\citenamefont {Kantsler}, \citenamefont {Segre},\
  and\ \citenamefont {Steinberg}}]{KantslerSegreSteinberg2008_PRL}%
  \BibitemOpen
  \bibfield  {author} {\bibinfo {author} {\bibfnamefont {V.}~\bibnamefont
  {Kantsler}}, \bibinfo {author} {\bibfnamefont {E.}~\bibnamefont {Segre}}, \
  and\ \bibinfo {author} {\bibfnamefont {V.}~\bibnamefont {Steinberg}},\
  }\bibfield  {title} {\enquote {\bibinfo {title} {Critical dynamics of vesicle
  stretching transition in elongational flow},}\ }\href@noop {} {\bibfield
  {journal} {\bibinfo  {journal} {Physical Review Letters}\ }\textbf {\bibinfo
  {volume} {101}},\ \bibinfo {pages} {048101} (\bibinfo {year}
  {2008}{\natexlab{b}})}\BibitemShut {NoStop}%
\bibitem [{\citenamefont {Zhao}\ and\ \citenamefont
  {Shaqfeh}(2011)}]{ZhaoShaqfeh2011_JFM}%
  \BibitemOpen
  \bibfield  {author} {\bibinfo {author} {\bibfnamefont {H.}~\bibnamefont
  {Zhao}}\ and\ \bibinfo {author} {\bibfnamefont {E.~S.~G.}\ \bibnamefont
  {Shaqfeh}},\ }\bibfield  {title} {\enquote {\bibinfo {title} {The dynamics of
  a vesicle in simple shear flow},}\ }\href@noop {} {\bibfield  {journal}
  {\bibinfo  {journal} {Journal of Fluid Mechanics}\ }\textbf {\bibinfo
  {volume} {674}},\ \bibinfo {pages} {578--604} (\bibinfo {year}
  {2011})}\BibitemShut {NoStop}%
\bibitem [{\citenamefont {Spann}\ \emph {et~al.}(2014)\citenamefont {Spann},
  \citenamefont {Zhao},\ and\ \citenamefont
  {Shaqfeh}}]{SpannZhaoShaqfeh2014_PoF}%
  \BibitemOpen
  \bibfield  {author} {\bibinfo {author} {\bibfnamefont {A.~P.}\ \bibnamefont
  {Spann}}, \bibinfo {author} {\bibfnamefont {H.}~\bibnamefont {Zhao}}, \ and\
  \bibinfo {author} {\bibfnamefont {E.~S.~G.}\ \bibnamefont {Shaqfeh}},\
  }\bibfield  {title} {\enquote {\bibinfo {title} {Loop subdivision surface
  boundary integral method simulations of vesicles at low reduced volume ratio
  in shear and extensional flow},}\ }\href@noop {} {\bibfield  {journal}
  {\bibinfo  {journal} {Physics of Fluids}\ }\textbf {\bibinfo {volume} {16}},\
  \bibinfo {pages} {031902} (\bibinfo {year} {2014})}\BibitemShut {NoStop}%
\bibitem [{\citenamefont {Zhao}\ and\ \citenamefont
  {Shaqfeh}(2013)}]{ZhaoShaqfeh2013_JFM}%
  \BibitemOpen
  \bibfield  {author} {\bibinfo {author} {\bibfnamefont {H.}~\bibnamefont
  {Zhao}}\ and\ \bibinfo {author} {\bibfnamefont {E.~S.~G.}\ \bibnamefont
  {Shaqfeh}},\ }\bibfield  {title} {\enquote {\bibinfo {title} {The shape
  stability of a lipid vesicle in a uniaxial extensional flow},}\ }\href@noop
  {} {\bibfield  {journal} {\bibinfo  {journal} {Journal of Fluid Mechanics}\
  }\textbf {\bibinfo {volume} {719}},\ \bibinfo {pages} {345--361} (\bibinfo
  {year} {2013})}\BibitemShut {NoStop}%
\bibitem [{\citenamefont {Narsimhan}\ \emph {et~al.}(2014)\citenamefont
  {Narsimhan}, \citenamefont {Spann},\ and\ \citenamefont
  {Shaqfeh}}]{Narsimhan2014_JFM}%
  \BibitemOpen
  \bibfield  {author} {\bibinfo {author} {\bibfnamefont {V.}~\bibnamefont
  {Narsimhan}}, \bibinfo {author} {\bibfnamefont {A.~P.}\ \bibnamefont
  {Spann}}, \ and\ \bibinfo {author} {\bibfnamefont {E.~S.~G.}\ \bibnamefont
  {Shaqfeh}},\ }\bibfield  {title} {\enquote {\bibinfo {title} {The mechanism
  of shape instability for a vesicle in extensional flow},}\ }\href@noop {}
  {\bibfield  {journal} {\bibinfo  {journal} {Journal of Fluid Mechanics}\
  }\textbf {\bibinfo {volume} {750}},\ \bibinfo {pages} {144--190} (\bibinfo
  {year} {2014})}\BibitemShut {NoStop}%
\bibitem [{\citenamefont {Dahl}\ \emph {et~al.}(2016)\citenamefont {Dahl},
  \citenamefont {Narsimhan}, \citenamefont {Gouveia}, \citenamefont {Kumar},
  \citenamefont {Shaqfeh},\ and\ \citenamefont
  {Muller}}]{DahlNarsimhanGouveia2016_SoftMatt}%
  \BibitemOpen
  \bibfield  {author} {\bibinfo {author} {\bibfnamefont {J.~B.}\ \bibnamefont
  {Dahl}}, \bibinfo {author} {\bibfnamefont {V.}~\bibnamefont {Narsimhan}},
  \bibinfo {author} {\bibfnamefont {B.}~\bibnamefont {Gouveia}}, \bibinfo
  {author} {\bibfnamefont {S.}~\bibnamefont {Kumar}}, \bibinfo {author}
  {\bibfnamefont {E.~S.~G.}\ \bibnamefont {Shaqfeh}}, \ and\ \bibinfo {author}
  {\bibfnamefont {S.~J.}\ \bibnamefont {Muller}},\ }\bibfield  {title}
  {\enquote {\bibinfo {title} {Experimental observation of the asymmetric
  instability of intermediate-reduced-volume vesicles in extensional flow},}\
  }\href@noop {} {\bibfield  {journal} {\bibinfo  {journal} {Soft Matter}\
  }\textbf {\bibinfo {volume} {12}},\ \bibinfo {pages} {3787--3796} (\bibinfo
  {year} {2016})}\BibitemShut {NoStop}%
\bibitem [{\citenamefont {Ziherl}(2007)}]{Ziherl2007_PRL}%
  \BibitemOpen
  \bibfield  {author} {\bibinfo {author} {\bibfnamefont {P.}~\bibnamefont
  {Ziherl}},\ }\bibfield  {title} {\enquote {\bibinfo {title} {{Aggregates of
  Two-Dimensional Vesicles: Rouleaux, Sheets, and Convergent Extension}},}\
  }\href@noop {} {\bibfield  {journal} {\bibinfo  {journal} {Physical Review
  Letters}\ }\textbf {\bibinfo {volume} {99}},\ \bibinfo {pages} {128102}
  (\bibinfo {year} {2007})}\BibitemShut {NoStop}%
\bibitem [{\citenamefont {Ziherl}\ and\ \citenamefont
  {Svetina}(2007)}]{ZiherlSvetina2007_PNAS}%
  \BibitemOpen
  \bibfield  {author} {\bibinfo {author} {\bibfnamefont {P.}~\bibnamefont
  {Ziherl}}\ and\ \bibinfo {author} {\bibfnamefont {S.}~\bibnamefont
  {Svetina}},\ }\bibfield  {title} {\enquote {\bibinfo {title} {Flat and
  sigmoidally curved contact zones in vesicle-vesicle adhesion},}\ }\href@noop
  {} {\bibfield  {journal} {\bibinfo  {journal} {Proceedings of the National
  Academy of Sciences}\ }\textbf {\bibinfo {volume} {104}},\ \bibinfo {pages}
  {761--765} (\bibinfo {year} {2007})}\BibitemShut {NoStop}%
\bibitem [{\citenamefont {Svetina}\ and\ \citenamefont
  {Ziherl}(2008)}]{SvetinaZiherl2008_Bioelectrochemistry}%
  \BibitemOpen
  \bibfield  {author} {\bibinfo {author} {\bibfnamefont {S.}~\bibnamefont
  {Svetina}}\ and\ \bibinfo {author} {\bibfnamefont {P.}~\bibnamefont
  {Ziherl}},\ }\bibfield  {title} {\enquote {\bibinfo {title} {Morphology of
  small aggregates of red blood cells},}\ }\href@noop {} {\bibfield  {journal}
  {\bibinfo  {journal} {Bioelectrochemistry}\ }\textbf {\bibinfo {volume}
  {73}},\ \bibinfo {pages} {84--91} (\bibinfo {year} {2008})}\BibitemShut
  {NoStop}%
\bibitem [{\citenamefont {Gu}\ \emph {et~al.}(2016)\citenamefont {Gu},
  \citenamefont {Wang},\ and\ \citenamefont
  {Gunzburger}}]{GuWangGunzburger2016_MathBiol}%
  \BibitemOpen
  \bibfield  {author} {\bibinfo {author} {\bibfnamefont {R.}~\bibnamefont
  {Gu}}, \bibinfo {author} {\bibfnamefont {X.}~\bibnamefont {Wang}}, \ and\
  \bibinfo {author} {\bibfnamefont {M.}~\bibnamefont {Gunzburger}},\ }\bibfield
   {title} {\enquote {\bibinfo {title} {A two phase field model for tracking
  vesicle-vesicle adhesion},}\ }\href@noop {} {\bibfield  {journal} {\bibinfo
  {journal} {Math. Biol.}\ }\textbf {\bibinfo {volume} {73}},\ \bibinfo {pages}
  {1293--1319} (\bibinfo {year} {2016})}\BibitemShut {NoStop}%
\bibitem [{\citenamefont {Flormann}\ \emph {et~al.}(2017)\citenamefont
  {Flormann}, \citenamefont {Aouane}, \citenamefont {Kaestner}, \citenamefont
  {Ruloff}, \citenamefont {Misbah}, \citenamefont {Podgorski},\ and\
  \citenamefont {Wagner}}]{FlormannAouane2017_SciReports}%
  \BibitemOpen
  \bibfield  {author} {\bibinfo {author} {\bibfnamefont {D.}~\bibnamefont
  {Flormann}}, \bibinfo {author} {\bibfnamefont {O.}~\bibnamefont {Aouane}},
  \bibinfo {author} {\bibfnamefont {L.}~\bibnamefont {Kaestner}}, \bibinfo
  {author} {\bibfnamefont {C.}~\bibnamefont {Ruloff}}, \bibinfo {author}
  {\bibfnamefont {C.}~\bibnamefont {Misbah}}, \bibinfo {author} {\bibfnamefont
  {T.}~\bibnamefont {Podgorski}}, \ and\ \bibinfo {author} {\bibfnamefont
  {C.}~\bibnamefont {Wagner}},\ }\bibfield  {title} {\enquote {\bibinfo {title}
  {The buckling instability of aggregating red blood cells},}\ }\href@noop {}
  {\bibfield  {journal} {\bibinfo  {journal} {Scientific Reports}\ }\textbf
  {\bibinfo {volume} {7}},\ \bibinfo {pages} {7928} (\bibinfo {year}
  {2017})}\BibitemShut {NoStop}%
\bibitem [{\citenamefont {Hoore}\ \emph {et~al.}(2018)\citenamefont {Hoore},
  \citenamefont {Yaya}, \citenamefont {Podgorski}, \citenamefont {Wagner},
  \citenamefont {Gompper},\ and\ \citenamefont
  {Fedosov}}]{HooreYayaPodgorski2018_SoftMatt}%
  \BibitemOpen
  \bibfield  {author} {\bibinfo {author} {\bibfnamefont {M.}~\bibnamefont
  {Hoore}}, \bibinfo {author} {\bibfnamefont {F.}~\bibnamefont {Yaya}},
  \bibinfo {author} {\bibfnamefont {T.}~\bibnamefont {Podgorski}}, \bibinfo
  {author} {\bibfnamefont {C.}~\bibnamefont {Wagner}}, \bibinfo {author}
  {\bibfnamefont {G.}~\bibnamefont {Gompper}}, \ and\ \bibinfo {author}
  {\bibfnamefont {D.~A.}\ \bibnamefont {Fedosov}},\ }\bibfield  {title}
  {\enquote {\bibinfo {title} {Effect of spectrin network elasticity on the
  shapes of erythrocyte doublets},}\ }\href@noop {} {\bibfield  {journal}
  {\bibinfo  {journal} {Soft Matt.}\ }\textbf {\bibinfo {volume} {14}},\
  \bibinfo {pages} {6278--6289} (\bibinfo {year} {2018})}\BibitemShut {NoStop}%
\bibitem [{\citenamefont {Brust}\ \emph {et~al.}(2014)\citenamefont {Brust},
  \citenamefont {Aouane}, \citenamefont {Thi\'{e}baud}, \citenamefont
  {Flormann}, \citenamefont {Verdier}, \citenamefont {Kaestner}, \citenamefont
  {Laschke}, \citenamefont {Selmi}, \citenamefont {Benyoussef}, \citenamefont
  {Podgorski}, \citenamefont {Coupier}, \citenamefont {Misbah},\ and\
  \citenamefont
  {Wagner}}]{bru-aou-thi-flo-ver-Kae-las-sel-ben-pod-cou-mis-wag2014}%
  \BibitemOpen
  \bibfield  {author} {\bibinfo {author} {\bibfnamefont {M.}~\bibnamefont
  {Brust}}, \bibinfo {author} {\bibfnamefont {O.}~\bibnamefont {Aouane}},
  \bibinfo {author} {\bibfnamefont {M.}~\bibnamefont {Thi\'{e}baud}}, \bibinfo
  {author} {\bibfnamefont {D.}~\bibnamefont {Flormann}}, \bibinfo {author}
  {\bibfnamefont {C.}~\bibnamefont {Verdier}}, \bibinfo {author} {\bibfnamefont
  {L.}~\bibnamefont {Kaestner}}, \bibinfo {author} {\bibfnamefont {M.~W.}\
  \bibnamefont {Laschke}}, \bibinfo {author} {\bibfnamefont {H.}~\bibnamefont
  {Selmi}}, \bibinfo {author} {\bibfnamefont {A.}~\bibnamefont {Benyoussef}},
  \bibinfo {author} {\bibfnamefont {T.}~\bibnamefont {Podgorski}}, \bibinfo
  {author} {\bibfnamefont {G.}~\bibnamefont {Coupier}}, \bibinfo {author}
  {\bibfnamefont {C.}~\bibnamefont {Misbah}}, \ and\ \bibinfo {author}
  {\bibfnamefont {C.}~\bibnamefont {Wagner}},\ }\bibfield  {title} {\enquote
  {\bibinfo {title} {The plasma protein fibrinogen stabilizes clusters of red
  blood cells in microcapillary flows},}\ }\href@noop {} {\bibfield  {journal}
  {\bibinfo  {journal} {Scientific Reports}\ }\textbf {\bibinfo {volume} {4}},\
  \bibinfo {pages} {4348} (\bibinfo {year} {2014})}\BibitemShut {NoStop}%
\bibitem [{\citenamefont {Claver\'ia}\ \emph {et~al.}(2017)\citenamefont
  {Claver\'ia}, \citenamefont {Aouane}, \citenamefont {Thi\'ebaud},
  \citenamefont {Abkarian}, \citenamefont {Coupier}, \citenamefont {Misbah},
  \citenamefont {John},\ and\ \citenamefont
  {Wagner}}]{cla-aou-thi-abk-cou-mis-joh-wag2017}%
  \BibitemOpen
  \bibfield  {author} {\bibinfo {author} {\bibfnamefont {Viviana}\ \bibnamefont
  {Claver\'ia}}, \bibinfo {author} {\bibfnamefont {Othmane}\ \bibnamefont
  {Aouane}}, \bibinfo {author} {\bibfnamefont {Marine}\ \bibnamefont
  {Thi\'ebaud}}, \bibinfo {author} {\bibfnamefont {Manouk}\ \bibnamefont
  {Abkarian}}, \bibinfo {author} {\bibfnamefont {Gwennou}\ \bibnamefont
  {Coupier}}, \bibinfo {author} {\bibfnamefont {Chaouqi}\ \bibnamefont
  {Misbah}}, \bibinfo {author} {\bibfnamefont {Thomas}\ \bibnamefont {John}}, \
  and\ \bibinfo {author} {\bibfnamefont {Christian}\ \bibnamefont {Wagner}},\
  }\bibfield  {title} {\enquote {\bibinfo {title} {Clusters of red blood cells
  in microcapillary flow: hydrodynamic versus macromolecule induced
  interaction},}\ }\href@noop {} {\bibfield  {journal} {\bibinfo  {journal}
  {Soft Matter}\ }\textbf {\bibinfo {volume} {12}},\ \bibinfo {pages}
  {8235--8245} (\bibinfo {year} {2017})}\BibitemShut {NoStop}%
\bibitem [{\citenamefont {Chien}\ \emph {et~al.}(1967)\citenamefont {Chien},
  \citenamefont {Usami}, \citenamefont {Dellenback}, \citenamefont {Gregersen},
  \citenamefont {Nanninga},\ and\ \citenamefont
  {Guest}}]{chi-usa-del-gre-nan-gue1967}%
  \BibitemOpen
  \bibfield  {author} {\bibinfo {author} {\bibfnamefont {Shu}\ \bibnamefont
  {Chien}}, \bibinfo {author} {\bibfnamefont {Shunichi}\ \bibnamefont {Usami}},
  \bibinfo {author} {\bibfnamefont {Robert~J.}\ \bibnamefont {Dellenback}},
  \bibinfo {author} {\bibfnamefont {Magnus~I.}\ \bibnamefont {Gregersen}},
  \bibinfo {author} {\bibfnamefont {Luddo~B.}\ \bibnamefont {Nanninga}}, \ and\
  \bibinfo {author} {\bibfnamefont {M.~Mason}\ \bibnamefont {Guest}},\
  }\bibfield  {title} {\enquote {\bibinfo {title} {{Blood Viscosity: Influence
  of Erythrocyte Aggregation}},}\ }\href@noop {} {\bibfield  {journal}
  {\bibinfo  {journal} {Science}\ }\textbf {\bibinfo {volume} {157}},\ \bibinfo
  {pages} {829--831} (\bibinfo {year} {1967})}\BibitemShut {NoStop}%
\bibitem [{\citenamefont {Rahimian}\ \emph {et~al.}(2010)\citenamefont
  {Rahimian}, \citenamefont {Veerapaneni},\ and\ \citenamefont
  {Biros}}]{RahimianVeerapaneniBiros2010_JCP}%
  \BibitemOpen
  \bibfield  {author} {\bibinfo {author} {\bibfnamefont {A.}~\bibnamefont
  {Rahimian}}, \bibinfo {author} {\bibfnamefont {S.~K.}\ \bibnamefont
  {Veerapaneni}}, \ and\ \bibinfo {author} {\bibfnamefont {G.}~\bibnamefont
  {Biros}},\ }\bibfield  {title} {\enquote {\bibinfo {title} {Dynamic
  simulation of locally inextensible vesicles suspended in an arbitrary
  two-dimensional domain, a boundary integral method},}\ }\href@noop {}
  {\bibfield  {journal} {\bibinfo  {journal} {Journal of Computational
  Physics}\ }\textbf {\bibinfo {volume} {229}},\ \bibinfo {pages} {6466--6484}
  (\bibinfo {year} {2010})}\BibitemShut {NoStop}%
\bibitem [{\citenamefont {Neu}\ and\ \citenamefont
  {Meiselman}(2002)}]{NeuMeiselman2002_BJ}%
  \BibitemOpen
  \bibfield  {author} {\bibinfo {author} {\bibfnamefont {B.}~\bibnamefont
  {Neu}}\ and\ \bibinfo {author} {\bibfnamefont {H.~J.}\ \bibnamefont
  {Meiselman}},\ }\bibfield  {title} {\enquote {\bibinfo {title}
  {{Depletion-Mediated Red Blood Cell Aggregation in Polymer Solutions}},}\
  }\href@noop {} {\bibfield  {journal} {\bibinfo  {journal} {Biophysical
  Journal}\ }\textbf {\bibinfo {volume} {83}},\ \bibinfo {pages} {2482--2490}
  (\bibinfo {year} {2002})}\BibitemShut {NoStop}%
\bibitem [{\citenamefont {Evans}\ and\ \citenamefont
  {Metcalfe}(1984)}]{EvansMetcalfe1984_BJ}%
  \BibitemOpen
  \bibfield  {author} {\bibinfo {author} {\bibfnamefont {E.}~\bibnamefont
  {Evans}}\ and\ \bibinfo {author} {\bibfnamefont {M.}~\bibnamefont
  {Metcalfe}},\ }\bibfield  {title} {\enquote {\bibinfo {title} {{Free energy
  potential for aggregation of giant, neutral lipid bilayer vesicles by Van der
  Waals attraction}},}\ }\href@noop {} {\bibfield  {journal} {\bibinfo
  {journal} {Biophysical Journal}\ }\textbf {\bibinfo {volume} {46}},\ \bibinfo
  {pages} {423--426} (\bibinfo {year} {1984})}\BibitemShut {NoStop}%
\bibitem [{\citenamefont {Evans}(1988)}]{Book_PhysicalBasisCellAdhesion}%
  \BibitemOpen
  \bibfield  {author} {\bibinfo {author} {\bibfnamefont {E.}~\bibnamefont
  {Evans}},\ }\href@noop {} {\emph {\bibinfo {title} {Physical basis of
  cell-cell adhesion}}},\ edited by\ \bibinfo {editor} {\bibfnamefont
  {P.}~\bibnamefont {Bongrand}}\ (\bibinfo  {publisher} {CRC Press},\ \bibinfo
  {year} {1988})\BibitemShut {NoStop}%
\bibitem [{\citenamefont
  {Israelachvili}(1991)}]{Book_IntermolecularSurfaceForces}%
  \BibitemOpen
  \bibfield  {author} {\bibinfo {author} {\bibfnamefont {J.}~\bibnamefont
  {Israelachvili}},\ }\href@noop {} {\emph {\bibinfo {title} {Intermolecular
  and surface forces}}}\ (\bibinfo  {publisher} {Academic Press Inc., San
  Diego},\ \bibinfo {year} {1991})\BibitemShut {NoStop}%
\bibitem [{\citenamefont {Perutkova}\ \emph {et~al.}(2013)\citenamefont
  {Perutkova}, \citenamefont {Frank-Bertoncelj}, \citenamefont {Rozman},
  \citenamefont {Kralj-Iglic},\ and\ \citenamefont
  {Iglic}}]{PerutkovaFrank-Bertoncelij2013_CSB}%
  \BibitemOpen
  \bibfield  {author} {\bibinfo {author} {\bibfnamefont {S.}~\bibnamefont
  {Perutkova}}, \bibinfo {author} {\bibfnamefont {M.}~\bibnamefont
  {Frank-Bertoncelj}}, \bibinfo {author} {\bibfnamefont {B.}~\bibnamefont
  {Rozman}}, \bibinfo {author} {\bibfnamefont {V.}~\bibnamefont {Kralj-Iglic}},
  \ and\ \bibinfo {author} {\bibfnamefont {A.}~\bibnamefont {Iglic}},\
  }\bibfield  {title} {\enquote {\bibinfo {title} {Influence of ionic strength
  and beta2-glycoprotein i concentration on agglutination of like-charged
  phospholipid membranes},}\ }\href@noop {} {\bibfield  {journal} {\bibinfo
  {journal} {Colloids and Surfaces B: Biointerfaces}\ }\textbf {\bibinfo
  {volume} {111}},\ \bibinfo {pages} {699--706} (\bibinfo {year}
  {2013})}\BibitemShut {NoStop}%
\bibitem [{\citenamefont {Seifert}\ and\ \citenamefont
  {Lipowsky}(1990)}]{Seifert1990_PRA}%
  \BibitemOpen
  \bibfield  {author} {\bibinfo {author} {\bibfnamefont {U.}~\bibnamefont
  {Seifert}}\ and\ \bibinfo {author} {\bibfnamefont {R.}~\bibnamefont
  {Lipowsky}},\ }\bibfield  {title} {\enquote {\bibinfo {title} {Adhesion of
  vesicles},}\ }\href@noop {} {\bibfield  {journal} {\bibinfo  {journal}
  {Physical Review A}\ }\textbf {\bibinfo {volume} {42}},\ \bibinfo {pages}
  {4768} (\bibinfo {year} {1990})}\BibitemShut {NoStop}%
\bibitem [{\citenamefont {Bernard}\ \emph {et~al.}(2000)\citenamefont
  {Bernard}, \citenamefont {Guedeau-Boudeville}, \citenamefont {Jullien},\ and\
  \citenamefont {di~Meglio}}]{BernardGuedeau-Boudeville2000_Langmuir}%
  \BibitemOpen
  \bibfield  {author} {\bibinfo {author} {\bibfnamefont {A.-L.}\ \bibnamefont
  {Bernard}}, \bibinfo {author} {\bibfnamefont {M.-A.}\ \bibnamefont
  {Guedeau-Boudeville}}, \bibinfo {author} {\bibfnamefont {L.}~\bibnamefont
  {Jullien}}, \ and\ \bibinfo {author} {\bibfnamefont {J.-M.}\ \bibnamefont
  {di~Meglio}},\ }\bibfield  {title} {\enquote {\bibinfo {title} {Strong
  adhesion of giant vesicles on surface and permeability},}\ }\href@noop {}
  {\bibfield  {journal} {\bibinfo  {journal} {Langmuir}\ }\textbf {\bibinfo
  {volume} {16}},\ \bibinfo {pages} {6809--6820} (\bibinfo {year}
  {2000})}\BibitemShut {NoStop}%
\bibitem [{\citenamefont {Shi}\ \emph {et~al.}(2006)\citenamefont {Shi},
  \citenamefont {Feng},\ and\ \citenamefont
  {Gao}}]{ShiFengGao2006_ActaMechSin}%
  \BibitemOpen
  \bibfield  {author} {\bibinfo {author} {\bibfnamefont {W.}~\bibnamefont
  {Shi}}, \bibinfo {author} {\bibfnamefont {X.~Q.}\ \bibnamefont {Feng}}, \
  and\ \bibinfo {author} {\bibfnamefont {H.}~\bibnamefont {Gao}},\ }\bibfield
  {title} {\enquote {\bibinfo {title} {Two-dimensional model of vesicle
  adhesion on curved substrates},}\ }\href@noop {} {\bibfield  {journal}
  {\bibinfo  {journal} {Acta Mechanica Sinica}\ }\textbf {\bibinfo {volume}
  {22}},\ \bibinfo {pages} {529--535} (\bibinfo {year} {2006})}\BibitemShut
  {NoStop}%
\bibitem [{\citenamefont {Lin}\ and\ \citenamefont
  {Freund}(2007)}]{LinFreund2007_IntJSolidsStructures}%
  \BibitemOpen
  \bibfield  {author} {\bibinfo {author} {\bibfnamefont {Y.}~\bibnamefont
  {Lin}}\ and\ \bibinfo {author} {\bibfnamefont {L.~B.}\ \bibnamefont
  {Freund}},\ }\bibfield  {title} {\enquote {\bibinfo {title} {Forced
  detachment of a vesicle in adhesive contact with a substrate},}\ }\href@noop
  {} {\bibfield  {journal} {\bibinfo  {journal} {International Journal of
  Solids and Structures}\ }\textbf {\bibinfo {volume} {44}},\ \bibinfo {pages}
  {1927--1938} (\bibinfo {year} {2007})}\BibitemShut {NoStop}%
\bibitem [{\citenamefont {Gruhn}\ \emph {et~al.}(2007)\citenamefont {Gruhn},
  \citenamefont {Franke}, \citenamefont {Dimova},\ and\ \citenamefont
  {Lipowsky}}]{GruhnFrankeDimova2007_Langmuir}%
  \BibitemOpen
  \bibfield  {author} {\bibinfo {author} {\bibfnamefont {T.}~\bibnamefont
  {Gruhn}}, \bibinfo {author} {\bibfnamefont {T.}~\bibnamefont {Franke}},
  \bibinfo {author} {\bibfnamefont {R.}~\bibnamefont {Dimova}}, \ and\ \bibinfo
  {author} {\bibfnamefont {R.}~\bibnamefont {Lipowsky}},\ }\bibfield  {title}
  {\enquote {\bibinfo {title} {Novel method for measuring the adhesion energy
  of vesicles},}\ }\href@noop {} {\bibfield  {journal} {\bibinfo  {journal}
  {Langmuir}\ }\textbf {\bibinfo {volume} {23}},\ \bibinfo {pages} {5423--5429}
  (\bibinfo {year} {2007})}\BibitemShut {NoStop}%
\bibitem [{\citenamefont {Das}\ and\ \citenamefont
  {Du}(2008)}]{das2008adhesion}%
  \BibitemOpen
  \bibfield  {author} {\bibinfo {author} {\bibfnamefont {S.}~\bibnamefont
  {Das}}\ and\ \bibinfo {author} {\bibfnamefont {Q.}~\bibnamefont {Du}},\
  }\bibfield  {title} {\enquote {\bibinfo {title} {Adhesion of vesicles to
  curved substrates},}\ }\href@noop {} {\bibfield  {journal} {\bibinfo
  {journal} {Physical Review E}\ }\textbf {\bibinfo {volume} {77}},\ \bibinfo
  {pages} {011907} (\bibinfo {year} {2008})}\BibitemShut {NoStop}%
\bibitem [{\citenamefont {Keh}\ \emph {et~al.}(2014)\citenamefont {Keh},
  \citenamefont {Walter},\ and\ \citenamefont {Leal}}]{KehWalterLeal2014_PoF}%
  \BibitemOpen
  \bibfield  {author} {\bibinfo {author} {\bibfnamefont {M.~P.}\ \bibnamefont
  {Keh}}, \bibinfo {author} {\bibfnamefont {J.}~\bibnamefont {Walter}}, \ and\
  \bibinfo {author} {\bibfnamefont {L.~G.}\ \bibnamefont {Leal}},\ }\bibfield
  {title} {\enquote {\bibinfo {title} {Hydrodynamic interaction between a
  capsule and a solid boundary in unbounded stokes flow},}\ }\href@noop {}
  {\bibfield  {journal} {\bibinfo  {journal} {Phys. Fluids}\ ,\ \bibinfo
  {pages} {111903}} (\bibinfo {year} {2014})}\BibitemShut {NoStop}%
\bibitem [{\citenamefont {Zhang}\ \emph {et~al.}(2009)\citenamefont {Zhang},
  \citenamefont {Das},\ and\ \citenamefont {Du}}]{zhang2009phase}%
  \BibitemOpen
  \bibfield  {author} {\bibinfo {author} {\bibfnamefont {J.}~\bibnamefont
  {Zhang}}, \bibinfo {author} {\bibfnamefont {S.}~\bibnamefont {Das}}, \ and\
  \bibinfo {author} {\bibfnamefont {Q.}~\bibnamefont {Du}},\ }\bibfield
  {title} {\enquote {\bibinfo {title} {A phase field model for
  vesicle-substrate adhesion},}\ }\href@noop {} {\bibfield  {journal} {\bibinfo
   {journal} {Journal of Computational Physics}\ }\textbf {\bibinfo {volume}
  {228}},\ \bibinfo {pages} {7837--7849} (\bibinfo {year} {2009})}\BibitemShut
  {NoStop}%
\bibitem [{\citenamefont {Agudo-Canalejo}\ and\ \citenamefont
  {Lipowsky}(2015{\natexlab{a}})}]{Agudo-Canalejo2015_ACSNano}%
  \BibitemOpen
  \bibfield  {author} {\bibinfo {author} {\bibfnamefont {J.}~\bibnamefont
  {Agudo-Canalejo}}\ and\ \bibinfo {author} {\bibfnamefont {R.}~\bibnamefont
  {Lipowsky}},\ }\bibfield  {title} {\enquote {\bibinfo {title} {{Critical
  particle sizes for the Engulfment of Nanoparticles by Membranes and Vesicles
  with Bilayer Asymmetry}},}\ }\href@noop {} {\bibfield  {journal} {\bibinfo
  {journal} {ACS Nano Lett.}\ }\textbf {\bibinfo {volume} {9}},\ \bibinfo
  {pages} {3704--3720} (\bibinfo {year} {2015}{\natexlab{a}})}\BibitemShut
  {NoStop}%
\bibitem [{\citenamefont {Agudo-Canalejo}\ and\ \citenamefont
  {Lipowsky}(2015{\natexlab{b}})}]{Agudo-CanalejoLipowsky2015_NanoLett}%
  \BibitemOpen
  \bibfield  {author} {\bibinfo {author} {\bibfnamefont {J.}~\bibnamefont
  {Agudo-Canalejo}}\ and\ \bibinfo {author} {\bibfnamefont {R.}~\bibnamefont
  {Lipowsky}},\ }\bibfield  {title} {\enquote {\bibinfo {title} {{Adhesive
  Nanoparticles as Local Probes of Membrane Curvature}},}\ }\href@noop {}
  {\bibfield  {journal} {\bibinfo  {journal} {Nano Lett.}\ }\textbf {\bibinfo
  {volume} {15}},\ \bibinfo {pages} {7168--7173} (\bibinfo {year}
  {2015}{\natexlab{b}})}\BibitemShut {NoStop}%
\bibitem [{\citenamefont {Steinkuhler}\ \emph {et~al.}(2016)\citenamefont
  {Steinkuhler}, \citenamefont {Agudo-Canalejo}, \citenamefont {Lipowsky},\
  and\ \citenamefont {Dimova}}]{SteinkuhlerAgudo-Canalejo2016_BJ}%
  \BibitemOpen
  \bibfield  {author} {\bibinfo {author} {\bibfnamefont {J.}~\bibnamefont
  {Steinkuhler}}, \bibinfo {author} {\bibfnamefont {J.}~\bibnamefont
  {Agudo-Canalejo}}, \bibinfo {author} {\bibfnamefont {R.}~\bibnamefont
  {Lipowsky}}, \ and\ \bibinfo {author} {\bibfnamefont {R.}~\bibnamefont
  {Dimova}},\ }\bibfield  {title} {\enquote {\bibinfo {title} {Modulating
  vesicle adhesion by electric fields},}\ }\href@noop {} {\bibfield  {journal}
  {\bibinfo  {journal} {Biophysical Journal}\ }\textbf {\bibinfo {volume}
  {111}},\ \bibinfo {pages} {1454--1464} (\bibinfo {year} {2016})}\BibitemShut
  {NoStop}%
\bibitem [{\citenamefont {Keh}\ and\ \citenamefont
  {Leal}(2016)}]{KehLeal2016_PRF}%
  \BibitemOpen
  \bibfield  {author} {\bibinfo {author} {\bibfnamefont {M.~P.}\ \bibnamefont
  {Keh}}\ and\ \bibinfo {author} {\bibfnamefont {L.~G.}\ \bibnamefont {Leal}},\
  }\bibfield  {title} {\enquote {\bibinfo {title} {Adhesion and detachment of a
  capsule in axisymmetric flow},}\ }\href@noop {} {\bibfield  {journal}
  {\bibinfo  {journal} {Phys. Rev. Fluids}\ ,\ \bibinfo {pages} {013201}}
  (\bibinfo {year} {2016})}\BibitemShut {NoStop}%
\bibitem [{\citenamefont {Agudo-Canalejo}\ and\ \citenamefont
  {Lipowsky}(2017)}]{Agudo-CanalejoLipowsky2017_SoftMatt}%
  \BibitemOpen
  \bibfield  {author} {\bibinfo {author} {\bibfnamefont {J.}~\bibnamefont
  {Agudo-Canalejo}}\ and\ \bibinfo {author} {\bibfnamefont {R.}~\bibnamefont
  {Lipowsky}},\ }\bibfield  {title} {\enquote {\bibinfo {title} {{Uniform and
  Janus-like nanoparticles in contact with vesicles: energy landscapes and
  curvature-induced forces}},}\ }\href@noop {} {\bibfield  {journal} {\bibinfo
  {journal} {Soft Matt.}\ }\textbf {\bibinfo {volume} {13}},\ \bibinfo {pages}
  {2155} (\bibinfo {year} {2017})}\BibitemShut {NoStop}%
\bibitem [{\citenamefont {Cantat}\ and\ \citenamefont
  {Misbah}(1999)}]{cantat1999lift}%
  \BibitemOpen
  \bibfield  {author} {\bibinfo {author} {\bibfnamefont {I.}~\bibnamefont
  {Cantat}}\ and\ \bibinfo {author} {\bibfnamefont {C.}~\bibnamefont
  {Misbah}},\ }\bibfield  {title} {\enquote {\bibinfo {title} {Lift force and
  dynamical unbinding of adhering vesicles under shear flow},}\ }\href@noop {}
  {\bibfield  {journal} {\bibinfo  {journal} {Physical Review Letters}\
  }\textbf {\bibinfo {volume} {83}},\ \bibinfo {pages} {880} (\bibinfo {year}
  {1999})}\BibitemShut {NoStop}%
\bibitem [{\citenamefont {Sukumaran}\ and\ \citenamefont
  {Seifert}(2001)}]{suk-sei2001}%
  \BibitemOpen
  \bibfield  {author} {\bibinfo {author} {\bibfnamefont {S.}~\bibnamefont
  {Sukumaran}}\ and\ \bibinfo {author} {\bibfnamefont {U.}~\bibnamefont
  {Seifert}},\ }\bibfield  {title} {\enquote {\bibinfo {title} {{Influence of
  shear flow on vesicles near a wall: A numerical study}},}\ }\href@noop {}
  {\bibfield  {journal} {\bibinfo  {journal} {Physical Review E}\ }\textbf
  {\bibinfo {volume} {64}} (\bibinfo {year} {2001})}\BibitemShut {NoStop}%
\bibitem [{\citenamefont {Blount}\ \emph {et~al.}(2013)\citenamefont {Blount},
  \citenamefont {Miksis},\ and\ \citenamefont
  {Davis}}]{BlountMiksisDavis2013_PRSa}%
  \BibitemOpen
  \bibfield  {author} {\bibinfo {author} {\bibfnamefont {M.~J.}\ \bibnamefont
  {Blount}}, \bibinfo {author} {\bibfnamefont {M.~J.}\ \bibnamefont {Miksis}},
  \ and\ \bibinfo {author} {\bibfnamefont {S.~H.}\ \bibnamefont {Davis}},\
  }\bibfield  {title} {\enquote {\bibinfo {title} {The equilibria of vesicles
  adhered to substrates by short-ranged potentials},}\ }\href@noop {}
  {\bibfield  {journal} {\bibinfo  {journal} {Proceedings of the Royal Society
  A}\ }\textbf {\bibinfo {volume} {469}},\ \bibinfo {pages} {20120729}
  (\bibinfo {year} {2013})}\BibitemShut {NoStop}%
\bibitem [{\citenamefont {Ramachandran}\ \emph {et~al.}(2010)\citenamefont
  {Ramachandran}, \citenamefont {Anderson}, \citenamefont {Leal},\ and\
  \citenamefont
  {Israelachvili}}]{RamachandranAndersonLealIsraelachvili2010_Langmuir}%
  \BibitemOpen
  \bibfield  {author} {\bibinfo {author} {\bibfnamefont {A.}~\bibnamefont
  {Ramachandran}}, \bibinfo {author} {\bibfnamefont {T.~H.}\ \bibnamefont
  {Anderson}}, \bibinfo {author} {\bibfnamefont {L.~G.}\ \bibnamefont {Leal}},
  \ and\ \bibinfo {author} {\bibfnamefont {J.~N.}\ \bibnamefont
  {Israelachvili}},\ }\bibfield  {title} {\enquote {\bibinfo {title} {{Adhesive
  Interactions between Vesicles in the Strong Adhesion Limit}},}\ }\href@noop
  {} {\bibfield  {journal} {\bibinfo  {journal} {Langmuir}\ }\textbf {\bibinfo
  {volume} {27}},\ \bibinfo {pages} {59--73} (\bibinfo {year}
  {2010})}\BibitemShut {NoStop}%
\bibitem [{\citenamefont {Mares}\ \emph {et~al.}(2012)\citenamefont {Mares},
  \citenamefont {Daniel}, \citenamefont {Iglic}, \citenamefont {Kralj-Iglic},\
  and\ \citenamefont {Fosnaric}}]{MaresDanielIglic2012_SciWorldJ}%
  \BibitemOpen
  \bibfield  {author} {\bibinfo {author} {\bibfnamefont {T.}~\bibnamefont
  {Mares}}, \bibinfo {author} {\bibfnamefont {M.}~\bibnamefont {Daniel}},
  \bibinfo {author} {\bibfnamefont {A.}~\bibnamefont {Iglic}}, \bibinfo
  {author} {\bibfnamefont {V.}~\bibnamefont {Kralj-Iglic}}, \ and\ \bibinfo
  {author} {\bibfnamefont {M.}~\bibnamefont {Fosnaric}},\ }\bibfield  {title}
  {\enquote {\bibinfo {title} {Determination of the strength of adhesion
  between lipid vesicles},}\ }\href@noop {} {\bibfield  {journal} {\bibinfo
  {journal} {Scientific World Journal}\ }\textbf {\bibinfo {volume} {2012}},\
  \bibinfo {pages} {146804} (\bibinfo {year} {2012})}\BibitemShut {NoStop}%
\bibitem [{\citenamefont {Frostad}\ \emph {et~al.}(2014)\citenamefont
  {Frostad}, \citenamefont {Seth}, \citenamefont {Bernasek},\ and\
  \citenamefont {Leal}}]{FrostadSethBernasekLeal2014_SoftMatt}%
  \BibitemOpen
  \bibfield  {author} {\bibinfo {author} {\bibfnamefont {J.~M.}\ \bibnamefont
  {Frostad}}, \bibinfo {author} {\bibfnamefont {M.}~\bibnamefont {Seth}},
  \bibinfo {author} {\bibfnamefont {S.~M.}\ \bibnamefont {Bernasek}}, \ and\
  \bibinfo {author} {\bibfnamefont {L.~G.}\ \bibnamefont {Leal}},\ }\bibfield
  {title} {\enquote {\bibinfo {title} {Direct measurement of interaction forces
  between charged multilamellar vesicles},}\ }\href@noop {} {\bibfield
  {journal} {\bibinfo  {journal} {Soft Matt.}\ }\textbf {\bibinfo {volume}
  {10}},\ \bibinfo {pages} {7769} (\bibinfo {year} {2014})}\BibitemShut
  {NoStop}%
\bibitem [{\citenamefont {Agrawal}(2011)}]{Agrawal2011_MathMechSolids}%
  \BibitemOpen
  \bibfield  {author} {\bibinfo {author} {\bibfnamefont {A.}~\bibnamefont
  {Agrawal}},\ }\bibfield  {title} {\enquote {\bibinfo {title} {Mechanics of
  membrane-membrane adhesion},}\ }\href@noop {} {\bibfield  {journal} {\bibinfo
   {journal} {Math. Mech. Solids}\ }\textbf {\bibinfo {volume} {16}},\ \bibinfo
  {pages} {872} (\bibinfo {year} {2011})}\BibitemShut {NoStop}%
\bibitem [{\citenamefont {Gires}\ \emph {et~al.}(2012)\citenamefont {Gires},
  \citenamefont {Danker},\ and\ \citenamefont
  {Misbah}}]{GiresDankerMisbah2012_PRE}%
  \BibitemOpen
  \bibfield  {author} {\bibinfo {author} {\bibfnamefont {P.~Y.}\ \bibnamefont
  {Gires}}, \bibinfo {author} {\bibfnamefont {G.}~\bibnamefont {Danker}}, \
  and\ \bibinfo {author} {\bibfnamefont {C.}~\bibnamefont {Misbah}},\
  }\bibfield  {title} {\enquote {\bibinfo {title} {Hydrodynamic interaction
  between two vesicles in a linear shear flow},}\ }\href@noop {} {\bibfield
  {journal} {\bibinfo  {journal} {Physical Review E}\ }\textbf {\bibinfo
  {volume} {86}},\ \bibinfo {pages} {011408} (\bibinfo {year}
  {2012})}\BibitemShut {NoStop}%
\bibitem [{\citenamefont {Gires}\ \emph {et~al.}(2014)\citenamefont {Gires},
  \citenamefont {Srivastav}, \citenamefont {Misbah}, \citenamefont
  {Podgorski},\ and\ \citenamefont {Coupier}}]{gir-sri-mis-pod-cou2014}%
  \BibitemOpen
  \bibfield  {author} {\bibinfo {author} {\bibfnamefont {Pierre-Yves}\
  \bibnamefont {Gires}}, \bibinfo {author} {\bibfnamefont {Aparna}\
  \bibnamefont {Srivastav}}, \bibinfo {author} {\bibfnamefont {Chaouqi}\
  \bibnamefont {Misbah}}, \bibinfo {author} {\bibfnamefont {Thomas}\
  \bibnamefont {Podgorski}}, \ and\ \bibinfo {author} {\bibfnamefont {Gwennou}\
  \bibnamefont {Coupier}},\ }\bibfield  {title} {\enquote {\bibinfo {title}
  {Pairwise hydrodynamic interactions and diffusion in a vesicle suspension},}\
  }\href@noop {} {\bibfield  {journal} {\bibinfo  {journal} {Physics of
  Fluids}\ }\textbf {\bibinfo {volume} {26}} (\bibinfo {year}
  {2014})}\BibitemShut {NoStop}%
\bibitem [{\citenamefont {Quaife}\ and\ \citenamefont
  {Biros}(2014)}]{qua-bir2014}%
  \BibitemOpen
  \bibfield  {author} {\bibinfo {author} {\bibfnamefont {B.}~\bibnamefont
  {Quaife}}\ and\ \bibinfo {author} {\bibfnamefont {G.}~\bibnamefont {Biros}},\
  }\bibfield  {title} {\enquote {\bibinfo {title} {High-volume fraction
  simulations of two-dimensional vesicle suspensions},}\ }\href@noop {}
  {\bibfield  {journal} {\bibinfo  {journal} {Journal of Computational
  Physics}\ }\textbf {\bibinfo {volume} {274}},\ \bibinfo {pages} {245--267}
  (\bibinfo {year} {2014})}\BibitemShut {NoStop}%
\bibitem [{\citenamefont {Quaife}\ and\ \citenamefont
  {Biros}(2016)}]{quaife2016adaptive}%
  \BibitemOpen
  \bibfield  {author} {\bibinfo {author} {\bibfnamefont {B.}~\bibnamefont
  {Quaife}}\ and\ \bibinfo {author} {\bibfnamefont {G.}~\bibnamefont {Biros}},\
  }\bibfield  {title} {\enquote {\bibinfo {title} {Adaptive time stepping for
  vesicle suspensions},}\ }\href@noop {} {\bibfield  {journal} {\bibinfo
  {journal} {Journal of Computational Physics}\ }\textbf {\bibinfo {volume}
  {306}},\ \bibinfo {pages} {478--499} (\bibinfo {year} {2016})}\BibitemShut
  {NoStop}%
\bibitem [{\citenamefont {Young}\ and\ \citenamefont
  {Stone}(2017)}]{YoungStone2017_PRF}%
  \BibitemOpen
  \bibfield  {author} {\bibinfo {author} {\bibfnamefont {Y.-N.}\ \bibnamefont
  {Young}}\ and\ \bibinfo {author} {\bibfnamefont {H.~A.}\ \bibnamefont
  {Stone}},\ }\bibfield  {title} {\enquote {\bibinfo {title} {Long-wave
  dynamics of an elastic sheet lubricated by a thin liquid film on a wetting
  substrate},}\ }\href@noop {} {\bibfield  {journal} {\bibinfo  {journal}
  {Physical Review Fluids}\ }\textbf {\bibinfo {volume} {2}},\ \bibinfo {pages}
  {064001} (\bibinfo {year} {2017})}\BibitemShut {NoStop}%
\bibitem [{\citenamefont {Ramachandran}\ and\ \citenamefont
  {Leal}(2010)}]{RamachandranLeal2010_PoF}%
  \BibitemOpen
  \bibfield  {author} {\bibinfo {author} {\bibfnamefont {A.}~\bibnamefont
  {Ramachandran}}\ and\ \bibinfo {author} {\bibfnamefont {G.}~\bibnamefont
  {Leal}},\ }\bibfield  {title} {\enquote {\bibinfo {title} {A scaling theory
  for the hydrodynamic interaction between a pair of vesicles or capsules},}\
  }\href@noop {} {\bibfield  {journal} {\bibinfo  {journal} {Physics of
  Fluids}\ }\textbf {\bibinfo {volume} {22}},\ \bibinfo {pages} {091702}
  (\bibinfo {year} {2010})}\BibitemShut {NoStop}%
\bibitem [{\citenamefont {Kantsler}\ \emph {et~al.}(2007)\citenamefont
  {Kantsler}, \citenamefont {Segre},\ and\ \citenamefont
  {Steinberg}}]{KantslerSegreSteinberg2007_PRL}%
  \BibitemOpen
  \bibfield  {author} {\bibinfo {author} {\bibfnamefont {V.}~\bibnamefont
  {Kantsler}}, \bibinfo {author} {\bibfnamefont {E.}~\bibnamefont {Segre}}, \
  and\ \bibinfo {author} {\bibfnamefont {V.}~\bibnamefont {Steinberg}},\
  }\bibfield  {title} {\enquote {\bibinfo {title} {Vesicle dynamics in
  time-dependent elongation flow: Wrinkling instability},}\ }\href@noop {}
  {\bibfield  {journal} {\bibinfo  {journal} {Physical Review Letters}\
  }\textbf {\bibinfo {volume} {99}},\ \bibinfo {pages} {178102} (\bibinfo
  {year} {2007})}\BibitemShut {NoStop}%
\bibitem [{\citenamefont {Jansen}\ \emph {et~al.}(1997)\citenamefont {Jansen},
  \citenamefont {Boon},\ and\ \citenamefont
  {Agterof}}]{JanssenBoonAgterof1997_AIChE}%
  \BibitemOpen
  \bibfield  {author} {\bibinfo {author} {\bibfnamefont {J.~J.~M.}\
  \bibnamefont {Jansen}}, \bibinfo {author} {\bibfnamefont {A.}~\bibnamefont
  {Boon}}, \ and\ \bibinfo {author} {\bibfnamefont {W.~G.~M.}\ \bibnamefont
  {Agterof}},\ }\bibfield  {title} {\enquote {\bibinfo {title} {{Influence of
  Dynamic Interfacial Properties on Droplet Breakup in Plane Hyperbolic
  Flow}},}\ }\href@noop {} {\bibfield  {journal} {\bibinfo  {journal} {AIChE
  J.}\ }\textbf {\bibinfo {volume} {43}},\ \bibinfo {pages} {1436--1447}
  (\bibinfo {year} {1997})}\BibitemShut {NoStop}%
\bibitem [{\citenamefont {Hu}\ \emph {et~al.}(2000)\citenamefont {Hu},
  \citenamefont {Pine},\ and\ \citenamefont {Leal}}]{HuPineLeal2000_PoF}%
  \BibitemOpen
  \bibfield  {author} {\bibinfo {author} {\bibfnamefont {Y.~T.}\ \bibnamefont
  {Hu}}, \bibinfo {author} {\bibfnamefont {D.~J.}\ \bibnamefont {Pine}}, \ and\
  \bibinfo {author} {\bibfnamefont {L.~Gary}\ \bibnamefont {Leal}},\ }\bibfield
   {title} {\enquote {\bibinfo {title} {Drop deformation, breakup, and
  coalescence with compatibilizer},}\ }\href@noop {} {\bibfield  {journal}
  {\bibinfo  {journal} {Phys. Fluids}\ }\textbf {\bibinfo {volume} {12}},\
  \bibinfo {pages} {484} (\bibinfo {year} {2000})}\BibitemShut {NoStop}%
\bibitem [{\citenamefont {Frostad}\ \emph {et~al.}(2013)\citenamefont
  {Frostad}, \citenamefont {Walter},\ and\ \citenamefont
  {Leal}}]{FrostadWalterLeal2013_PoF}%
  \BibitemOpen
  \bibfield  {author} {\bibinfo {author} {\bibfnamefont {J.~M.}\ \bibnamefont
  {Frostad}}, \bibinfo {author} {\bibfnamefont {J.}~\bibnamefont {Walter}}, \
  and\ \bibinfo {author} {\bibfnamefont {L.~G.}\ \bibnamefont {Leal}},\
  }\bibfield  {title} {\enquote {\bibinfo {title} {A scaling relation for the
  capillary-pressure driven drainage of thin films},}\ }\href@noop {}
  {\bibfield  {journal} {\bibinfo  {journal} {Phys. Fluids}\ }\textbf {\bibinfo
  {volume} {25}},\ \bibinfo {pages} {052108} (\bibinfo {year}
  {2013})}\BibitemShut {NoStop}%
\bibitem [{\citenamefont {Spjut}(2010)}]{Spjut2010_MSThesis_Chapter3}%
  \BibitemOpen
  \bibfield  {author} {\bibinfo {author} {\bibfnamefont {J.~E.}\ \bibnamefont
  {Spjut}},\ }\emph {\bibinfo {title} {Trapping, deformation, and dynamics of
  phospholipid vesicles}},\ \href@noop {} {Master's thesis},\ \bibinfo
  {school} {University of California, Berkeley} (\bibinfo {year}
  {2010})\BibitemShut {NoStop}%
\bibitem [{\citenamefont {Bentley}\ and\ \citenamefont
  {Leal}(1986)}]{BentleyLeal1986_JFMa}%
  \BibitemOpen
  \bibfield  {author} {\bibinfo {author} {\bibfnamefont {B.~J.}\ \bibnamefont
  {Bentley}}\ and\ \bibinfo {author} {\bibfnamefont {L.~G.}\ \bibnamefont
  {Leal}},\ }\bibfield  {title} {\enquote {\bibinfo {title} {A
  computer-controlled four-roll mill for investigations of particle and drop
  dynamics in two-dimensional linear shear flows},}\ }\href@noop {} {\bibfield
  {journal} {\bibinfo  {journal} {J. Fluid Mech.}\ }\textbf {\bibinfo {volume}
  {167}},\ \bibinfo {pages} {219--240} (\bibinfo {year} {1986})}\BibitemShut
  {NoStop}%
\bibitem [{\citenamefont {Johnson-Chavarria}\ \emph {et~al.}(2011)\citenamefont
  {Johnson-Chavarria}, \citenamefont {Tanyeri},\ and\ \citenamefont
  {Schroeder}}]{Johnson-Chavarria2011_EMJ}%
  \BibitemOpen
  \bibfield  {author} {\bibinfo {author} {\bibfnamefont {E.M.}\ \bibnamefont
  {Johnson-Chavarria}}, \bibinfo {author} {\bibfnamefont {M.}~\bibnamefont
  {Tanyeri}}, \ and\ \bibinfo {author} {\bibfnamefont {C.M.}\ \bibnamefont
  {Schroeder}},\ }\bibfield  {title} {\enquote {\bibinfo {title} {{A
  Microfluidic-Based Hydrodynamic Trap for Single Particles}},}\ }\href@noop {}
  {\bibfield  {journal} {\bibinfo  {journal} {Journal of Visualized
  Experiments}\ }\textbf {\bibinfo {volume} {47}} (\bibinfo {year}
  {2011})}\BibitemShut {NoStop}%
\bibitem [{\citenamefont {Breyiannis}\ and\ \citenamefont
  {Pozrikidis}(2000)}]{Breyiannis-Pozrikidis:2000}%
  \BibitemOpen
  \bibfield  {author} {\bibinfo {author} {\bibfnamefont {G.}~\bibnamefont
  {Breyiannis}}\ and\ \bibinfo {author} {\bibfnamefont {C.}~\bibnamefont
  {Pozrikidis}},\ }\bibfield  {title} {\enquote {\bibinfo {title} {Simple shear
  flow of suspensions of elastic capsules},}\ }\href@noop {} {\bibfield
  {journal} {\bibinfo  {journal} {Theoret. Comput. Fluid Dynamics}\ }\textbf
  {\bibinfo {volume} {13}},\ \bibinfo {pages} {327--347} (\bibinfo {year}
  {2000})}\BibitemShut {NoStop}%
\bibitem [{\citenamefont {Lac}\ \emph {et~al.}(2007)\citenamefont {Lac},
  \citenamefont {Morel},\ and\ \citenamefont
  {Barthes-Biesel}}]{LacMorelBarthes-Biesel2007_JFM}%
  \BibitemOpen
  \bibfield  {author} {\bibinfo {author} {\bibfnamefont {E.}~\bibnamefont
  {Lac}}, \bibinfo {author} {\bibfnamefont {A.}~\bibnamefont {Morel}}, \ and\
  \bibinfo {author} {\bibfnamefont {D.}~\bibnamefont {Barthes-Biesel}},\
  }\bibfield  {title} {\enquote {\bibinfo {title} {Hydrodynamic interaction
  between two identical capsules in simple shear flow},}\ }\href@noop {}
  {\bibfield  {journal} {\bibinfo  {journal} {J. Fluid Mech.}\ }\textbf
  {\bibinfo {volume} {573}},\ \bibinfo {pages} {149--169} (\bibinfo {year}
  {2007})}\BibitemShut {NoStop}%
\bibitem [{\citenamefont {Lac}\ and\ \citenamefont
  {Barthes-Biesel}(2008)}]{LacBarthes-Biesel2008_PoF}%
  \BibitemOpen
  \bibfield  {author} {\bibinfo {author} {\bibfnamefont {E.}~\bibnamefont
  {Lac}}\ and\ \bibinfo {author} {\bibfnamefont {D.}~\bibnamefont
  {Barthes-Biesel}},\ }\bibfield  {title} {\enquote {\bibinfo {title} {Pairwise
  interaction of capsules in simple shear flow: Three-dimensional effects},}\
  }\href@noop {} {\bibfield  {journal} {\bibinfo  {journal} {Phys. Fluids}\
  }\textbf {\bibinfo {volume} {20}},\ \bibinfo {pages} {040801} (\bibinfo
  {year} {2008})}\BibitemShut {NoStop}%
\bibitem [{\citenamefont {Omori}\ \emph {et~al.}(2013)\citenamefont {Omori},
  \citenamefont {Ishikawa}, \citenamefont {Imai},\ and\ \citenamefont
  {Yamaguchi}}]{OmoriIshikawaImaiYamaguchi2013_JBiomechics}%
  \BibitemOpen
  \bibfield  {author} {\bibinfo {author} {\bibfnamefont {T.}~\bibnamefont
  {Omori}}, \bibinfo {author} {\bibfnamefont {T.}~\bibnamefont {Ishikawa}},
  \bibinfo {author} {\bibfnamefont {Y.}~\bibnamefont {Imai}}, \ and\ \bibinfo
  {author} {\bibfnamefont {T.}~\bibnamefont {Yamaguchi}},\ }\bibfield  {title}
  {\enquote {\bibinfo {title} {Membrane tension of red blood cells pairwisely
  interacting in simple shear flow},}\ }\href@noop {} {\bibfield  {journal}
  {\bibinfo  {journal} {J. Biomechanics}\ }\textbf {\bibinfo {volume} {46}},\
  \bibinfo {pages} {548--553} (\bibinfo {year} {2013})}\BibitemShut {NoStop}%
\bibitem [{\citenamefont {Rahimian}\ \emph {et~al.}(2015)\citenamefont
  {Rahimian}, \citenamefont {Veerapaneni}, , \citenamefont {Zorin},\ and\
  \citenamefont {Biros}}]{RahimianVeerapaneniZorinBiros2015_JCP}%
  \BibitemOpen
  \bibfield  {author} {\bibinfo {author} {\bibfnamefont {A.}~\bibnamefont
  {Rahimian}}, \bibinfo {author} {\bibfnamefont {S.~K.}\ \bibnamefont
  {Veerapaneni}}, , \bibinfo {author} {\bibfnamefont {D.}~\bibnamefont
  {Zorin}}, \ and\ \bibinfo {author} {\bibfnamefont {G.}~\bibnamefont
  {Biros}},\ }\bibfield  {title} {\enquote {\bibinfo {title} {Boundary integral
  method for the flow of vesicles with viscosity contrast in three
  dimensions},}\ }\href@noop {} {\bibfield  {journal} {\bibinfo  {journal}
  {Journal of Computational Physics}\ }\textbf {\bibinfo {volume} {298}},\
  \bibinfo {pages} {766--786} (\bibinfo {year} {2015})}\BibitemShut {NoStop}%
\bibitem [{\citenamefont {Jeffery}(1922)}]{jef1922}%
  \BibitemOpen
  \bibfield  {author} {\bibinfo {author} {\bibfnamefont {G.~B.}\ \bibnamefont
  {Jeffery}},\ }\bibfield  {title} {\enquote {\bibinfo {title} {{The Motion of
  Ellipsoidal Particles Immersed in a Viscous Fluid}},}\ }\href@noop {}
  {\bibfield  {journal} {\bibinfo  {journal} {Proceedings of the Royal Society
  A}\ }\textbf {\bibinfo {volume} {102}},\ \bibinfo {pages} {161--179}
  (\bibinfo {year} {1922})}\BibitemShut {NoStop}%
\bibitem [{\citenamefont {Fenz}\ \emph {et~al.}(2017)\citenamefont {Fenz},
  \citenamefont {Bihr}, \citenamefont {Schmidt}, \citenamefont {Merkel},
  \citenamefont {Seifert}, \citenamefont {Sengupta},\ and\ \citenamefont
  {Smith}}]{FenzBihrSchmidt2017_NaturePhys}%
  \BibitemOpen
  \bibfield  {author} {\bibinfo {author} {\bibfnamefont {S.~F.}\ \bibnamefont
  {Fenz}}, \bibinfo {author} {\bibfnamefont {T.}~\bibnamefont {Bihr}}, \bibinfo
  {author} {\bibfnamefont {D.}~\bibnamefont {Schmidt}}, \bibinfo {author}
  {\bibfnamefont {R.}~\bibnamefont {Merkel}}, \bibinfo {author} {\bibfnamefont
  {U.}~\bibnamefont {Seifert}}, \bibinfo {author} {\bibfnamefont
  {K.}~\bibnamefont {Sengupta}}, \ and\ \bibinfo {author} {\bibfnamefont
  {A.-S.}\ \bibnamefont {Smith}},\ }\bibfield  {title} {\enquote {\bibinfo
  {title} {Membrane fluctuations mediate lateral interactions between cadherin
  bonds},}\ }\href@noop {} {\bibfield  {journal} {\bibinfo  {journal} {Nature
  Phys.}\ }\textbf {\bibinfo {volume} {13}},\ \bibinfo {pages} {906--913}
  (\bibinfo {year} {2017})}\BibitemShut {NoStop}%
\bibitem [{\citenamefont {Liu}\ \emph {et~al.}(2019)\citenamefont {Liu},
  \citenamefont {Chu}, \citenamefont {Newby}, \citenamefont {Read},
  \citenamefont {Lowengrub},\ and\ \citenamefont
  {Allard}}]{LiuChuNewbyRead2018_bioRxiv}%
  \BibitemOpen
  \bibfield  {author} {\bibinfo {author} {\bibfnamefont {K.}~\bibnamefont
  {Liu}}, \bibinfo {author} {\bibfnamefont {B.}~\bibnamefont {Chu}}, \bibinfo
  {author} {\bibfnamefont {J.}~\bibnamefont {Newby}}, \bibinfo {author}
  {\bibfnamefont {E.~L.}\ \bibnamefont {Read}}, \bibinfo {author}
  {\bibfnamefont {J.}~\bibnamefont {Lowengrub}}, \ and\ \bibinfo {author}
  {\bibfnamefont {J.}~\bibnamefont {Allard}},\ }\bibfield  {title} {\enquote
  {\bibinfo {title} {{Hydrodynamics of transient cell-cell contact: The role of
  membrane permeability and active protrusion length}},}\ }\href@noop {}
  {\bibfield  {journal} {\bibinfo  {journal} {PLOS Computational Biology}\
  }\textbf {\bibinfo {volume} {15}},\ \bibinfo {pages} {e1006352} (\bibinfo
  {year} {2019})}\BibitemShut {NoStop}%
\bibitem [{\citenamefont {Alpert}(1999)}]{alp1999}%
  \BibitemOpen
  \bibfield  {author} {\bibinfo {author} {\bibfnamefont {B.~K.}\ \bibnamefont
  {Alpert}},\ }\bibfield  {title} {\enquote {\bibinfo {title} {{Hybrid
  Gauss-Trapezoidal Quadrature Rules}},}\ }\href@noop {} {\bibfield  {journal}
  {\bibinfo  {journal} {SIAM Journal on Scientific Computing}\ }\textbf
  {\bibinfo {volume} {20}},\ \bibinfo {pages} {1551--1584} (\bibinfo {year}
  {1999})}\BibitemShut {NoStop}%
\bibitem [{\citenamefont {Trefethen}\ and\ \citenamefont
  {Weideman}(2014)}]{tre-wei2014}%
  \BibitemOpen
  \bibfield  {author} {\bibinfo {author} {\bibfnamefont {L.~N.}\ \bibnamefont
  {Trefethen}}\ and\ \bibinfo {author} {\bibfnamefont {J.~A.~C.}\ \bibnamefont
  {Weideman}},\ }\bibfield  {title} {\enquote {\bibinfo {title} {{The
  Exponentially Convergent Trapezoidal Rule}},}\ }\href@noop {} {\bibfield
  {journal} {\bibinfo  {journal} {SIAM Review}\ }\textbf {\bibinfo {volume}
  {56}},\ \bibinfo {pages} {385--458} (\bibinfo {year} {2014})}\BibitemShut
  {NoStop}%
\end{thebibliography}%

\end{document}